\documentclass[11pt]{article}
\usepackage[margin=1in]{geometry}
\usepackage[T1]{fontenc}
\usepackage{lmodern,microtype,amsmath,amssymb,amsthm,mathtools}
\usepackage{booktabs,array,xcolor,enumitem,hyperref,caption,tikz,placeins,float}
\floatstyle{ruled}
\newfloat{algorithm}{tbp}{loa}
\floatstyle{plain}
\floatname{algorithm}{Algorithm}

\newlist{pseudolines}{enumerate}{1}
\setlist[pseudolines]{label=\arabic*:,ref=\arabic*,leftmargin=2.5em,
  labelsep=.6em,itemsep=0pt,parsep=0pt,topsep=3pt}
\newcommand{\codeguide}{\vrule width .35pt\hspace{\dimexpr1em-.35pt\relax}}
\newcommand{\codeline}[2]{\item\leavevmode\hbox{%
  \ifnum#1>0 \codeguide\fi\ifnum#1>1 \codeguide\fi
  \parbox[t]{\dimexpr\linewidth-#1em\relax}{\raggedright\strut #2\strut}}}
\newcommand{\codephase}[1]{\item[]\textbf{\underline{#1}}}

\usetikzlibrary{arrows.meta,positioning,calc}
\hypersetup{pdftitle={The Complexity of Submodular Matching in One-Pass Streams},
 pdfauthor={Mohammad Hajiaghayi, Morteza Monemizadeh, Xinchen Yang},colorlinks=true,hypertexnames=false,
 linkcolor=blue!45!black,citecolor=blue!45!black,urlcolor=blue!45!black}
\setlist{topsep=4pt,itemsep=3pt}
\newtheorem{theorem}{Theorem}[section]
\newtheorem{lemma}[theorem]{Lemma}
\newtheorem{proposition}[theorem]{Proposition}
\newtheorem{corollary}[theorem]{Corollary}
\theoremstyle{definition}
\theoremstyle{remark}
\newcommand{\OPT}{\operatorname{OPT}}
\newcommand{\Ex}[2][]{\mathbb{E}\if\relax\detokenize{#1}\relax\else_{#1}\fi\!\left[#2\right]}
\newcommand{\Prb}{\mathbb P}

\newcommand{\Pjoint}{\mathcal P_D}

\title{\textbf{The Complexity of Submodular Matching \\in One-Pass Streams}}

\usepackage{tcolorbox}
\newtcolorbox{mainresult}{colback=blue!2!white,colframe=blue!35!black,
  boxrule=.55pt,arc=1pt,left=8pt,right=8pt,top=0pt,bottom=0pt,
  before skip=9pt,after skip=9pt}

\author{
MohammadTaghi Hajiaghayi\thanks{\texttt{hajiagha@umd.edu}}\\
UMD\\
\and
Morteza Monemizadeh\thanks{\texttt{m.monemizadeh@tue.nl}}\\
TU/e
\and
Xinchen Yang\thanks{\texttt{xcyang@umd.edu}}\\
UMD
}

\date{\today}

\newcommand{\NMFactor}{4.727}
\newcommand{\NMDegree}{\tfrac{13}{40}}
\newcommand{\NMVertexTangent}{\tfrac15}
\newcommand{\NMEdgeTangent}{\tfrac{13}{20}}
\newcommand{\NMSlacks}{(1325,\,1741,\,2261,\,22477)}

\begin{document}
\date{}
\maketitle

\begin{abstract}
We study maximum monotone submodular matching when the edges of a graph arrive in a single pass in adversarial order. On general simple graphs, we give a randomized $4.525$-approximation, improving the factor $3+2\sqrt{2}\approx 5.828$ of Levin and Wajc~\cite{LW}. The algorithm uses $O((\nu+1)\log^2 n)$ words, where $\nu$ is the maximum matching cardinality. It runs in polynomial time, makes polynomially many value queries, and achieves the stated factor both in expectation and with high probability.

Our lower bound separates monotone submodular matching from cardinality
and weighted matching in the one-pass semi-streaming model. For maximum
cardinality matching, Assadi, Jiang, and Xiang~\cite{ajx2} proved that
the greedy algorithm's factor of $2$ is optimal. For maximum weight
matching, Paz and Schwartzman~\cite{pazschwartzman} achieve a factor of
$2+\varepsilon$ for every fixed $\varepsilon>0$. For monotone submodular
matching, Feldman et al.~\cite{fnsz} proved a lower bound of $2.692$
on bipartite multigraphs. We prove that even on simple bipartite graphs,
no randomized one-pass semi-streaming algorithm can achieve any fixed
approximation factor below $3$.
\end{abstract}

\section{Introduction}
\label{sec:introduction}

Maximum submodular matching (MSM) models one-to-one assignments with
diminishing returns, with applications in machine learning and resource
allocation. In a recommender system, for example, a platform may use
learned relevance scores to assign distinct items to users while also
rewarding the range of topics represented. Two items may each be highly
relevant to their respective users, yet recommending both adds less
value when their topics overlap. Thus, the value of an assignment depends on both its individual relevance and what the other selected assignments already contribute.

To formalize this dependence, let $G=(V,E)$ be an undirected graph and
let $f:2^E\to\mathbb{R}_{\geq 0}$ assign a value to each set of edges.
A \emph{matching} is a set of edges with pairwise disjoint endpoints.
The \emph{marginal gain} of adding an edge $e$ to a set $A$ is
$f_A(e)=f(A\cup\{e\})-f(A)$. In the recommendation example, this gain
includes the item's relevance score and a reward for any topics it adds
to those already represented. A function $f$ is \emph{submodular} if
$f_A(e)\geq f_B(e)$ whenever $A\subseteq B$ and $e\notin B$, and
\emph{monotone} if $f(A)\leq f(B)$ whenever $A\subseteq B$. In
\emph{maximum monotone submodular matching}, the goal is to find a
matching $M$ maximizing $f(M)$.

When the objective is additive, MSM specializes to \emph{maximum weight
matching}, a classical problem in combinatorial optimization that models
assignments such as pairing workers with jobs or tasks with machines.
Each edge $e$ has a nonnegative weight $w(e)$, and the goal is to find a
matching $M$ maximizing $\sum_{e\in M}w(e)$.  Unit weights give \emph{maximum cardinality
matching}. Unlike additivity—where selecting an edge only rules out incident edges while keeping every remaining edge's marginal gain fixed at $w(e)$—a submodular objective allows selecting an edge to shift the relative value of remaining candidates, even edges disjoint from it. For example, suppose two ads compete for the same slot, and the second
has a smaller total audience than the first. If an ad chosen for another
slot already reaches many of the first ad's viewers but few of the
second's, the second ad may reach more new viewers than the first.

We study maximum monotone submodular matching in the one-pass
semi-streaming setting~\cite{FKMSZ}. The edges of an $n$-vertex graph
arrive one at a time in adversarial order, and the algorithm processes
them in a single pass using $O(n\,\mathrm{polylog}\,n)$ bits of memory.
This space suffices to store a matching, but a dense graph can have
$\Theta(n^2)$ edges. The algorithm must therefore decide which edges to
retain without knowing how later arrivals will affect their marginal
gains. In the advertising example, it may discard one of the two
candidate placements before seeing the placement that reverses their
ranking.

For maximum cardinality matching, the greedy algorithm achieves a factor
of $2$ in one pass. Assadi, Jiang, and Xiang~\cite{ajx2} proved that
this factor is optimal even for randomized semi-streaming algorithms on
bipartite graphs. Taking all edge weights equal to one gives the same
lower bound for maximum weight matching. Conversely, Paz and
Schwartzman~\cite{pazschwartzman} gave a deterministic one-pass
$(2+\varepsilon)$-approximation for every fixed $\varepsilon>0$.
Thus weighted matching admits approximation factors arbitrarily close
to the optimum factor for cardinality matching. Ghaffari and
Wajc~\cite{GW} subsequently simplified the analysis of the weighted
algorithm and reduced its space usage.

For general monotone submodular objectives, Chakrabarti and Kale~\cite{CK}
gave a $7.75$-approximation in one pass and a $(3+\varepsilon)$-approximation
using multiple passes. Levin and Wajc~\cite{LW} improved the one-pass
factor to $3+2\sqrt{2}\approx 5.828$.

We improve the one-pass approximation factor for monotone submodular
matching on general simple graphs to $4.525$. Our randomized
algorithm achieves this factor both in expectation and with high
probability.

The one-pass approximation threshold changes when the objective admits
diminishing returns. Feldman et al.~\cite{fnsz} proved a lower bound of
$2.692$ on bipartite multigraphs. We prove a lower bound of $3$ for
randomized one-pass semi-streaming algorithms even on simple bipartite
graphs. On this same graph class a factor of $2$ suffices for
cardinality matching, and factors arbitrarily close to $2$ suffice for
weighted matching, whereas every fixed factor below $3$ is impossible
for monotone submodular matching. Table~\ref{tab:bounds} summarizes the
bounds for the submodular problem.

\begin{table}[htbp]
\centering\small
\caption{Selected one-pass bounds for monotone submodular matching under edge arrivals. An upper bound $R$ guarantees a $1/R$ fraction of optimum in the stated sense. A lower threshold $R$ rules out every fixed factor smaller than $R$ under the stated assumptions.}
\label{tab:bounds}
\begin{tabular}{@{}p{.34\linewidth}p{.17\linewidth}p{.425\linewidth}@{}}
\toprule
Result & Factor & Guarantee or scope\\
\midrule
\multicolumn{3}{@{}l}{\emph{Algorithmic guarantees on general simple graphs}}\\
Chakrabarti and Kale~\cite{CK} & $7.75$ & Deterministic\\
Levin and Wajc~\cite{LW} & $3+2\sqrt2$ & Deterministic and randomized (in expectation)\\
This work (Thm.~\ref{thm:main-upper}) & $4.525$ & Randomized (in expectation and with high probability)\\
\midrule
\multicolumn{3}{@{}l}{\emph{Lower bounds}}\\
Levin and Wajc~\cite{LW} & $1.914$ & Simple graphs, polynomial time, conditional on ETH\\
Assadi et al.~\cite{ajx2} & $2$ & Simple bipartite graphs, already for cardinality\\
Feldman et al.~\cite{fnsz} & $2.692$ & Bipartite multigraphs, element storage\\
This work (Thm.~\ref{thm:main-lower}) & $3$ & Simple bipartite graphs, arbitrary encoded state\\
\bottomrule
\end{tabular}
\end{table}

\subsection{Our results}

Let $G=(V,E)$ be a finite undirected graph, let $n=\max\{2,|V|\}$, and let $\nu$ denote its
maximum matching cardinality. A simple graph has no loops and at most one edge between each
pair of vertices. In a multigraph, distinct edges may have the same endpoints. For a
nonnegative monotone submodular function $f:2^E\to\mathbb R_{\ge0}$, write
$\OPT=\max\{f(M):M\text{ is a matching in }G\}$. An $R$-approximation returns a matching of
value at least $\OPT/R$ in the probabilistic sense specified in each theorem.

\paragraph{Adversary.}
Before the algorithm draws any random bits, the adversary fixes the
graph, the objective, and the order of the edges. The algorithm then
sees each edge exactly once in that order. In particular, the adversary
cannot change later arrivals in response to the algorithm's choices.
For each fixed input, the expectation and success probability in
Theorem~\ref{thm:main-upper} are over the algorithm's random choices.

\paragraph{Value-oracle access.}
The algorithm learns about $f$ by specifying a set $A$ of edges that
have already arrived and receiving the exact value $f(A)$. It may query
sets containing edges that share an endpoint. Its output must always be
a matching of arrived edges. All working storage counts toward the
space bound, including the workspace for queries, encoded summaries,
and candidate outputs. In the value-oracle arithmetic model used for
the upper bound, each original objective scale occupies one arithmetic
word, and indices relative to it use $O(\log n)$ bits. When each oracle
answer fits in $O(1)$ words of $O(\log n)$ bits each, both theorems
use the same finite-word model.

Our two main theorems bound the approximation factor achievable in one
pass. Theorem~\ref{thm:main-upper} gives a randomized
$4.525$-approximation on general simple graphs.
Theorem~\ref{thm:main-lower} rules out every fixed factor below $3$
even when the graph is simple and bipartite.

\newsavebox{\matchingresultbox}
\renewenvironment{mainresult}{%
  \par\addvspace{9pt}\noindent
  \begin{lrbox}{\matchingresultbox}%
  \begin{minipage}{\dimexpr\linewidth-19pt\relax}%
}{%
  \end{minipage}\end{lrbox}%
  \begin{tikzpicture}
    \node[fill=black!7,draw=black,inner xsep=7pt,inner ysep=5pt,
          line width=.5pt,outer sep=0pt] (result) {\usebox{\matchingresultbox}};
    \fill[black] (result.south west) -- ++(1.5pt,-1.5pt)
      -- ([shift={(1.5pt,-1.5pt)}]result.south east)
      -- ([shift={(1.5pt,-1.5pt)}]result.north east)
      -- (result.north east) -- (result.south east) -- cycle;
  \end{tikzpicture}\par\addvspace{9pt}%
}

\begin{mainresult}
\begin{theorem}[One-pass algorithm]\label{thm:main-upper}
On every simple graph, a randomized one-pass algorithm uses $O((\nu+1)\log^2 n)$ words
and returns a matching $M$ satisfying
\[
 \Ex{f(M)}\ge\frac{\OPT}{4.525}.
\]
For every fixed $c>0$, the same space bound supports
\[
 \Prb\!\left[f(M)\ge\frac{\OPT}{4.525}\right]\ge1-n^{-c}.
\]
The algorithm runs in polynomial time and makes polynomially many value queries.
It requires no advance knowledge of $\nu$ or of the range of $f$.
\end{theorem}
\end{mainresult}

Theorem~\ref{thm:main-upper} holds for general simple graphs, not
necessarily bipartite, and arbitrary nonnegative monotone submodular
functions. More generally, for any fixed $\varepsilon>0$,
Corollary~\ref{cor:fixed-source-family} establishes a
$(4.5246551+\varepsilon)$-approximation.
While the randomized algorithm of Levin and Wajc~\cite{LW} runs
in linear time, our algorithm achieves an improved approximation
factor through polynomial-time postprocessing of the stored edges,
without requiring an additional pass over the stream.

Our lower bound holds even on simple bipartite graphs with exact
oracle answers that fit in a constant number of words.
It applies to algorithms with unrestricted computation and an
arbitrary number of legal value-oracle queries, regardless of
how they encode the information stored in memory.

\begin{mainresult}
\begin{theorem}[One-pass lower bound]\label{thm:main-lower}
Fix $1\le R<3$ and $a\ge0$. No randomized one-pass algorithm using $O(n\log^a n)$ words of
$O(\log n)$ bits returns an $R$-approximate matching with fixed positive probability on every
simple bipartite graph. Nor can it guarantee $\Ex{f(M)}\ge\OPT/R$ on every input.

The hard instances have $f(\varnothing)=0$, exact oracle answers of $O_R(1)$ words, and
$\nu=\Theta(n)$ on every realization. Their vertex count and stream length are fixed and may be
announced in advance.
\end{theorem}
\end{mainresult}

Feldman et al.~\cite[Theorem~4]{fnsz} proved a lower bound of $2.692$
on bipartite multigraphs. Their construction uses parallel edges that can
make different contributions to $f$. Merging them would remove a choice
available to the optimum matching. Theorem~\ref{thm:main-lower} raises the
lower bound to $3$ even on simple bipartite graphs. Whether an algorithm
can achieve a factor of exactly $3$ remains open.

When the algorithm accepts an edge, it records the expected value that
the edge adds to a random sample of the previously accepted edges. It
also updates a potential at each endpoint. For each discarded optimum
edge, the sum of its endpoint potentials bounds its remaining expected
marginal value. These bounds can be summed because edges in the optimum
matching have disjoint endpoints.

After the pass, the algorithm uses the retained edges to build one
matching in two steps. It first chooses a matching $T$ for the value it
adds to a random sample of the retained edges. It then uses the recorded
weights to select additional edges on vertices left free by $T$.
An edge may now add less than its recorded weight suggests because $T$
already supplies some of its value. We track these decreases in arrival
order to bound how much value the two steps count twice.

The lower-bound construction hides a bipartite matching in a stream that
arrives in stages. Each stage has labeled edges. Edges with one
distinguished label contribute value while preserving the contribution of
later stages. During the stage, value queries cannot reveal which label
has this property. To represent the labels on a simple graph, we replace
each vertex by copies and assign labels to distinct edges between copies.
We sample the original edges and independently sample the edges between
copies. Even if the algorithm remembers that an original edge $uv$
survived, it may later need to output a particular copy edge $u_1v_2$.
Its memory must then establish that this particular edge arrived.

\subsection{Related work}

The semi-streaming model permits nearly linear space in the number of vertices~\cite{FKMSZ}.
For maximum weight matching, Paz and Schwartzman~\cite{pazschwartzman} gave a one-pass
$(2+\varepsilon)$-approximation using local ratio. Ghaffari and Wajc~\cite{GW} simplified its
analysis and reduced its space use. For submodular matching, Chakrabarti and Kale~\cite{CK}
showed how to adapt weighted streaming methods to diminishing returns. Levin and Wajc~\cite{LW}
used endpoint potentials to decide which edges to keep. Our postprocessing combines the weights obtained from this type of selection rule with fresh value queries on the retained edges. Related results cover $b$-matching~\cite{HS}, two matroid
constraints~\cite{GJS}, and more general or nonmonotone objectives~\cite{CGQ,FKK}.

Levin and Wajc~\cite{LW} derived a lower bound for polynomial-time streaming algorithms under
the Exponential Time Hypothesis (ETH). Building on one-way communication lower bounds for
submodular maximization~\cite{FNSZ2023}, Feldman et al.~\cite{fnsz} combined a hard
cardinality-matching instance with a submodular objective on parallel copies of each edge. In
their element-storage model, the algorithm queries sets of explicitly stored elements. Their
reduction adds nearly one to the cardinality hardness factor. Assadi, Jiang, and Xiang
introduced the blueprint framework~\cite{ajx1} and used it to prove that the greedy factor of
two for cardinality matching is optimal~\cite{ajx2}. Our construction must also pay for the
additional vertices needed to realize different labels in a simple graph, while permitting
queries on arbitrary arrived edges.

The multilinear extension and continuous greedy turn submodular maximization into an
optimization problem over fractional vectors~\cite{CCPV2011,CVZ}. We use Feldman's
time-dependent weight~\cite{FeldmanDistorted} to keep the full linear contribution. For
bipartite graphs, dependent rounding can preserve edge marginals and control correlations at a
vertex~\cite{GKPS2006}. Here the proof needs a bound on the expected value of an arbitrary
monotone submodular function on a general simple graph. Such comparisons and correlations of
unmatched vertices have also been studied in~\cite{QS,AASV}. We use the distribution from
Shearer's lemma~\cite{Shearer,GJL2019} for the first matching, and a Gibbs distribution for the
matching on the remaining vertices. The latter uses the matching-sampling framework
of~\cite{JS1989,JSV2004,CLV2026} and Sinclair's chain~\cite{sinclair1992}. Our samplers may
choose any stored edge. A contention resolution scheme has the additional requirement that the
output lie in a previously sampled set~\cite{CVZ,bz,NV}.

With offline access, local search gives a $(2+\varepsilon)$-approximation via $2$-exchange
systems~\cite{FNSW}. Multiple passes also help: Chakrabarti and Kale~\cite{CK} gave a
$(3+\varepsilon)$-approximation, and Huang, Thiery, and Ward~\cite{HTW} obtained this factor in
$O(1/\varepsilon)$ passes. Random-order streams allow different
algorithms~\cite{RandomOrder2026}. For cardinality matching in dynamic streams, Assadi et
al.~\cite{ABKNS2026} determine the pass complexity of constant-factor approximation. With edge
insertions and deletions, Banihashem et al.~\cite{DynamicMSM} maintain an expected
$(8+\varepsilon)$-approximation for submodular matching with polylogarithmic expected amortized
update time against an oblivious adversary.

Section~\ref{sec:techniques} outlines the proofs of the two main theorems.
Section~\ref{sec:upper-framework} gives the upper-bound algorithm and
its three principal comparisons. Section~\ref{sec:lb-lower} proves
the simple-graph lower bound. The appendices supply the streaming implementation, both matching samplers, the fractional optimization,
and the finite error analysis. All numerical choices are made together
in Appendix~\ref{sec:joint-exact-parameters}.

\section{Overview of the proofs}
\label{sec:techniques}
For the upper bound, we compare the optimum with two quantities defined on the stored graph.
The main step is to recover both in a single matching. For the lower bound, we bound the
number of edges whose presence the algorithm can certify after a stream prefix. If two
prefixes yield the same memory state, the output under a common continuation and common
random bits must be feasible for both.

\subsection{The upper bound: two functions on the stored edges}
\label{sec:upper-techniques}

An edge can have a large marginal value at arrival and contribute little to the final stored
set. Its arrival value still matters because it determined the endpoint potentials used for
subsequent decisions. We therefore retain these weights and define a second objective that
measures the additional value of a matching relative to a sample of the final stored set.

For the overview assume $f(\varnothing)=0$ and exact expectations.
Section~\ref{sec:upper-framework} accounts for finite-precision errors. Fix a small constant
$p>0$. Write $S_{<e}$ for the stored edges preceding $e$, and $R_p(A)$ for a subset formed by
retaining each member of $A$ independently with probability $p$. For an accepted edge $e$,
denote its expected marginal value at arrival by $w_e$. Telescoping the contributions of the
sampled edges gives
\[
 w_e=\Ex{f(e\mid R_p(S_{<e}))},
 \qquad G_S=p\sum_{e\in S}w_e=\Ex{f(R_p(S))}.
\]
The sum follows the arrival order. The stored set $S$ need not be a matching. Define
\[
 g(T)=\Ex{f(T\cup R_p(S))-f(R_p(S))}.
\]
Thus $G_S$ measures the expected value of the random subset, and $g(T)$
measures what $T$ adds to it.  The function $g$ is again monotone and
submodular, and its oracle can be simulated using stored edges.

\paragraph{Comparison with the optimum.}
Fix an optimal matching and let $O$ be its stored part.  The streaming
analysis will identify a nonnegative quantity $X(O)$ such that
\[
 \OPT\le X(O)+g(O).
 \tag{AT1}
\]
Our goal is therefore a single matching whose expected value is at least
$\sigma X(O)+\beta g(O)$.  This would give approximation factor
$1/\min\{\sigma,\beta\}$.  The streaming analysis determines $\sigma$;
the coefficient of $g(O)$ is $\beta=1-e^{-D}$ for a fractional-degree bound $D$ just below
$1/4$.  Choosing $\sigma$ slightly below $\beta$ leaves room for the
finite losses and gives the stated factor $4.525$.  The exact
parameter choice appears in Appendix~\ref{sec:joint-exact-parameters}.

\paragraph{Charging discarded edges.}
As in Levin and Wajc~\cite{LW}, the algorithm maintains a nonnegative potential at each vertex.
It accepts $e=uv$ only if $w_e$ is at least the sum of the potentials at $u$ and $v$. Upon
acceptance, it increases each potential by a multiple of $pw_e$ determined by the relative
sizes of the two potentials. The update uses only information available at arrival.

For a rejected optimum edge, the marginal value at the end of the stream is at most the sum of its endpoint potentials at rejection.  Because the rejected edges form a matching and
avoid every vertex used by $O$, their total charge uses only vertices unmatched by $O$.  If $\psi_v$ is the accumulated increase in vertex potential at $v$, then
\[
 X(O)=G_S+\sum_{v\notin V(O)}\psi_v.
\]
This proves (AT1).  It also explains why the distinction between vertices matched and unmatched by $O$ matters later: the stored part of the optimum
uses exactly the vertices whose terms are absent from this sum.

The same vertex potentials bound memory.  Every accepted incidence increases a
positive endpoint potential by a fixed multiplicative factor.  At one
objective scale, only $O(\log n)$ such incidences are possible at a
vertex.  The endpoints of a maximum-cardinality matching cover all
input edges, so $S$ has $O(\nu\log n)$ edges.  Standard overlapping
scales remove the need to know the largest singleton value, using
$O((\nu+1)\log^2 n)$ words in total.

\paragraph{A multilinear relaxation.} Write $\delta(v)$ for the edges incident to $v$ and $y(\delta(v))=\sum_{e\ni v}y_e$. Let $y$ be a fractional solution for the function $g$, with fractional degree $\ell_v=y(\delta(v))\le D$ at every vertex. Let $x$ be a second edge vector for the linear function $w\cdot x$.  We require
\[
 x(\delta(v))+\ell_v\le1,
 \qquad x_{uv}\le p(1-\ell_u-\ell_v).
 \tag{AT2}
\]
The first condition shares the vertex capacity.  The second anticipates
rounding: both endpoints of $uv$ will be unmatched by $T$ with probability at least
$1-\ell_u-\ell_v$, so dividing by that probability will keep the rescaled coordinate at most $p$.

Optimize the objective $w\cdot x+F_g(y)$ over these constraints,
where $R_y$ includes each stored edge $e$ independently with probability $y_e$ and $F_g(y)=\Ex{g(R_y)}$.
For comparison, fix $y=D\mathbf1_O$ and let $W_D(O)$ be the largest
value of the linear term allowed by (AT2). Continuous greedy with Feldman's weighting~\cite{FeldmanDistorted} gives
\[
 w\cdot x+F_g(y)\ge W_D(O)+(1-e^{-D})g(O).
 \tag{AT3}
\]
Only the submodular term incurs the usual continuous-greedy loss.  The
linear term keeps coefficient one.  A proof by linear programming duality
shows that our updates to vertex potentials guarantee $W_D(O)\ge\sigma X(O)$.
There are four local cases, according to which endpoints of an accepted
edge belong to $V(O)$.  Two auxiliary potential functions account for $pw_e$ and for the increases in vertex potentials at endpoints unmatched by $O$.  The local inequalities
telescope as the stored edges are processed in arrival order.  Their symbolic form and
proof precede the numerical choice of rates.

\paragraph{An identity for marginal values.}
Adding a matching $T$ chosen for the function $g$ may reduce the marginal values of many stored edges. We retain the weights recorded during the stream and bound the total decrease. For this purpose, every coordinate used in the second rounding step must stay below $p$.  For \emph{every} set $T$ and every
vector $0\le z\le p\mathbf1_S$, the original arrival order gives
\[
 f(T)+F_{f_T}(z)\ge g(T)+w\cdot z,
 \qquad f_T(A)=f(T\cup A)-f(T).
 \tag{AT4}
\]
By submodularity, adding $T$ can only decrease each expected marginal value.  The sum of these decreases, each multiplied by
$p$, is exactly $f(T)-g(T)$.  If every coordinate is at most $p$, the decrease in the linear term is at most this total.  Thus $f(T)-g(T)$ bounds the full loss in the recorded weights after the first rounding step.

\paragraph{Two-stage rounding.}
First sample a matching $T$ with edge marginals $y$ and
$\Ex{g(T)}\ge F_g(y)$.  When the fractional degrees are at most $D<1/4$, the distribution associated with Shearer's lemma has the required marginals; disjoint edges have their product joint inclusion probabilities. It can be sampled to any prescribed accuracy in polynomial time. For a stored edge $uv$, write $q_{uv}$ for the probability that both endpoints remain unmatched by $T$.
Give it mass $x_{uv}/q_{uv}$ when they do.  This preserves its expected
mass and, by (AT2), its coordinate bound $p$.

Some fractional degrees may now exceed one. We set the coordinates incident to those vertices to zero and scale the remaining coordinates slightly. The expected fractional degree is
at most $1+O(\sqrt p)$ and its variance is $O(p)$, including in graphs
with triangles.
For any edge being charged, only its two endpoints can cause deletion.
Consequently the resulting feasible vector $z(T)$ satisfies
$\Ex{z_e(T)}\ge(1-o(1))x_e$, with no loss depending on the graph size.

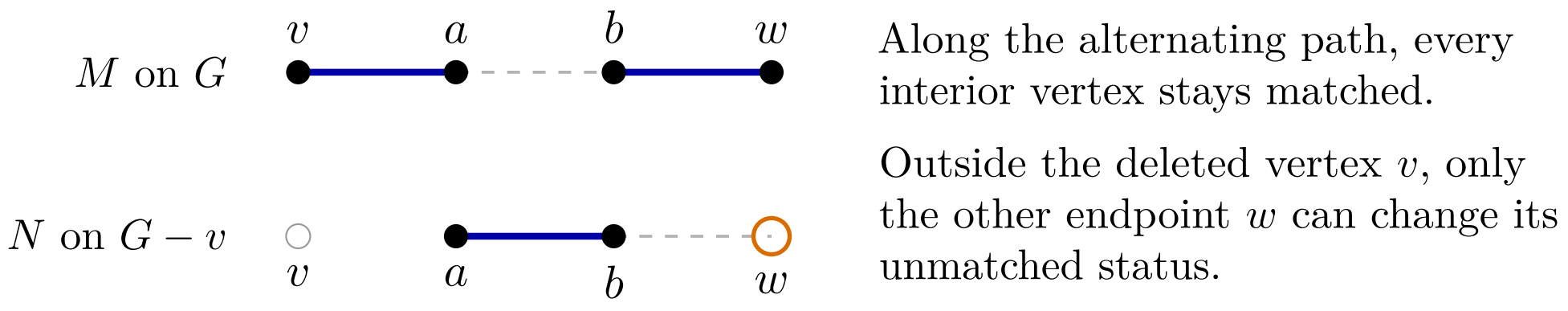
\begin{figure}[!htbp]
\centering
\begin{tikzpicture}[x=1cm,y=1cm,
 vertex/.style={circle,fill=black,inner sep=2pt},
 selected/.style={draw=blue!65!black,line width=1.6pt},
 unused/.style={draw=black!30,dashed,line width=.7pt}]
 \node[anchor=east,font=\small] at (-.45,1) {$M$ on $G$};
 \node[anchor=east,font=\small] at (-.45,-.35) {$N$ on $G-v$};
 \draw[selected] (0,1)--(1.3,1);
 \draw[unused] (1.3,1)--(2.6,1);
 \draw[selected] (2.6,1)--(3.9,1);
 \draw[selected] (1.3,-.35)--(2.6,-.35);
 \draw[unused] (2.6,-.35)--(3.9,-.35);
 \foreach \x/\lab in {0/v,1.3/a,2.6/b,3.9/w}{
   \node[vertex,label=above:$\lab$] at (\x,1) {};
 }
 \foreach \x/\lab in {1.3/a,2.6/b}{
   \node[vertex,label=below:$\lab$] at (\x,-.35) {};
 }
 \node[circle,draw=black!40,inner sep=2pt,label=below:$v$]
       at (0,-.35) {};
 \node[circle,draw=orange!85!black,line width=1pt,inner sep=3pt,
       label=below:$w$] at (3.9,-.35) {};
 \node[anchor=west,text width=5.6cm,align=left,font=\small]
       at (4.65,.35)
 {Along the alternating path, every interior vertex stays matched.\\[5pt]
 Outside the deleted vertex $v$, only the other endpoint $w$ can
 change its unmatched status.};
\end{tikzpicture}
\caption{The vertex-deletion coupling for the Gibbs distribution.
The symmetric difference can be a longer alternating path; the same
observation applies.  Edges outside the path agree.}
\label{fig:upper-deletion}
\end{figure}

It remains to round $z(T)$ while retaining almost all of
$F_{f_T}(z(T))$.  We prove that any fractional matching in a simple
graph whose individual coordinates are sufficiently small admits such
rounding for every monotone submodular objective.  We use a Gibbs distribution on matchings, assigning each matching probability proportional to the product of its edge activities~\cite{JS1989,CLV2026}.  Deleting a vertex can be coupled to change the
unmatched status of at most one other vertex
(Figure~\ref{fig:upper-deletion}).  This elementary observation controls
the variance of weighted sums of unmatched indicators.  It then controls
how much conditioning on one edge changes the relevant submodular marginal values.  Choosing the edge activities gives the desired comparison
with the fractional vector.

The graph must be simple in this rounding statement.  Replacing each
edge of a triangle by many parallel copies would make all coordinates
small while preserving a fractional cardinality of $3/2$ and an
integral optimum of one.  On a simple graph, small coordinates instead
make the dependence error vanish.  Applying this rounding on the vertices unmatched by $T$ gives a matching $M$ containing $T$.  Applying (AT4) to the remaining vector bounds the decrease in marginal values. Equations (AT1)--(AT3) then give the approximation guarantee.

\FloatBarrier

\subsection{The lower bound on simple graphs}
\label{sec:lb-techniques}
Two properties of the construction determine the factor $3$. In each stage, a small memory
can recover only about half as many edges as the hidden matching provides. Moreover, most
recovered edges reduce the contribution of later stages. The first property follows from a
cardinality-matching construction. The second uses a hidden edge label, which we realize on
a simple graph by replacing each vertex with copies.

\paragraph{A stage with a hidden large matching.}
The blueprint construction of Assadi, Jiang, and Xiang
\cite{ajx1,ajx2} supplies a bipartite graph with $N$ vertices on each
side and a canonical matching of size $H\simeq 2N/3$.  Its edges arrive
in a constant number of phases.  An edge seen in one phase has only a
small probability of having both endpoints matched by the eventual
canonical matching; which edges have this property is determined by later
random choices.  Attach a private leaf to each original vertex outside
the endpoint sets of this matching.  There are $B=2(N-H)$ such vertices.
The canonical matching and the leaf edges form a matching of size
$U=H+B\simeq4N/3$.
On the other hand, every edge not contained in these endpoint sets
uses one of the $B$ outside vertices.  If the algorithm
recovers few edges inside those sets, its output has size at most
$B+o(N)\simeq U/2$.

\begin{figure}[t]
\centering
\begin{tikzpicture}[x=0.85cm,y=0.85cm,
  clone/.style={circle,fill=black,inner sep=1.5pt},
  other/.style={draw=black!25,line width=0.65pt},
  useful/.style={draw=blue!65!black,line width=1.7pt},
  absent/.style={draw=red!65!black,dashed,line width=0.8pt}]
  \node at (1.5,2.45) {All $K^2$ potential edges};
  \node at (8.0,2.45) {A fixed-size random subset};
  \foreach \a in {0,1,2} {
    \coordinate (lu\a) at (0,1.5-1.5*\a);
    \coordinate (lv\a) at (3,1.5-1.5*\a);
    \coordinate (ru\a) at (6.5,1.5-1.5*\a);
    \coordinate (rv\a) at (9.5,1.5-1.5*\a);
  }
  \foreach \a in {0,1,2} {
    \foreach \b in {0,1,2} {
      \draw[other] (lu\a)--(lv\b);
    }
    \draw[useful] (lu\a)--(lv\a);
  }
  \foreach \a/\b in {0/1,1/0,1/2,2/0,2/1} {
    \draw[other] (ru\a)--(rv\b);
  }
  \draw[useful] (ru0)--(rv0);
  \draw[useful] (ru2)--(rv2);
  \draw[absent] (ru0)--(rv2);
  \draw[absent] (ru1)--(rv1);
  \foreach \a in {0,1,2} {
    \node[clone,label=left:$u_{\a}$] at (lu\a) {};
    \node[clone,label=right:$v_{\a}$] at (lv\a) {};
    \node[clone,label=left:$u_{\a}$] at (ru\a) {};
    \node[clone,label=right:$v_{\a}$] at (rv\a) {};
  }
  \draw[->,line width=0.8pt] (4.0,0)--(5.5,0);
  \node[align=center,font=\small] at (4.75,0.65) {fixed-size\\ sampling};
  \node[font=\small] at (1.5,-2.2) {$\operatorname{label}(u_av_b)=b-a\pmod K$};
  \node[font=\small] at (8,-2.2) {Dashed edges do not arrive.};
\end{tikzpicture}
\caption{One base edge becomes a simple bipartite graph on two sets of
vertex copies.  Here $K=3$ and $o_i=0$; the horizontal edges are precisely
the edges with label $o_i$.  Every label likewise defines a matching.
For each base pair, we independently choose a uniform subset of fixed
size from the edges between its endpoint copies.  This prevents an
algorithm from recovering a particular edge merely by remembering its
base edge.  Across the base edges of a matching, the surviving edges
with label $o_i$ remain a matching.}
\label{fig:lb-cyclic-lift}
\end{figure}
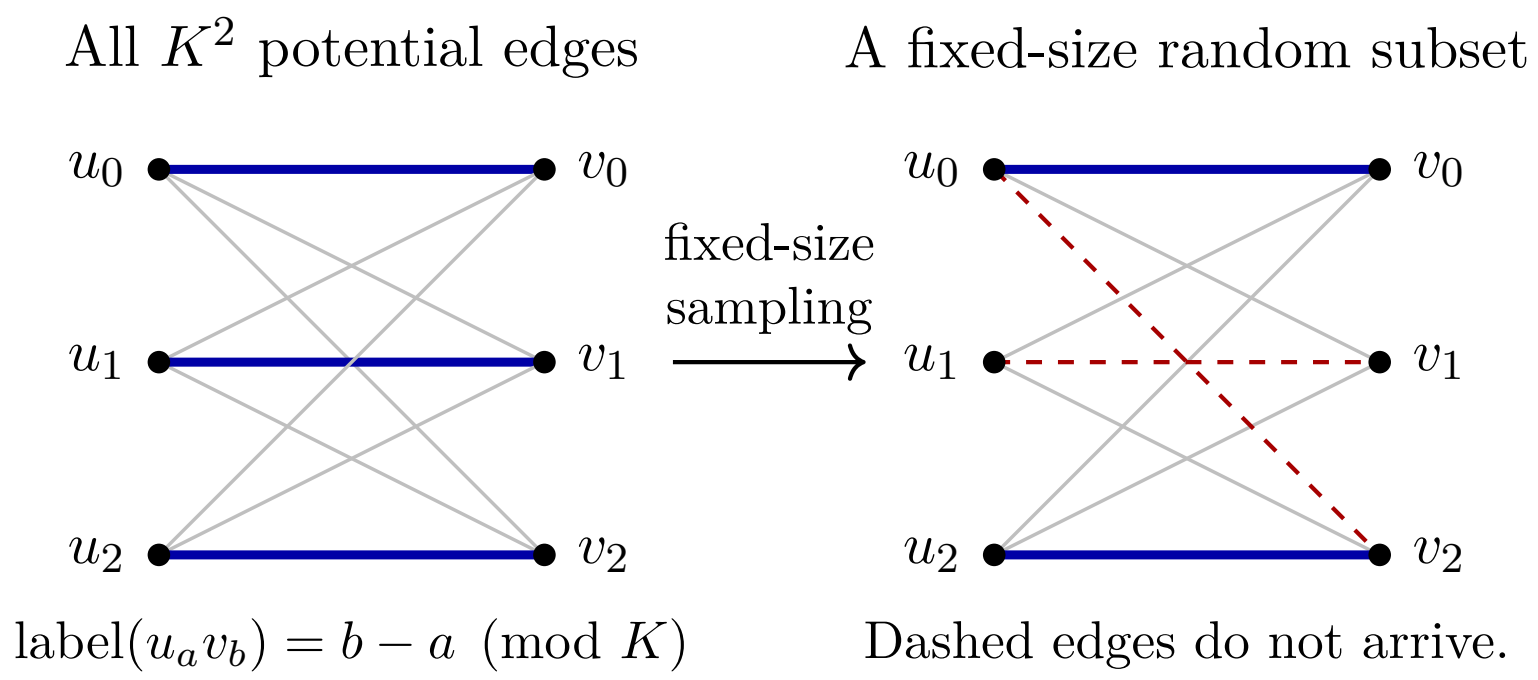

\paragraph{Replacing parallel edges by edges between vertex copies.}
The recursive objective of Feldman et al.~\cite{fnsz} uses many
parallel copies of each edge, with one distinguished index whose effect
is not visible while its stage is being processed.  We instead replace
each vertex by $K$ copies and each base edge $uv$ by the distinct
edges $u_a v_b$, for $a,b\in\mathbb Z_K$.  Give this edge the public
label $b-a$.  For each fixed label, the edges corresponding to a base
matching form $K$ vertex-disjoint copies of that matching.  The graph is
simple, and its vertex count increases by a factor $K$
(Figure~\ref{fig:lb-cyclic-lift}).

This replacement alone would reveal too much: remembering $uv$ would allow
an algorithm to reconstruct every edge between its endpoint copies.
We therefore independently delete a small fixed fraction of the base
edges in each phase and a small fixed fraction of the edges between
the endpoint copies of every base edge.  Each retained subset has a fixed
size, so even the exact stream length reveals no information.

The information argument is elementary.  If a random retained set
contains every fixed $r$-element set with probability at most $q^r$,
an $s$-bit memory state can determine with certainty that at most
$s/\log_2(1/q)$ elements belong to the set, in expectation.  The presence
of any earlier edge used after the relevant phase or stage ends must be
determined by the memory state at that time: otherwise the same state is
compatible with an input on which that edge never arrived.  Later oracle
answers cannot distinguish the two inputs, since the objective is
independent of the deletions.  This argument covers arbitrary encodings
of memory.

The two independent sampling steps give complementary conclusions.
Sampling the base edges keeps the output size near one half of the
constructed matching in each stage.  Sampling the edges between vertex
copies ensures that few output edges have the distinguished label,
relative to the size of that matching.  A polylogarithmic $K$ is large
enough for the second conclusion and small enough for the first, even
after charging the algorithm for the increased vertex count.

\paragraph{Value across stages.}
Take $p$ vertex-disjoint stages and choose a distinguished label $o_i$
independently and uniformly for each stage $i$.  An edge with label $o_i$
contributes to its stage without reducing the value of later stages; an
edge with a different label contributes locally but discounts the remaining
suffix.  The objective is arranged so that all current-stage edges have
exactly the same effect until a later stage arrives.  Thus queries on
edges that have already arrived do not reveal $o_i$.  If every stage
supplies one normalized unit of edges with its distinguished label, the
objective is $p$.  If every stage supplies only one half unit of edges
with other labels, its value is at most $(p+1)/3$.  The ratio approaches
three as the fixed number of stages grows.  The formal proof verifies
submodularity on all subsets, bounds the number of recovered edges with
label $o_i$, and chooses all constant losses before letting the graph
size grow.

\FloatBarrier
\section{The upper-bound algorithm and its analysis}
\label{sec:upper-framework}
We prove Theorem~\ref{thm:main-upper} by comparing the output to an optimal matching $M^*$.
First we bound the loss from discarding edges of $M^*$. This leaves two quantities to recover
from the stored graph. A fractional optimization assigns capacity to each, and two successive
rounding steps turn the fractional solution into one feasible matching. We specify the degree and accuracy parameters after deriving the two coefficients in the approximation bound.

\subsection{Notation and the algorithm}
Only the selection step reads the stream. Once it ends, all queries and all rounding steps use
the retained edges. Algorithm~\ref{alg:upper-overview} summarizes the two parts.

Recall that $f(e\mid A)=f(A\cup\{e\})-f(A)$, which is zero if $e\in A$. We work with the
normalized objective $f-f(\varnothing)$. A factor $R\ge1$ for this objective gives the same
factor for $f$ because $f(\varnothing)\ge0$. For an edge vector $z\in[0,1]^E$, let $R_z$
include each edge $e$ independently with probability $z_e$. Its expected value
$F_f(z)=\Ex{f(R_z)}$ is the multilinear extension~\cite{CCPV2011,CVZ}. Write $F=F_f$. The
notation $R_p(S)$ sets the probability to $p$ on $S$ and zero elsewhere. For a vertex $v$,
$\delta(v)$ is the set of incident edges, and $z(\delta(v))=\sum_{e\ni v}z_e$ is its fractional
degree~\cite{GKPS2006}.

The streaming selection rule uses a constant $p>0$, ratio boundaries
$1/2=c_0<\cdots<c_k=1$, and positive rational rates
$A_j\ge B_j$ satisfying the direction condition
\eqref{eq:source-drift}.  This condition makes the ratio of the larger endpoint potential to their sum move toward $1/2$ during each ideal update.
Fix also a rational $r\in(1/2,1)$ and set
$D=r(1-r)<1/4$.  The lower bound on $W_D(O)$ below will have coefficient $\sigma>0$.  All these quantities are constants independent of the
input size.

For each guess of the largest singleton value, the algorithm stores an ordered set $S$ and nonnegative vertex potentials $\phi_v$, as in~\cite{LW}. Only vertices incident to $S$ need stored potentials; all other potentials are zero. For an arriving edge $e=uv$, define its weight to be the expected marginal value
\[
 w_e=\Ex{f(e\mid R_p(S_{<e}))}.
\]
The algorithm computes a conservative estimate $\widehat w_e$ and
accepts $e$ only if this estimate exceeds the sum of its endpoint
vertex potentials and a small threshold.  Upon acceptance it stores $e$ and
$\widehat w_e$ permanently in this instance.  It then raises the
endpoint vertex potentials during an interval of length $p\widehat w_e$.
When the higher potential is a fraction $c$ of the potential sum, the two
rates are $(A_j,B_j)$ for the interval containing $c$.  The interval index only decreases during this ideal update: at a boundary the
algorithm continues in the interval for $c$ on its left.  The first interval uses equal rates for the remaining duration of the update.  Appendix~\ref{sec:joint-source-path}
proves that this rule is well defined, including at zero vertex potentials.
The complete increments are then rounded upward once.

At the end of the stream, define
\begin{equation}\label{eq:main-residual}
 G_S=p\sum_{e\in S}\widehat w_e,
 \qquad
 g(T)=\Ex{f(T\cup R_p(S))-f(R_p(S))}.
\end{equation}
The function $g$ is normalized, monotone, and submodular.  All its
queries use stored edges.  The estimates $\widehat w_e$ are recorded at arrival and never recomputed using a later stored set.

\begin{algorithm}[!htbp]
\caption{One-pass submodular matching: main steps}
\label{alg:upper-overview}
\centering
\begin{minipage}{0.97\linewidth}
\textit{Input:} the edge stream, a value oracle, and fixed admissible
parameters.  The detailed finite routines are in
Appendix~\ref{sec:upper-algorithm-specification}.
\begin{pseudolines}
\codephase{During the stream}
\codeline{0}{For each arriving edge, update the best singleton and process the edge for each active guess (Algorithms~\ref{alg:upper-stream} and~\ref{alg:upper-source}). Maintain $O(\log n)$ geometric guesses, recording accepted edges and their weight estimates and updating endpoint potentials.}
\codephase{After the pass}
\codeline{0}{For each retained set, run continuous greedy with time-dependent weights to compute a feasible pair $(x,y)$ for $\widehat w\cdot x+F_g(y)$ (Algorithm~\ref{alg:upper-flow}). Finish all estimates before rounding any pair.}
\codeline{0}{\label{line:overview-first}For the current pair $(x,y)$, round $y$ to a matching $T$ with edge marginals $y$. On the remaining vertices, rescale each $x_{uv}$ by the reciprocal of the probability that both endpoints remain unmatched. Zero the coordinates incident to vertices of excessive fractional degree and scale the rest to obtain a fractional matching $z$ with $z_e\le p$ (Algorithm~\ref{alg:upper-completion}).}
\codeline{0}{\label{line:overview-second}Sample a Gibbs matching $N$ with activities determined by $z$ and keep $T\cup N$ as a candidate. For the high-probability guarantee, repeat lines~\ref{line:overview-first}--\ref{line:overview-second} independently for the same pair before moving to the next pair.}
\codeline{0}{\textbf{return} the best candidate, including the best singleton, using the finite comparison grid of Algorithm~\ref{alg:upper-stream}.}
\end{pseudolines}
\end{minipage}
\end{algorithm}

\paragraph{Oracle and arithmetic conventions.}
During the stream, queries use the current arrival and stored edges;
afterward they use only stored edges.  We count arithmetic operations
and value queries separately.  The state stores a bounded number of
original objective scales and integer indices of $O(\log n)$ bits
relative to those scales.  Thus the word bound is in the value-oracle
arithmetic model; the representation of an arbitrary original real
scale is additional if it is charged.  When oracle values occupy a
constant number of words, this is the same finite-word convention as
in the lower bound.  All persistent state and query workspace count
toward memory.  Explicit running-time bounds appear in
Appendix~\ref{sec:explicit-upper-runtime}.

\subsection{Two bounds obtained from the streaming algorithm}
\label{sec:main-history}
The streaming rule yields two inequalities. The first bounds the contribution of discarded
optimum edges by the final endpoint potentials. The second bounds the decrease in stored
marginal values when the rounding procedure has already selected a matching.

Write $V(O)$ for the vertices incident to a matching $O$. Let $\psi_v$ be the sum of ideal
increases in the potential at $v$, before upward rounding. For a matching $O\subseteq S$, put
\[
 X(O)=G_S+\sum_{v\notin V(O)}\psi_v.
\]
For example, if two discarded edges of an optimal matching are $uv$ and $wx$, their four
endpoints are distinct. Each can be charged once, and none belongs to the vertex set of the
retained part $O$ of that matching. Only these unmatched vertices enter the sum defining $X(O)$.

\begin{lemma}[Charging the optimum]\label{lem:main-charge}
Let $B=\max_{e\in E}f(\{e\})>0$ and $0<\varepsilon_0<1/4$. The algorithm uses $O((\nu+1)\log^2
n)$ words for $O(\log n)$ guesses on every execution. Except with the prescribed estimation
failure probability, the final guess $\tau\in[B,2B)$ has a matching $O\subseteq S$ such that
\begin{equation}\label{eq:main-charge}
 \OPT\le X(O)+g(O)+6\varepsilon_0\tau.
\end{equation}
\end{lemma}
Here and below, $S,G_S,g$ refer to the instance for the indicated guess. We prove the charge
first with exact weights, then account for the number of retained edges and the finite errors.
\begin{proof}
First ignore numerical errors and thresholds, and suppose the guess was present from the start.  Then $\widehat w=w$,
$G_S=F(p\mathbf1_S)$, and $\psi=\phi$.
Let $M^*$ be an optimum and $O=M^*\cap S$.
Monotonicity and submodularity give
\[
 f(M^*)\le F(p\mathbf1_S)+g(M^*)
 \le G_S+g(O)+\sum_{e\in M^*\setminus S}g(\{e\}).
\]
For a rejected edge $e$, the value $g(\{e\})$ is at most its expected marginal value at arrival, hence at most the sum of endpoint potentials that caused its rejection.
The rejected optimum edges have distinct endpoints outside $V(O)$.
Their total charge is therefore at most
$\sum_{v\notin V(O)}\psi_v$.

For space, each accepted incidence multiplies a positive endpoint
potential by a fixed factor greater than one.  At a fixed guess the
potential range has polynomial ratio, so a vertex has $O(\log n)$ accepted
incidences.  The endpoints of a maximum-cardinality matching form a
vertex cover of size at most $2\nu$; every retained edge meets this
cover.  Thus one stored set has $O(\nu\log n)$ edges.

The finite implementation uses geometric guesses with ratio two, conservative
marginal estimates, and upward potential rounding.  The decisive guess
loses at most $6\varepsilon_0\tau$ from these operations and its
ignored prefix.  Appendix~\ref{sec:joint-source-foundations} proves
these bounds and the deterministic space guarantee, including on
inaccurate estimation transcripts.
\end{proof}

Suppose a later choice $T$ already covers a topic counted in an earlier weight. Keeping the
original weight might then credit the topic twice. The next identity bounds the total
overcount, even if $T$ was chosen using $S$. Here $p$ is the probability used to record the
original weights and $z_e$ is the probability assigned to $e$ in a new independent sample.
\begin{lemma}[Comparison of marginal values]\label{lem:main-history}
Suppose $0\le\widehat w_e\le w_e$ for the weights $w_e$ defined at arrival.
For every $T\subseteq E$ and every $0\le z\le p\mathbf1_S$,
\begin{equation}\label{eq:main-history}
 f(T)+F_{f_T}(z)\ge g(T)+\widehat w\cdot z,
 \qquad f_T(A)=f(T\cup A)-f(T).
\end{equation}
\end{lemma}
The proof couples the two samples, then sums the changes in marginal value in the original order.
\begin{proof}
Let $w_e^T=\Ex{f(e\mid T\cup R_p(S_{<e}))}$. Submodularity gives $0\le w_e^T\le w_e$. For each
earlier edge $a$, draw one independent uniform number $U_a\in[0,1]$. Include $a$ in the
$z$-sample if $U_a\le z_a$ and in the $p$-sample if $U_a\le p$. Since $z_a\le p$, the former is
a subset of the latter. Thus the expected marginal of $e$ on the $z$-sample with $T$ already
present is at least $w_e^T$. Telescoping in arrival order gives
\[
 F_{f_T}(z)\ge\sum_e z_e w_e^T.
\]
Telescoping in arrival order, with and without $T$, gives the exact identity
\[
 p\sum_e(w_e-w_e^T)=f(T)-g(T).
\]
Since $z_e\le p$ and each difference is nonnegative,
\[
 \widehat w\cdot z-F_{f_T}(z)
 \le\sum_e z_e(w_e-w_e^T)
 \le f(T)-g(T).
\]
Rearrange to obtain the claim.
\end{proof}

The inequality applies even when $T$ contains stored edges whose marginal values drop to zero.
For example, suppose all $k$ stored edges cover the same topic of value one, and $T$ consists
of one of these edges. The $j$th arrival in $S$ has weight $(1-p)^{j-1}$ because it adds value
precisely when no earlier stored edge was sampled. Once $T$ is present, all these weights fall
to zero. Their total decrease after multiplication by $p$ is $1-(1-p)^k$. This is exactly
$f(T)-g(T)$, since $T$ adds value to the final random sample only when the sample is empty.

\subsection{Optimization of the multilinear relaxation}
\label{sec:main-joint}
The two comparisons suggest separate fractional variables: $y$ selects edges for the residual
function $g$, and $x$ assigns mass to the recorded weights. Both will use the same vertices.
For instance, if the first matching uses a vertex with probability $\ell$, at most $1-\ell$ of
its expected capacity remains for the second. Let $\Pjoint$ consist of the nonnegative pairs on
$S$ satisfying
\begin{align}
 y(\delta(v))&\le D &&(v\in V),\label{eq:main-joint-a}\\
 x(\delta(v))+y(\delta(v))&\le1 &&(v\in V),\label{eq:main-joint-b}\\
 x_{uv}+p[y(\delta(u))+y(\delta(v))]&\le p &&(uv\in S).
 \label{eq:main-joint-c}
\end{align}
The last constraint ensures that, after conditioning on both endpoints
being unmatched, the rescaled coordinate of $x$ still obeys the bound in
Lemma~\ref{lem:main-history}.

For a matching $O\subseteq S$, put $I_v=\mathbf1_{v\in V(O)}$.
The largest value of the linear term when $y=D\mathbf1_O$ is
\begin{equation}\label{eq:main-comparator}
 \begin{split}
 W_D(O)=\max\ &\widehat w\cdot z\\
 \text{subject to }&z\ge0,\quad z(\delta(v))\le1-DI_v,\\
 &z_{uv}\le p[1-D(I_u+I_v)].
 \end{split}
\end{equation}
If $z^O$ attains this maximum, then
$(z^O,D\mathbf1_O)\in\Pjoint$.  Neither $O$ nor $z^O$ is known to
the algorithm; they are used only in the analysis.

We can compare with any matching $O$ by reserving $D$ units of capacity on each of its edges
for $g$. For a single edge $uv$ in $O$, this leaves $1-D$ units at both $u$ and $v$ for $x$.
The lemma makes all these comparisons with one fractional pair, so the algorithm never has to
guess $O$.
\begin{lemma}[Continuous greedy]\label{lem:main-optimize}
Given a stored set and its estimates made at arrival, one can compute
$(x,y)\in\Pjoint$ in polynomial time and linear working space such
that, except with a prescribed failure probability,
\begin{equation}\label{eq:main-flow}
 \widehat w\cdot x+F_g(y)
 \ge W_D(O)+(1-e^{-D})g(O)-\varepsilon_0\tau/2
\end{equation}
simultaneously for every matching $O\subseteq S$.
\end{lemma}
We choose the time-dependent weight of $g$ so that differentiating it cancels the loss from $g$
already collected. The remaining derivative can then be integrated explicitly.
\begin{proof}
Start at $(x(0),y(0))=(0,0)$.  At time $t\in[0,1]$, choose a
feasible direction $(v_x,v_y)\in\Pjoint$ maximizing
\[
 \widehat w\cdot v_x+e^{D(t-1)}\nabla F_g(y)\cdot v_y,
\]
and follow this direction~\cite{CCPV2011,FeldmanDistorted}. Adding the elements of
$O$ one at a time to $R_y$ shows that their expected total gain is at least $g(O)-F_g(y)$, by
monotonicity. Submodularity bounds this gain from above by the sum of their initial marginals,
which is at most $\sum_{e\in O}\partial_e F_g(y)$ (the derivative measures the marginal with
$e$ absent). Comparing the maximizing direction with the feasible pair $(z^O,D\mathbf1_O)$ now
gives
\[
 \widehat w\cdot v_x+e^{D(t-1)}\nabla F_g(y)\cdot v_y
 \ge W_D(O)+De^{D(t-1)}[g(O)-F_g(y)].
\]
Consequently the potential
$\Psi(t)=\widehat w\cdot x(t)+e^{D(t-1)}F_g(y(t))$ satisfies
\[
 \Psi'(t)\ge W_D(O)+De^{D(t-1)}g(O).
\]
The derivative of the exponential cancels the negative term involving $F_g(y)$.
Since $\Psi(0)=0$, integration gives coefficient one for $W_D(O)$
and coefficient $1-e^{-D}$ for $g(O)$.

Appendix~\ref{sec:joint-distorted-flow-oracle} implements this
comparison by a finite Euler scheme and gradient estimates obtained by queries to $f$.  Its error bounds hold against all matchings $O$ on one common
accuracy event.  Computing a direction reduces to a packing linear program;
Appendix~\ref{sec:linear-space-positive-packing} solves this program in linear
workspace using adjacency scans.
\end{proof}

\subsection{A lower bound on the linear objective}
\label{sec:main-source-certificate}
To apply the preceding lemma, we need a lower bound on $W_D(O)$ in terms of $X(O)$. Available capacity at each endpoint depends on whether $O$ uses it. For an accepted edge
$uv$ with $u\in V(O)$ and $v\notin V(O)$, the linear program leaves capacities $1-D$ and $1$,
respectively. The update at $v$ also enters $X(O)$, whereas the update at $u$ does not. This is
why the proof uses a different scalar function $H_i$ for each status $i=\mathbf1_{v\in V(O)}$.

Take nonnegative, bounded, nondecreasing functions $H_0,H_1$ that vanish near zero. The
integral condition~\eqref{eq:main-integral} bounds how much can be charged to a vertex. The
local condition~\eqref{eq:main-local} shows that a single edge update can pay its share. Both
are stated for the rates actually used by the algorithm.
\begin{lemma}[Lower bound on $W_D(O)$]\label{lem:main-certificate}
If the rates and $H_0,H_1$ satisfy these two conditions with coefficient $\sigma$, then $W_D(O)\ge\sigma X(O)$ for every matching $O\subseteq S$.
\end{lemma}
We fix arbitrary prices for the vertex constraints, check the inequality for one accepted edge,
and sum over the stream. The accumulated increases at each vertex telescope.
\begin{proof}
Fix $O$ and put $k_{uv}=1-D(I_u+I_v)$.  Eliminating variables for the coordinate bounds from the dual of~\eqref{eq:main-comparator} gives the
objective
\begin{equation}\label{eq:main-source-dual}
 \sum_v(1-DI_v)\lambda_v
 +p\sum_{uv\in S}k_{uv}(\widehat w_{uv}-\lambda_u-\lambda_v)_+,
 \qquad\lambda\ge0.
\end{equation}
We show that every such objective is at least $\sigma X(O)$.
The vector $\lambda$ is arbitrary and belongs only to this dual
argument; it is distinct from the streaming vertex potentials $\phi$.

Set $t_*=(1+p\max_j(A_j+B_j))^{-1}$.  For this fixed dual vector,
assign vertex $v$ the auxiliary potential function
\[
 \Phi_v(\phi)=
 (1-DI_v)\int_0^\phi
 H_{I_v}\!\left(\frac{\lambda_v}{t_*s}\right)\,ds.
\]
At zero arguments use the limiting values, and take $\Phi_v=0$ if
$\lambda_v=0$.  The integral condition on $H_0,H_1$ says exactly
that $0\le\Phi_v(\phi)\le(1-DI_v)\lambda_v$ for every $\phi\ge0$.
Thus the vertex part of the dual bounds the terminal values of the auxiliary potential functions.
As $\phi_v$ increases, the auxiliary potential $\Phi_v(\phi_v)$ increases at a nonincreasing rate and eventually becomes constant.  The edge slack accounts for the remaining cost of each update.

Consider an accepted edge $e=uv$.  During its ideal update to vertex potentials,
parameterize the update by elapsed time $s$ and write $\dot\phi_u,\dot\phi_v$
for the two rates.  The four local inequalities give
\begin{equation}\label{eq:main-source-increment}
 \frac{d}{ds}(\Phi_u+\Phi_v)
 +k_e\left(1-\frac{\lambda_u+\lambda_v}{\widehat w_e}\right)_+
 \ge\sigma\bigl[1+(1-I_u)\dot\phi_u+(1-I_v)\dot\phi_v\bigr].
\end{equation}
The right side accounts for the increase in $G_S$ and in the potentials at endpoints unmatched by $O$.  The left side uses the increase in $\Phi_u+\Phi_v$ and the edge's dual slack.  To obtain this inequality, the
only scaling observation is
$t_*(\phi_u+\phi_v)\le\widehat w_e$ throughout the update.
Indeed, the pre-update sum is at most $\widehat w_e$, and the
total ideal increase is at most $pC_{\max}\widehat w_e$, where
$C_{\max}=\max_j(A_j+B_j)$.
Both endpoint vertex potentials are positive in the interior of every update
interval, so it suffices that the differential inequality hold almost
everywhere.  Appendix~\ref{sec:joint-inherited-potential} writes
the local conditions explicitly and verifies the interval boundaries.
The indicator $I_v$ never changes during this argument, even
when a vertex switches between the high- and low-potential roles.

Integrate~\eqref{eq:main-source-increment} over the update interval
of length $p\widehat w_e$, then sum over accepted edges.  Rounding $\phi_v$ upward can only increase $\Phi_v(\phi_v)$, since $H_0,H_1$ are nonnegative. The terminal bounds on the auxiliary potential functions give
\[
 \sigma\left(G_S+\sum_{v\notin V(O)}\psi_v\right)
 \le\sum_v(1-DI_v)\lambda_v
      +p\sum_e k_e(\widehat w_e-\lambda_u-\lambda_v)_+.
\]
This holds for every dual vector.  Strong duality proves the claim.
\end{proof}

This is the only part of the approximation argument that requires a
numerical certificate.  There are two scalar functions and four cases for the indicators $I_u,I_v$.  The final appendix supplies rational,
piecewise linear functions $H_0,H_1$ and verifies the inequalities on their
entire continuous domains.  The algorithm uses the rates, but never
uses $O$ or the auxiliary potential functions $\Phi_v$.

\subsection{Rounding both contributions to one matching}
\label{sec:main-rounding}
A feasible fractional pair can assign positive mass to several edges at a vertex. Independent rounding could therefore violate the matching constraint. Instead, we first draw a matching $T$
for $y$. For each edge of $x$ we then compensate for the chance that $T$ occupied an endpoint.
This step needs two facts: how the first matching affects the remaining capacity, and how to
round the small coordinates left on those vertices.

\paragraph{A matching with prescribed marginals.}
If $y(\delta(v))\le D=r(1-r)$ with $r>1/2$, the distribution $\mu_y$ associated with Shearer's lemma~\cite{Shearer,GJL2019} is supported on matchings, has edge marginals $y$, and satisfies
\[
 \Pr(Q\subseteq T)=\prod_{e\in Q}y_e
 \quad\text{for every matching }Q.
\]
Conflicting edges are never selected together.  Conditioning on one
edge being selected leaves the distribution on disjoint edges unchanged and
removes its neighbors.  A one-coordinate hybrid therefore gives
$\Ex{g(T)}\ge F_g(y)$.  Appendix~\ref{sec:referee-shearer-sampler}
proves the properties of this distribution and its finite-accuracy sampler.

For the second step, degree at most one allows a vector to have large total mass at a vertex.
The coordinate bound limits the influence of any single edge. For example, a star with $d$
leaves can have $z_e=1/d$ on every edge. Choosing one edge uniformly already realizes those
marginals on the star. The lemma allows arbitrary simple graphs and compares the expected value
of the matching with the value of independent sampling for every monotone submodular objective.
\begin{lemma}[Rounding small coordinates]\label{lem:main-diffuse}
For every $0<\epsilon<1$ and sufficiently small $p>0$, let $z\ge0$ on a simple graph satisfy
$z_e\le p$ and $z(\delta(v))\le1$ for all edges and vertices. There is an explicit Gibbs
distribution on matchings $N$ with $\Ex{h(N)}\ge(1-\epsilon)F_h(z)$ for every normalized
monotone submodular $h$. It admits a polynomial-time, linear-space sampler to any prescribed
total variation accuracy.
\end{lemma}

The proof first bounds the effect of dependence between edges. It then chooses activities so
that the resulting marginals approximate the target vector. The distribution depends only on
$z$. It assigns each matching $N$ a probability proportional to
$\prod_{e\in N}\lambda_e$.  For $0<\gamma<1$ close to one, its
activities are
\[
 \lambda_{uv}=
 \frac{\gamma z_{uv}}
 {(1-\gamma z(\delta(u)))(1-\gamma z(\delta(v)))}.
\]
The vertex-deletion coupling in
Figure~\ref{fig:upper-deletion} bounds the variance of weighted sums
of unmatched indicators.  This shows that the restriction of a Gibbs
matching to $G-v$ is within total variation
$\sqrt{\max_e\lambda_e/8}$ of the Gibbs distribution on $G-v$
(Lemma~\ref{lem:gibbs-deletion-stability}).  Apply this comparison
twice to compare the remainders when an edge $uv$ is absent or present.
Any additional edges incident to $u$ or $v$ can only decrease a
submodular marginal value.

To compare with independent rounding, replace independent coordinates
by matching coordinates one
at a time.  For a fixed independent suffix $U$, the marginal
$h(e\mid U\cup A)$ is decreasing in $A$ and lies between zero and
$h(e\mid U)$.  The error in total variation distance can therefore be charged to
this marginal value with respect to the suffix.
Fix an order of the support and let $E_{>e}$ consist of
the edges after $e$.  Write $m_e=\Pr(e\in N)$.  The error charged to
edge $e$ is at most $d\,m_e\,\Ex{h(e\mid R_m(E_{>e}))}$, where
$d=\sqrt{\max_e\lambda_e/2}$.  These charges telescope:
\[
 \sum_e m_e\,\Ex{h(e\mid R_m(E_{>e}))}=F_h(m).
\]
Thus the total loss is at most $dF_h(m)$, independently of the number
of edges.  The displayed activities make $m$ multiplicatively close
to $\gamma z$ coordinatewise.  Monotonicity and concavity of $F_h$
along nonnegative rays then compare $F_h(m)$ with $F_h(z)$.
Appendix~\ref{sec:gibbs-small-coordinate} proves both comparisons;
$p\le\epsilon^5/5120$ is sufficient for $0<\epsilon<1$.
The sampler appears in Appendix~\ref{sec:gibbs-sampling-audit}.

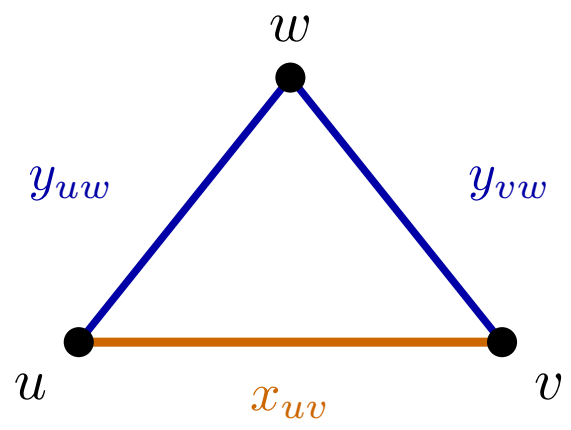
\begin{figure}[!htbp]
\centering
\begin{tikzpicture}[x=1cm,y=1cm,
 vertex/.style={circle,fill=black,inner sep=2pt},
 tentative/.style={draw=blue!65!black,line width=1.4pt},
 source/.style={draw=orange!80!black,line width=1.7pt}]
 \coordinate (u) at (0,0);
 \coordinate (v) at (2.8,0);
 \coordinate (w) at (1.4,1.75);
 \draw[source] (u)--(v);
 \draw[tentative] (u)--(w);
 \draw[tentative] (v)--(w);
 \node[vertex,label=below left:$u$] at (u) {};
 \node[vertex,label=below right:$v$] at (v) {};
 \node[vertex,label=above:$w$] at (w) {};
 \node[font=\small,text=orange!80!black] at (1.4,-.4) {$x_{uv}$};
 \node[font=\small,text=blue!65!black] at (-.05,1.05) {$y_{uw}$};
 \node[font=\small,text=blue!65!black] at (2.85,1.05) {$y_{vw}$};
 \node[anchor=west,text width=7.0cm,align=left,font=\small] at (3.75,.8)
 {The matching $T$ cannot use both blue edges.\\[5pt]
 Their common neighbor changes the joint probability that $u$ and
 $v$ are unmatched by $T$. We divide $x_{uv}$ by that exact probability.};
\end{tikzpicture}
\caption{Why the probabilities that the endpoints are unmatched cannot be multiplied.
Normalizing by the exact joint probability preserves each coordinate of $x$ in expectation, including on graphs with triangles.}
\label{fig:upper-triangle}
\end{figure}

It remains to use this distribution after choosing $T$. For example, if both endpoints of $e$
are free with probability $q$, assigning mass $x_e/q$ on that event recovers $x_e$ in
expectation. This calculation preserves each coordinate in expectation. A further argument bounds the mass lost at vertices whose rescaled fractional degree exceeds one.
\begin{lemma}[Rounding in two stages]\label{lem:main-round}
Fix $D=r(1-r)$, $1/2<r<1$, and $0<\varepsilon_R<1$. For sufficiently small $p>0$, every
$(x,y)\in\Pjoint$ and conservative arrival weights $\widehat w$ yield a distribution on
matchings $M\subseteq S$ satisfying
\[
 \Ex{f(M)}\ge(1-\varepsilon_R)
       [\widehat w\cdot x+F_g(y)].
\]
The statement concerns ideal sampling.  Finite sampling and arithmetic
introduce an arbitrarily small additive loss relative to $\tau$, using
polynomial time and linear working space.
\end{lemma}
\begin{proof}
Draw $T$ from the distribution $\mu_y$ above.  Write
$\ell_v=y(\delta(v))$, $X_v=\mathbf1_{v\notin V(T)}$, and
$q_{uv}=\Pr(X_u=X_v=1)$.  The distribution's one- and two-edge moments
express each $q_{uv}$ as an explicit quadratic in $y$, computable
by adjacency scans; see (GJ2) in
Appendix~\ref{sec:general-joint-completion}.  The union bound gives
$q_{uv}\ge1-\ell_u-\ell_v$.  Hence
\[
 a_{uv}=x_{uv}/q_{uv}\le p.
\]
The vector with coordinates $a_{uv}X_uX_v$ has expected coordinate $x_{uv}$ and uses only vertices unmatched by $T$. Its fractional degrees can exceed one,
so a small alteration is necessary.  Figure~\ref{fig:upper-triangle}
shows why $q_{uv}$ need not be the product of the two unmatched probabilities.

For each vertex define $Z_u=\sum_{uv\in S}a_{uv}X_v$; this sum
ignores whether $u$ itself is matched by $T$.  Put $\mu_u=\Ex{Z_u}$ and
$s_u=1-\ell_u$, taking $y_{vw}=0$ for absent edges.
For distinct $v,w$, the exact two-edge moments give
$\operatorname{Cov}(X_v,X_w)\le y_{vw}s_v$.  Since the graph is
simple, each neighbor has one coefficient at most $p$.  The diagonal
and off-diagonal terms in the variance are each at most $pD\mu_u$,
giving
\[
 \operatorname{Var}(Z_u)\le2pD\mu_u\le\frac12p\mu_u.
\]
To bound the mean, use the normalization itself:
$\Ex{X_uZ_u}=x(\delta(u))\le s_u$.  Cauchy--Schwarz then gives
\[
 s_u(\mu_u-1)\le-\operatorname{Cov}(X_u,Z_u)
 \le\sqrt{s_u\ell_u\operatorname{Var}(Z_u)}.
\]
Thus $(\mu_u-1)_+^2\le p\mu_u/6$.  Solving this quadratic
inequality yields $\mu_u\le1+\sqrt p$ for $p\le1$, and hence
$\operatorname{Var}(Z_u)<3p/4$ for $p\le1/16$.
For the fixed stored set and fractional solution, these calculations do not
condition on $u$ being unmatched by $T$.

For a small constant $\varepsilon_A>0$, set
\[
 z_{uv}(T)=\frac{a_{uv}X_uX_v}{1+\varepsilon_A}
   \mathbf1_{Z_u\le1+\varepsilon_A}
   \mathbf1_{Z_v\le1+\varepsilon_A}.
\]
This vector avoids $T$, has fractional degrees at most one, and retains
$z_e(T)\le p$.  If also $\sqrt p\le\varepsilon_A/2$, Chebyshev's
inequality at the two endpoints of each charged edge gives
\[
 \Ex{z_e(T)}\ge(1-\varepsilon_A-12p/\varepsilon_A^2)x_e.
\]
No union bound over all vertices is needed.  Taking
$\varepsilon_A=\varepsilon_R/4$ and $p$ sufficiently small makes
this coefficient at least $1-\varepsilon_R/2$.  The common bound
$p\le(\varepsilon_R/2)^5/5120$ suffices for these inequalities
and the subsequent small-coordinate rounding.

Conditional on $T$, apply Lemma~\ref{lem:main-diffuse} to $z(T)$
for the objective $f_T$, with error $\varepsilon_R/2$.  The resulting
matching $N$ avoids $T$, so $M=T\cup N$ is feasible.
Lemma~\ref{lem:main-history} gives, for every fixed $T$,
\[
 \Ex{f(T\cup N)\mid T}
 \ge(1-\varepsilon_R/2)
       [g(T)+\widehat w\cdot z(T)].
\]
Averaging, using $\Ex{g(T)}\ge F_g(y)$ and the coordinate retention
bound, proves the claim.  Appendix~\ref{sec:general-joint-completion}
gives the exact mean and variance estimates and finite arithmetic;
Algorithm~\ref{alg:upper-completion} specifies the sampler.
\end{proof}

\subsection{Choice of parameters and proof of Theorem~\ref{thm:main-upper}}
The ratio depends on two coefficients. The residual contribution $g(O)$ receives $1-e^{-D}$,
and the contribution $X(O)$ receives $\sigma$. An instance could concentrate its value in
either one, so only the smaller coefficient can be used for the whole optimum. Omitting the
finite additive errors for the moment, Lemmas~\ref{lem:main-charge}, \ref{lem:main-optimize},
\ref{lem:main-certificate}, and~\ref{lem:main-round} give
\[
 \begin{aligned}
 \Ex{f(M)}
 &\ge(1-\varepsilon_R)[\widehat w\cdot x+F_g(y)]\\
 &\ge(1-\varepsilon_R)[\sigma X(O)+(1-e^{-D})g(O)]\\
 &\ge(1-\varepsilon_R)\min\{\sigma,1-e^{-D}\}\,\OPT.
 \end{aligned}
\]
Thus the final coefficient is the minimum of the coefficients of $X(O)$ and $g(O)$, up to the chosen rounding allowance.
In particular, taking
$D$ up to $1/4$ would make the reciprocal of the coefficient of $g(O)$
approach $1/(1-e^{-1/4})\simeq4.5208$.  The update rates are tuned to
nearly match the coefficient of $g(O)$ at a slightly smaller fractional-degree bound.

For the finite statement, choose a rational lower bound
$b\le1-e^{-D}$ and
$\Theta=1-\varepsilon_R$.  Include the best singleton in the candidate
set.  Since the decisive guess satisfies $\tau<2B$, this candidate
absorbs the additive error relative to $\tau$.  Specifically, on an execution in the accuracy event, the conditional expected output value $Y$ satisfies
\[
 Y\ge B,\qquad
 Y+5\varepsilon_0B\ge\Theta\min\{\sigma,b\}\,\OPT.
\]
Hence $(1+5\varepsilon_0)Y$ is at least the right-hand side.  The accuracy event holds with probability at least $1-\varepsilon_0$, giving
\begin{equation}\label{eq:main-finite-factor}
 \Ex{f(M)}\ge
 \frac{(1-\varepsilon_0)\Theta\min\{\sigma,b\}}
 {1+5\varepsilon_0}\,\OPT.
\end{equation}
Appendix~\ref{sec:finite-joint-streaming-wrapper} accounts for these
finite errors.  Appendix~\ref{sec:joint-exact-parameters} supplies a simultaneous
rational choice with $\sigma=1/4.5246551$ and $b>\sigma$, certifies
all inequalities for the potential functions, and gives a finite expected factor below
$4.5246552$.  After amplification, the factor is at most $4.524656$, which implies the stated $4.525$-approximation.
The large arrays and exact verification records are confined to that
appendix and the accompanying verification files.

Finally, complete every streaming and optimization estimate before drawing
any rounding randomness.  Conditional on any transcript in the
common accuracy event, repeat
independent rounding trials on the same stored sets and
fractional points.  Each trial has value at most $\OPT$.  The strict
gap between its conditional expected factor and the stated target gives
a fixed positive success probability.  Sequentially repeating
$O(\log n)$ times and retaining the best candidate proves the
high-probability guarantee, without another pass or more asymptotic
space.  The finite implementation accounts separately for transcript failure
and the final comparison grid.  This proves
Theorem~\ref{thm:main-upper}.

\FloatBarrier
\section{A lower bound of three on simple bipartite graphs}
\label{sec:lb-lower}
Fix a target factor $R<3$. We will construct a distribution of inputs on which the optimum is
typically close to one unit per stage, while the expected value of the output is close to one
third of a unit per stage. The gap also forces the probability of meeting the target to tend to
zero. All parameters in this section are independent of the upper-bound construction.

We use the model of Section~\ref{sec:introduction}. In particular, a query on any set of
arrived edges is legal, and every output must be a matching of arrived edges, even in an
execution that misses the target value. The full memory state counts, including query
workspace, encoded summaries, and candidate outputs. There is no external edge-membership
oracle. These conventions matter because two prefixes reaching the same state may have
contained different edges.
\begin{theorem}[Hard distribution]\label{thm:lb-three}
For every fixed $1\le R<3$ and $a\ge0$, there are distributions of simple bipartite inputs of
increasing size on which each one-pass $O(n\log^a n)$-word algorithm has $o(1)$ probability of
returning an $R$-approximation. For all sufficiently large sizes, $\Ex{f(M_{\rm
out})}<\Ex{\OPT}/R$ on this distribution. The inputs have the graph sizes and oracle precision
stated in Theorem~\ref{thm:main-lower}.
\end{theorem}

We start with the bipartite blueprint expansion of Assadi, Jiang, and Xiang~\cite{ajx1,ajx2}.
It hides a large matching across successive phases. We then replace each vertex by copies and
adapt the recursive objective of Feldman et al.~\cite{fnsz} to the resulting simple graph. An
entropy bound is used twice: once for the edges of the original graph and once for the edges
between copies. Finally we compare the optimum and the algorithm's output under the same random
choices.

\subsection{Properties of the blueprint expansion}
The proof requires a distribution in which each early edge has small probability of having both endpoints among those of the eventual hidden matching, conditional on all preceding phases. The blueprint expansion supplies that property while
keeping the number of vertices and phase lengths fixed. For every fixed $\eta>0$, one may first
choose a finite
blueprint and a fixed density loss, and then obtain an infinite
sequence of simple bipartite graphs on $N+N$ vertices with the
following properties.  There is a constant number $P$ of phases,
a parameter $t=N^{\Omega(1/\log\log N)}$, and an independent uniform
tuple $J=(J_1,\ldots,J_P)\in[t]^P$.
\begin{enumerate}
\item Phase $j$ has an edge set $E_j^0(J_{<j})$ determined by the
      preceding tuple coordinates.  The phase sets are disjoint,
      and their cardinalities $m_j$ are deterministic.
\item A canonical matching $Q(J)$ has a deterministic size
      $H=hN$, where $2/3-\eta\le h<2/3$.  Its phase cardinalities
      $H_j$ are deterministic and sum to $H$.
\item Call an edge of the expansion \emph{special} if both its endpoints lie
      in the endpoint sets of $Q(J)$.  For every fixed phase prefix
      and every $e\in E_j^0(J_{<j})$,
      \begin{equation}\label{eq:lb-rarity}
       \Pr(e\text{ is special}\mid J_{<j})\le P/t.
      \end{equation}
\end{enumerate}
These are Proposition~4.10 and Lemma~5.3 of~\cite{ajx1}, together
with the blueprint family of Theorem~3 in~\cite{ajx2}.
For completeness, the deterministic phase counts follow directly
from the disjoint product matchings in Claims~4.6--4.7 of~\cite{ajx1}:
if a blueprint has phase edge set $\mathbb{E}_j$, and the underlying
extended Ruzsa--Szemer\'edi matchings have size $r$, then
\[
 m_j=|\mathbb{E}_j|r^P t^{P-j+1},\qquad H_j=|\mathbb{E}_j|r^P.
\]
The density loss is fixed before $N$ grows; it is not absorbed
into an asymptotic error.  Empty phases contribute zero and are
ignored in sums involving $m_j^{-1}$.

\subsection{An entropy bound}
The algorithm may compress many edges into one state. What matters here is which edges that
state certifies were present, since only such edges may safely be used later. Suppose, for
example, that each edge survives an independent fair coin. A state certifying $k$ specified
edges can occur with probability at most $2^{-k}$. The same counting argument works for
sampling without replacement and for states with unequal probabilities.
\begin{lemma}\label{lem:lb-certainty}
Let $X$ be a random subset of a finite universe such that
$\Pr(A\subseteq X)\le q^{|A|}$ for every fixed $A$, where
$0<q<1$.  Let $Y$ have at most $2^s$ states, and define
\[
 \mathcal C(y)=\{e:\Pr(e\in X\mid Y=y)=1\}.
\]
Then $\Ex{|\mathcal C(Y)|}\le s/\log_2(1/q)$.
\end{lemma}
We bound how rare each state must be and then average over the states.
\begin{proof}
For every positive-probability state $y$, the event $Y=y$ implies
$\mathcal C(y)\subseteq X$.  Hence
$\Pr(Y=y)\le q^{|\mathcal C(y)|}$, and therefore
\[
 |\mathcal C(y)|\log_2(1/q)\le-\log_2\Pr(Y=y).
\]
Average over $Y$ and use the Shannon entropy bound $H(Y)\le s$.
\end{proof}

The lemma applies to a uniform $k$-subset of an $m$-element set,
with $q=k/m$, because its joint inclusion probability is
$(k)_{|A|}/(m)_{|A|}\le(k/m)^{|A|}$.  It also applies to an
independent product of such samples when every retention fraction
is at most $q$.  No additive loss is paid for the number of groups.
Independent side information can be fixed before applying the lemma.

We will use the following consequence of the oracle model.
Consider the sampled edge set of one phase or stage, and let $Y$ be the
memory state at its end.  Suppose that no edge from the potential universe
of that sample arrives later.  Suppose also that, after fixing the relevant
side information, the future data are independent of the pair consisting
of the sampled edge set and $Y$, and that the oracle has one common
definition independent of that sample.  Then every edge from that universe
subsequently queried or output must belong to $\mathcal C(Y)$.
Indeed, if an edge is not in $\mathcal C(Y)$, it is absent from a stream
that reaches the same state with positive probability.  Couple the two
executions with the same future data and coins.  All subsequent permitted
queries receive the same answers.  The first query or output using that
edge would either query an edge that has not arrived or output an edge
absent from the input.  This reasoning does not assume that $Y$ stores an
explicit list of edges.  For randomized algorithms, fix coins for which
the query restriction and output feasibility hold for all inputs in the
finite support, and average afterward.

\subsection{The graph construction}
Build each stage in two steps. Sampling the original graph limits how much of its hidden
matching can be recovered. Replacing vertices by copies creates the labels needed by the
objective. We sample the edges between copies separately so that knowing the endpoints of a
base edge is insufficient to reconstruct its surviving copies. Both samples have fixed sizes,
which also fixes the length of the stream.
Fix small positive constants $\delta_b,\delta_\ell$ and construct
$p$ independent, vertex-disjoint stages, where $p$ will be fixed
later.  In each stage, start with the blueprint expansion above.  In phase $j$,
retain a uniform subset of exactly
$\lfloor(1-\delta_b)m_j\rfloor$ expansion edges, independently across
phases.  After these phases, attach a private leaf to each original
vertex outside the endpoint sets of the canonical matching.  The number
of such leaf edges is
\[
 B=2(N-H),\qquad U=H+B=(2-h)N.
\]
Give every original vertex a distinct potential private leaf in advance,
and pad the two leaf sets to $N$ vertices each.  We call the resulting
graph, consisting of the sampled expansion edges and the private-leaf
edges, the base graph of the stage.  It has exactly $4N$ vertices.
The leaf-edge set depends only on $J$, not on which expansion edges
are retained.

Choose an integer $A>a+1$ and let
\[
 K=\lceil(\log N)^A\rceil.
\]
Replace every base vertex by copies indexed by $\mathbb Z_K$.
For every present base edge $uv$, with $u$ on the left, its
potential edges between vertex copies are
\begin{equation}\label{eq:lb-cyclic}
 u_a v_b\quad(a,b\in\mathbb Z_K),\qquad
 \text{label}(u_av_b)=b-a\pmod K.
\end{equation}
For each potential base pair, independently sample exactly
$k_\ell=\lfloor(1-\delta_\ell)K^2\rfloor$ of these $K^2$
edges.  Such a subset is sampled for \emph{all} potential base
pairs, including absent expansion pairs and all potential private-leaf
pairs, independently of every $J$ and every sample of expansion edges.
Stream the sampled edges between vertex copies when their base edge
arrives, in a fixed order.  Sampling for every potential pair allows us
to condition on these subsets without disclosing base-edge membership.

The graph is simple: a pair of vertex copies uniquely determines its
base edge and the two indices of the copies.  Among the potential edges,
a fixed label gives $K$ vertex-disjoint copies of every base matching.
The total vertex count is
$n=4pNK$, and the stream length is $k_\ell$ times a deterministic
number of base edges.  Both may be announced in advance.

For each stage $i$, choose an independent uniform label
$o_i\in\mathbb Z_K$.  The objective will distinguish edges whose
public label is $o_i$ from edges with a different label.  The choices
of $o_i$ do not affect graph membership or stream order.

Set
\begin{equation}\label{eq:lb-normalizer}
 M=KU,\qquad c=B/U.
\end{equation}
Every matching in one stage of the final graph has at most $M$ edges.
To see this, if it uses $x$ edges corresponding to expansion edges and
$y$ edges corresponding to private-leaf edges, then
$2x+y\le2NK$ and $y\le KB$, so $x+y\le K(N+B/2)=KU$.
Moreover, if $Z_i$ counts its edges whose base pair is special,
then
\begin{equation}\label{eq:lb-slot-bound}
 |M_i|\le KB+Z_i=cM+Z_i.
\end{equation}
Every other edge uses one of the $KB$ copies of original vertices
outside the endpoint sets of the canonical matching, and a matching
uses each such copy at most once.

\subsection{The submodular objective}
The graph bounds the cardinality of the output in each stage. The objective determines how these stage contributions combine. A distinguished
edge contributes to its stage without affecting the value of later stages. An ordinary edge
contributes locally but discounts the remaining value. While a stage is the last one seen, the
suffix is empty and these effects are indistinguishable.
Let $\mathcal U_i$ be the fixed universe of all potential edges between
vertex copies in stage $i$.  The objective is defined on
$\bigcup_i\mathcal U_i$, independently of both sampling steps.
Let $g_i(S)$ and $b_i(S)$ be the numbers of stage-$i$ edges in $S$
with label $o_i$ and with a different label, respectively, divided by $M$.
Put $C_i=p-i+1$,
$t_i(S)=\min\{b_i(S),C_i\}$, and define
\begin{equation}\label{eq:lb-objective}
 F_{p+1}(S)=0,\qquad
 F_i(S)=\min\left\{C_i,
 g_i(S)+t_i(S)+\left(1-\frac{t_i(S)}{C_i}\right)F_{i+1}(S)
 \right\},\qquad f(S)=F_1(S).
\end{equation}
This is the recursive definition of~\cite{fnsz}, with $b_i(S)$ truncated
at $C_i$ so that diminishing returns are transparent on arbitrary,
possibly infeasible sets.

For example, with two stages remaining and one normalized unit of value in the suffix, an
ordinary edge whose addition leaves both caps inactive earns $(1-1/2)/M$, while a distinguished
edge earns $1/M$. If the suffix is empty, both earn $1/M$. The next lemma establishes label independence on the entire prefix, including queries whose edge counts exceed the caps.
\begin{lemma}[Value and information in a prefix]\label{lem:lb-objective}
On the full edge universe, $f$ is normalized, monotone and submodular, with values in $[0,p]$,
marginals at most $1/M$, and denominators dividing $p!M^p$. On sets drawn from stages
$1,\ldots,i$, the value of $f$ is independent of $o_i,\ldots,o_p$.
\end{lemma}
We first examine the three kinds of single-edge additions in the recursion. Then we evaluate a
prefix with an empty suffix and bound the denominators of all values.
\begin{proof}
Induct downward from $F_{p+1}=0$. Write $C=C_i$, $z=F_{i+1}$, $g=g_i$, and $t=t_i$, so $0\le
z\le C-1$. Adding a distinguished stage-$i$ edge, an ordinary stage-$i$ edge, or a later edge,
respectively, changes the expression before the outer cap by
\[
 1/M,\qquad \Delta t(1-z/C),\qquad (1-t/C)\Delta z.
\]
All factors are nonnegative and nonincreasing with the context.
Thus the expression is monotone and submodular; truncating it by
$C$ preserves these properties.  The base case is immediate.
For fixed stage counts, its dependence on $z$ is nondecreasing
and $1$-Lipschitz.  Hence the local bound $1/M$ also bounds the
global marginal contribution.

On a set restricted to stages through $i$, the suffix value is
zero and
\[
 F_i(S)=\min\{C_i,g_i(S)+t_i(S)\}
       =\min\{C_i,|S\cap\mathcal U_i|/M\}.
\]
The value is therefore independent of $o_i$, even for large queries.
Earlier recursion steps depend only on earlier labels.  Finally, each
step multiplies a possible denominator by at most $MC_i$; truncation
at an integer adds no denominator.  Their product
is $p!M^p$, and for fixed $p$ each value occupies $O_p(1)$ words.
\end{proof}

If a matching uses at most half a stage's capacity and all its edges have ordinary labels, the
recursion gives an upper bound of $(p+1)/3$ over $p$ stages. We also need to allow a few extra
edges and a few distinguished labels. Removing these exceptions costs at most $1/M$ per edge by
Lemma~\ref{lem:lb-objective}.
\begin{lemma}[Value from ordinary and distinguished edges]\label{lem:lb-profile}
Let $S$ have at most $cM+e_i$ edges in stage $i$, where
$0\le c\le1$, $cM$ and all $e_i$ are integers, and $e_i\ge0$.
If $G_i$ is its number of stage-$i$ edges with label $o_i$, then
\begin{equation}\label{eq:lb-profile}
 f(S)\le\frac{c(p+1)}{1+c}
             +\frac1M\sum_i(G_i+e_i).
\end{equation}
\end{lemma}
\begin{proof}
Remove all edges with label $o_i$ from each stage $i$ and then at most
$e_i$ additional edges from that stage.  The bound on marginal contributions
in Lemma~\ref{lem:lb-objective} bounds the lost value by the sum in
\eqref{eq:lb-profile}.  The remaining set has at most $cM$ edges per
stage, all with labels different from the stage's $o_i$.  By monotonicity,
its suffix value with $C$ stages is at most the recurrence
$V_C=c+(1-c/C)V_{C-1}$, with $V_0=0$.
Induction gives $V_C\le c(C+1)/(1+c)$: the step follows from
$c+(1-c/C)cC/(1+c)=c(C+1)/(1+c)$, and the first step uses $c\le1$.
\end{proof}

A matching with between $(1-\xi)M$ and $M$ edges in each stage, all
with that stage's label $o_i$, has value at least $p(1-\xi)$: every
suffix is a sum of at most one normalized unit per remaining stage,
so no truncation reduces its value.

\subsection{Bounding the number of output edges}
Lemma~\ref{lem:lb-profile} reduces the value bound to two counts: how far the output can exceed
the ordinary capacity of a stage, and how many output edges have its distinguished label.
Inequality~\eqref{eq:lb-slot-bound} bounds the first excess by the number of edges on special
base pairs. We bound these quantities separately, at the times when the stream stops revealing
information about each sample.
Fix a streaming algorithm whose complete memory state has $s$ bits.
Let $Z_i$ be the number of its final stage-$i$ edges whose base pair is
special, and $G_i$ its number of stage-$i$ edges with label $o_i$.  Write
$d_b=\log_2(1/(1-\delta_b))$ and
$d_\ell=\log_2(1/(1-\delta_\ell))$.

There are $K$ public labels per stage. For example, if a state certifies $m$ surviving copy
edges while the distinguished label is still hidden, only $m/K$ of them have that label on
average. The analogous calculation on base edges uses the special-edge probability $P/t$
from~\eqref{eq:lb-rarity} instead.
\begin{lemma}[Recoverable edges in one stage]\label{lem:lb-two-bounds}
For each stage $i$,
\begin{equation}\label{eq:lb-two-bounds}
 \Ex{Z_i}\le\frac{KsP^2}{td_b},\qquad
 \Ex{G_i}\le\frac{s}{Kd_\ell}.
\end{equation}
\end{lemma}
\begin{proof}
The two bounds use different conditioning arguments.

For the first, fix the sampled subsets between vertex copies for all
potential base pairs, all labels $o_i$, and the data of other stages.
These are independent of the current stage's tuple $J$ and samples of
expansion edges.  Before phase $j$, fix its preceding transcript and
$J_{<j}$ as well.  The current sample of expansion edges remains uniform
of fixed size.  Let $Y$ be the $s$-bit memory state at the end of the
phase.  Lemma~\ref{lem:lb-certainty} bounds the expected number of
expansion edges in $\mathcal C(Y)$ by $s/d_b$.  Both $Y$ and
$\mathcal C(Y)$ are independent of $J_{\ge j}$: the edge set of phase
$j$ depends only on $J_{<j}$, and the objective depends on stage counts
and labels $o_i$ but not on $J$.  By \eqref{eq:lb-rarity}, the expected
number of special edges in $\mathcal C(Y)$ from this phase is at most
$sP/(td_b)$.

Every returned edge corresponding to an expansion edge from phase $j$
must have its base edge in $\mathcal C(Y)$.  Indeed, after fixing the full
tuple $J$, later samples of expansion edges are independent of the
present one; leaf edges depend only on $J$, and the common oracle is
independent of which expansion edges were sampled.  The
indistinguishability argument following Lemma~\ref{lem:lb-certainty}
therefore applies to the memory state at the end of the phase.  Sum
over $P$ phases, and observe that a matching between vertex copies
uses a fixed base pair at most $K$ times.  This proves the first bound.

For the second, instead fix the actual base graphs and all data preceding
stage $i$, while leaving the sampled subsets between vertex copies and
$o_i$ random.  Those subsets are independent uniform samples of fixed
size.  Let $Y$ be the memory state at the end of the stage.
Lemma~\ref{lem:lb-certainty} gives $\Ex{|\mathcal C(Y)|}\le s/d_\ell$.
By Lemma~\ref{lem:lb-objective}, the oracle values on the stream prefix
are independent of $o_i$, so $Y$ and $\mathcal C(Y)$ are independent
of this uniform label.  Every edge has one public label, so an expected
$1/K$ fraction of $\mathcal C(Y)$ has label $o_i$.  Later graph data
are independent of the current samples, no earlier edge is reintroduced,
and the common oracle does not depend on those samples.  Hence every
stage-$i$ edge used in the final output must belong to $\mathcal C(Y)$,
proving the second bound.

Both arguments condition on fixed algorithmic coins and then
average.  They do not require $Z_i$ and $G_i$ to be independent.
\end{proof}

\subsection{Completeness, soundness, and parameter choices}
We compare two matchings on the same random graph. The first consists of surviving
distinguished copies of a large base matching. The second is the algorithm's output, whose two
exceptional edge counts we have just bounded. We choose the constant number of stages first to
leave a strict gap for the proposed $R$. Increasing the graph size then makes each error
smaller than this gap.
\begin{proof}[Proof of Theorem~\ref{thm:lb-three}]
Fix a proposed algorithm, and let $s$ be its space in bits.
The word-space assumption gives $s=O(n\log^{a+1}n)$.
All parameters other than $N$ and its derived quantities will
be fixed before the size limit.

Choose $p$ so that $R<3p/(p+1)$.  Then choose fixed
$\bar c>1/2$ and $\xi>0$ with $\bar c<1$ such that
\begin{equation}\label{eq:lb-gap}
 R<\frac{p(1-\xi)}{\bar c(p+1)/(1+\bar c)}.
\end{equation}

Indeed, setting $\bar c=1/2$ and $\xi=0$ would make the right side $3p/(p+1)$. Since the
inequality is strict and its right side depends continuously on $\bar c$ and $\xi$, small fixed
deviations preserve it. To obtain the chosen $\bar c$, use $h$ sufficiently close to $2/3$ from
below.
Choose the blueprint and a fixed expansion density loss so that
$c=(2-2h)/(2-h)\le\bar c$.  Choose small positive deletion
fractions and concentration slack, as detailed below, and an
integer $A>a+1$.  Finally let $N$ grow along the expansion sequence.

For soundness, combine \eqref{eq:lb-slot-bound}, the integral
identity $cM=KB$, and Lemma~\ref{lem:lb-profile} to obtain the
pointwise inequality
\[
 f(M_{\rm out})\le\frac{c(p+1)}{1+c}+\mathcal E_N,\qquad
 \mathcal E_N=\frac1M\sum_i(Z_i+G_i).
\]
Since $n=4pNK$, $M=\Theta(NK)$, and $p$ is fixed,
Lemma~\ref{lem:lb-two-bounds} gives
\[
 \Ex{\mathcal E_N}
 =O\left(\frac{K\log^{a+1}N}{t}
            +\frac{\log^{a+1}N}{K}\right)=o(1).
\]
The first limit uses the superpolylogarithmic $t$, and the second
uses $A>a+1$.  Markov's inequality also gives
$\mathcal E_N=o(1)$ with probability $1-o(1)$.

For completeness, keep the sampled edges of the canonical matching and
all private-leaf base edges.  In phase $j$, the expected retained
canonical count is at least
$(1-\delta_b)H_j-H_j/m_j$.  Summing gives at least
$(1-\delta_b)H-P$.  Conditional on the tuple $J$, sampling without replacement
has nonpositive pairwise covariances and the samples from different phases
are independent.
The conditional mean is deterministic, since $H_j$ and $m_j$ are fixed.
Thus the unconditional variance is at most $H/4$.  Chebyshev's inequality
therefore gives, for every fixed $\gamma>0$, a base matching of size at least
$(1-\delta_b-\gamma)H+B\ge(1-\delta_b-\gamma)U$
with probability $1-o(1)$.

Denote this base matching in stage $i$ by $Q_i$.  For each of its edges,
take every surviving edge between the endpoint copies with label $o_i$.
Conditional on the full base construction and label, their count has
mean $(k_\ell/K^2)K|Q_i|$ and variance at most $K|Q_i|/4$.
For sufficiently large $N$, it is therefore at least
\[
 \left[(1-\delta_\ell-\gamma)(1-\delta_b-\gamma)-\gamma\right]M
 \ge (1-\delta_b-\delta_\ell-3\gamma)M
\]
with probability $1-o(1)$.  Choose the fixed positive constants
so that $\delta_b+\delta_\ell+3\gamma\le\xi$.
The resulting set is a matching, its stage counts are at most $M$, and
the $p$ stages are vertex-disjoint.  The observation after
Lemma~\ref{lem:lb-profile} gives
\[
 \Pr(\mathrm{OPT}\ge p(1-\xi))=1-o(1),\qquad
 \Ex{\mathrm{OPT}}\ge p(1-\xi)(1-o(1)).
\]
Set $D=p(1-\xi)/R-\bar c(p+1)/(1+\bar c)>0$.
On the completeness event, an $R$-approximate output has value
at least $p(1-\xi)/R$, and the pointwise soundness bound then
forces $\mathcal E_N\ge D$.  Consequently its success probability,
averaged over the hard distribution, is at most
$o(1)+\Ex{\mathcal E_N}/D=o(1)$ by Markov's inequality.
This contradicts any fixed positive every-input success guarantee.
Averaging an every-input expected guarantee instead contradicts
\[
 \frac{p(1-\xi)}R-o(1)
 \le\Ex{f(M_{\rm out})}
 \le\frac{\bar c(p+1)}{1+\bar c}+o(1).
\]

It remains to verify the claimed instance-sensitive space scale.
For each private-leaf base edge, the graph between its endpoint copies
contains $k_\ell$ edges and has maximum degree at most $K$.  A maximal
matching in it has at least $k_\ell/(2K)$ edges.  The $pB$ such graphs are
vertex-disjoint, so every realization has $\nu\ge pB k_\ell/(2K)=\Omega(n)$.
The converse $\nu\le n/2$ is immediate.  The deterministic counts
and rational oracle representation were established above.
\end{proof}

\section{Extensions}
\label{sec:extensions}

We extend the upper bound in two ways. For monotone objectives, a one-pass
reduction from $b$-matching to ordinary matching preserves the optimum
and hence the $4.525$ approximation factor. For nonnegative nonmonotone
objectives, we retain the streaming selection framework and develop a
different fractional optimization procedure and a two-layer rounding
scheme. These give a \NMFactor{}-approximation for both matching and
$b$-matching.

\begin{theorem}\label{thm:extensions}
Let $\nu_b$ be the maximum cardinality of a $b$-matching on a simple
graph, with endpoint capacities available when each edge is processed.
For a nonnegative submodular objective, there are randomized
one-pass $b$-matching algorithms with approximation factor
$4.525$ when the objective is monotone and \NMFactor{} for
general nonnegative submodular objectives. Both algorithms use
$O((\nu_b+1)\log^2 n)$ words and run in polynomial time with
polynomially many value queries. Both guarantees hold in expectation
and with probability at least $1-n^{-c}$ for any fixed $c>0$.
Taking all capacities equal to one gives the same \NMFactor{}
guarantee for ordinary nonmonotone matching.
\end{theorem}

If capacities are announced only on the first appearance of each vertex,
an additional $O(n)$ words suffice to store them.

These factors improve on $3+2\sqrt{2}$ for monotone $b$-matching
and $4+2\sqrt{3}$ for nonmonotone matching and $b$-matching~\cite{LW,HS}.
Appendix~\ref{sec:nm-two-layers} gives the nonmonotone algorithm for
ordinary matching, and Appendix~\ref{sec:nm-direct-b} extends it to
vertex capacities. For monotone objectives, the following reduction
gives the guarantee directly from Theorem~\ref{thm:main-upper}.

\begin{proposition}[Exact monotone capacity reduction]\label{prop:exact-monotone-capacity}
For a monotone submodular objective on a simple graph with arbitrary
vertex capacities, a deterministic one-pass reduction to ordinary matching
preserves the optimum. Any one-pass $\alpha$-approximation for ordinary
monotone submodular matching therefore yields an $\alpha$-approximation
for monotone submodular $b$-matching. Under the capacity access of
Theorem~\ref{thm:extensions}, the reduction preserves polynomial time,
polynomial query complexity, and a space bound of
$O((\nu+1)\log^a n)$ words with $\nu_b$ in place of $\nu$.
\end{proposition}

The proof is in Appendix~\ref{sec:exact-monotone-reduction-proof}.

\section{Open questions}
The optimal one-pass approximation factor for monotone submodular matching on simple graphs
lies between $3$ and $4.525$. The lower bound permits a factor of exactly $3$ and already
applies on bipartite graphs. It remains open whether such a factor is attainable.

The coefficient of the residual term identifies one limitation of our upper bound. The analysis
recovers a $1-e^{-D}$ fraction of this term, while the first rounding step requires $D<1/4$.
Even as $D$ approaches $1/4$, this gives a factor of $1/(1-e^{-1/4})\approx4.5208$ on the
residual contribution. A substantially smaller factor would require a stronger estimate for
this term, a rounding procedure allowing larger fractional degrees, or a tighter comparison
of the two contributions. A separate question is whether the same approximation can be
obtained with faster postprocessing, particularly in the packing and sampling steps.

\section*{AI Use Statement}

This work resulted from extensive interaction between the authors and
OpenAI Astra, with the authors providing suggestions and critically
checking Astra's responses through iterative exchanges. OpenAI Astra
was also used in writing and revising the manuscript through detailed
prompts from the authors, with Gemini Argon assisting in some parts.
The authors directed the process and evaluated the proposed ideas and
arguments. Lean files formalizing the main theorem are available
at: https://github.com/xinchen-yang/Submodular-Matching-in-Streaming.

\newcommand{\etalchar}[1]{$^{#1}$}

\clearpage
\appendix

\section{Detailed algorithm specification}
\label{sec:upper-algorithm-specification}
Algorithm~\ref{alg:upper-overview} postpones its optimization until the stream ends. Here we
specify exactly what it must retain for that optimization: one ordered list of accepted edges
per active scale, their recorded weights, the incident vertex potentials, and the best
singleton. We give the choice of scales and the update for an arriving edge first. The
remaining routines read only these lists. The order in which the routines draw random bits will
matter when the success probability is amplified.
The fixed inputs are a bound $p$ on individual coordinates, a bound
$D=r(1-r)$ on fractional degrees with rational $r\in(1/2,1)$,
interval boundaries and positive update rates $(c_j,A_j,B_j)$
satisfying \eqref{eq:source-drift}, and accuracies
$\varepsilon_0,\varepsilon_R$.
Their admissibility conditions are stated before their numerical values.
Write $L=2+\lceil\log_2(n/\varepsilon_0)\rceil$ and
$W=n/\varepsilon_0$.  All oracle values below refer to
$f-f(\varnothing)$.  We use fractional degree for the sum of the
coordinates incident to a vertex~\cite{GKPS2006} and vertex potentials
for the quantities $\phi_v$ maintained by the streaming algorithm,
following Levin and Wajc~\cite{LW}.  Ties use a fixed ordering of edge and
vertex identifiers.

The total allowed probability of estimation failure is
$\omega=\varepsilon_0$ for the expected
guarantee.  To obtain overall failure probability $\delta$ as well, use
$\omega=\min\{\varepsilon_0,\delta/2\}$.  For the guarantee in
Theorem~\ref{thm:main-upper}, take $\delta=n^{-\lceil c\rceil}$;
all assigned failure probabilities are then rational.
In each trial, round the fractional pair computed for every surviving
set of stored edges and select the best candidate, including the best
singleton, using the same rounding of objective values.  Trials are
independent conditional on all estimates made during the stream and
the optimization.

Assign failure probability $\omega/(2n^2L)$ to each estimate made
during the stream.  There are at most $n^2L$ estimates, including
those made in discarded instances.  Each sample uses fresh independent
inclusion indicators with the probabilities specified below.

\subsection{Geometric guesses and updates of vertex potentials}

Before the stream, choose the least integers $K_{\rm src},b_{\rm hist}$ satisfying
\[
 2^{K_{\rm src}}\ge\frac{4n^2L}{\omega},\qquad
 2^{b_{\rm hist}}\ge\frac{64n^4}{\varepsilon_0},
\]
and set
\[
 N_{\rm src}=\left\lceil\frac{2n^4K_{\rm src}}{\varepsilon_0^2}\right\rceil,
 \qquad
 p_{\rm bit}=2^{-b_{\rm hist}}\lfloor2^{b_{\rm hist}}p\rfloor.
\]
Doubling and rational comparisons determine these constants.  They
apply to every guess and arrival.  Hoeffding's inequality bounds the
failure probability of each estimate by $2^{1-K_{\rm src}}\le\omega/(2n^2L)$.
Assign failure probability $\delta_{\rm flow}=\omega/(2L)$ to the
optimization for each surviving set of stored edges.  The probability that any estimate made during the stream or the
optimization is inaccurate is then at most $\omega$.

\begin{algorithm}[!htbp]
\caption{One-pass submodular matching}
\label{alg:upper-stream}
\centering
\begin{minipage}{0.97\linewidth}
\textit{Input:} a stream of distinct edges of a simple graph, vertex bound
$n$, value oracle $f$, and fixed admissible parameters.\par
\textit{State:} the best singleton, a reference value denoted by
$\mathrm{anchor}$, and at most $L$ streaming instances.
\begin{pseudolines}
\codephase{Initialization}
\codeline{0}{$W\gets n/\varepsilon_0$, $L\gets 2+\lceil\log_2 W\rceil$, $V\gets0$.}
\codeline{0}{Initialize the set of active instances to $\varnothing$.}
\codephase{Edge stream}
\codeline{0}{\textbf{for each} arriving edge $e=uv$ \textbf{do}}
\codeline{1}{Query $f(\{e\})$ and update the best singleton and its value $V$.}
\codeline{1}{\textbf{if} $V=0$ \textbf{then continue}.}
\codeline{1}{\textbf{if} this is the first positive singleton or $V>W$ times the preceding maximum \textbf{then}}
\codeline{2}{Discard all instances and set $\mathrm{anchor}\gets V$.}
\codeline{1}{Keep exactly the guesses $\tau=2^j\,\mathrm{anchor}$ in $[V,WV]$. Initialize each new instance with empty storage and zero potentials.}
\codeline{1}{\textbf{for each} active guess $\tau$ \textbf{do}}
\codeline{2}{Run $\textsc{ProcessEdge}(\tau,e)$ (Algorithm~\ref{alg:upper-source}).}
\codephase{Postprocessing}
\codeline{0}{\textbf{if} $V=0$ \textbf{then return} $\varnothing$.}
\codeline{0}{$B\gets V$.}
\codeline{0}{\textbf{for each} surviving instance $\tau$ \textbf{do}}
\codeline{1}{Compute $(x_\tau,y_\tau)$ on $S_\tau$ by Algorithm~\ref{alg:upper-flow}.}
\codeline{0}{Initialize the best candidate to the best singleton.}
\codeline{0}{\textbf{for each} surviving instance $\tau$ \textbf{do}}
\codeline{1}{Generate independent candidates one at a time by Algorithm~\ref{alg:upper-completion} as specified in Appendix~\ref{sec:finite-joint-streaming-wrapper}.}
\codeline{1}{For each candidate $M$, retain a maximizer of $\lfloor f(M)/(\varepsilon_0B/2)\rfloor$ among all candidates seen.}
\codeline{0}{\textbf{return} the best candidate.}
\end{pseudolines}
{\small The edge loop makes the only scan of the input.  All later work
uses only stored edges.  A newly created instance processes
the edge that caused its creation.\par}
\end{minipage}
\end{algorithm}

\begin{algorithm}[!htbp]
\caption{ProcessEdge$(\tau,e=uv)$}
\label{alg:upper-source}
\centering
\begin{minipage}{0.97\linewidth}
\textit{Instance state:} stored edges $S$ in arrival order, their
weight estimates, and vertex potentials $\phi$, initially zero.
\begin{pseudolines}
\codephase{Estimate the marginal gain}
\codeline{0}{$\theta\gets\varepsilon_0\tau/n$, $d\gets\varepsilon_0\tau/n^2$, $\zeta\gets p\varepsilon_0^2\tau/n^4$, $J\gets0$.}
\codeline{0}{\textbf{for} $i=1,\ldots,N_{\rm src}$ \textbf{do}}
\codeline{1}{Draw a fresh independent sample $R_i\sim R_{p_{\rm bit}}(S)$.}
\codeline{1}{$\Delta_i\gets f(R_i\cup\{e\})-f(R_i)$.}
\codeline{1}{$J\gets J+\lfloor\Delta_i/d\rfloor$ and discard $R_i$.}
\codeline{0}{$\widehat w\gets d\max\{0,\lfloor J/(4N_{\rm src})\rfloor-1\}$.}
\codephase{Decide and update}
\codeline{0}{\textbf{if} $\widehat w<\theta$ or $\widehat w<\phi_u+\phi_v+d$ \textbf{then return}.}
\codeline{0}{Append $e$ to $S$ with permanent weight estimate $\widehat w$.}
\codeline{0}{Label the endpoints $H,L$ in decreasing order of potential, breaking ties by vertex identifier. Fix these labels for this update.}
\codeline{0}{$(\Delta_H,\Delta_L)\gets\textsc{UpdatePotentials}(\phi_H,\phi_L,p\widehat w)$.}
\codeline{0}{\textbf{for each} $v\in\{H,L\}$ \textbf{do}}
\codeline{1}{$\phi_v\gets\phi_v+\zeta\lceil\Delta_v/\zeta\rceil$.}
\end{pseudolines}
{\small The weight estimate is computed at arrival from independent
samples of the earlier stored edges.  Each complete potential increment
is rounded once; boundary crossings are evaluated exactly.\par}
\end{minipage}
\end{algorithm}
\paragraph{Updating the vertex potentials.}
The procedure below is denoted \textsc{UpdatePotentials}.  Its input
consists of potentials $P_H\ge P_L\ge0$ and an update length $b\ge0$;
the update rates satisfy \eqref{eq:source-drift}.  Save the starting potentials, and keep
the endpoint labels fixed.  All operations within the procedure are exact.
\begin{enumerate}[leftmargin=2em,itemsep=3pt]
\item If $b=0$, return zero increments.  Otherwise choose the smallest
      $j\in\{1,\ldots,k\}$ such that $P_H\le c_j(P_H+P_L)$.
\item If $j=1$, add $b(A_1,B_1)$ to the potentials and finish.
\item For $j\ge2$, set $c=c_{j-1}$ and compute
      \[
       s=\frac{P_H-c(P_H+P_L)}{c(A_j+B_j)-A_j}.
      \]
      Both numerator and denominator are positive.  If $b\le s$,
      add $b(A_j,B_j)$ to the potentials and finish.
\item Otherwise add $s(A_j,B_j)$, replace $b$ by $b-s$, decrease
      $j$ by one, and return to Step 2.
\end{enumerate}
Return the differences between the final and starting potentials.
The initial comparison uses cross-multiplication, so it selects interval 1
even when both potentials are zero.  At an internal boundary it selects
the interval on the left.  The interval index decreases at every crossing;
the procedure takes $O(k)$ rational operations.  The caller rounds
the two complete increments only after this procedure returns.

\subsection{Optimizing the multilinear relaxation}
For each surviving guess, the list $S$ and the weights $\widehat w$ are now fixed. The linear
part of a direction's value can be computed from these weights. For the submodular part we
estimate the benefit of increasing one coordinate by sampling on $S$. The algorithm below
repeats this calculation for a bounded number of steps and returns a feasible pair that can be
rounded more than once.

\begin{algorithm}[!htbp]
\caption{Continuous greedy on the stored edges}
\label{alg:upper-flow}
\centering
\begin{minipage}{0.97\linewidth}
\textit{Input:} stored edges $S$, their weight estimates $\widehat w$,
guess $\tau$, accuracy $\varepsilon>0$, and failure probability
$\delta_{\rm flow}$.  Write $s=|S|$.
\begin{pseudolines}
\codephase{Initialization}
\codeline{0}{\textbf{if} $s=0$ \textbf{then return} $(0,0)$.}
\codeline{0}{$K\gets\max\{1,\lceil2D^2s^2/\varepsilon\rceil\}$, $h\gets1/K$, $\xi\gets\min\{1/4,\varepsilon/(16Ds)\}$.}
\codeline{0}{Set $Q$ to the least power of two at least $2K/\xi$.}
\codeline{0}{$(x,y)\gets(0,0)$.}
\codephase{Continuous greedy iterations}
\codeline{0}{\textbf{for} $j=0,\ldots,K-1$ \textbf{do}}
\codeline{1}{$q\gets Q$.}
\codeline{1}{\textbf{repeat} $K-j-1$ times}
\codeline{2}{$q\gets\lfloor(1-D/K)q\rfloor$.}
\codeline{1}{$\widehat a_{j+1}\gets q/Q$ (within $\xi/2$ of $a_{j+1}=(1-D/K)^{K-j-1}$).}
\codeline{1}{\textbf{for each} $e\in S$ \textbf{do}}
\codeline{2}{Estimate $\partial_e F_g(y)$ using samples $f(R\cup Z\cup\{e\})-f(R\cup Z)$ with independent $R\sim R_p(S)$ and $Z\sim R_y(S\setminus\{e\})$.}
\codeline{2}{Multiply the empirical gradient coordinate by $\widehat a_{j+1}$ and store the result as $\widehat d_e$, within $\xi\tau$ of $a_{j+1}\partial_e F_g(y)$.}
\codeline{1}{\label{line:upper-flow-packing}Use the linear packing routine below to find an exactly feasible $(v_x,v_y)\in\mathcal P_D$ maximizing $\widehat w\cdot v_x+\widehat d\cdot v_y$ to additive error $\varepsilon\tau/8$.}
\codeline{1}{$(x,y)\gets(x,y)+h(v_x,v_y)$.}
\codeline{0}{\textbf{return} $(x,y)$.}
\end{pseudolines}
{\small The feasible directions share one rational grid.  No matrix
with quadratically many stored entries is constructed.\par}
\end{minipage}
\end{algorithm}
\paragraph{Solving the linear packing problem.}
After scaling $v_x=p t_x$ and $v_y=D t_y$, each direction problem is
$\max\{c\cdot t:At\le\mathbf1,\ 0\le t\le\mathbf1\}$, after dividing its objective by $\tau$, with
$N=2s$ columns, $M=O(s)$ rows, and all coefficients in $[0,1]$.
Its objective coefficients are $p\widehat w_e/\tau$ and
$D\widehat d_e/\tau$, respectively. Given a dyadic additive accuracy $\epsilon$, set
\[
 R=\lceil4N^2/\epsilon\rceil,
 \eta=\epsilon/(2MN^2),
 J=\lceil2R^2M/(\eta\epsilon)\rceil,
 \delta=\epsilon/(2N).
\]
Initialize $u=0$ and $b=0$.  Repeat $J$ times: form
$t_i=\mathbf1\{c_i>(A^\top u)_i\}$, add $t$ to $b$, and set
$u\leftarrow\operatorname{clip}_{[0,R]^M}(u+\eta(At-\mathbf1))$.
Return $b/[J(1+\delta)]$.  Adjacency scans implement both matrix
products; the row types and their products are listed in the proof.
The projection argument proves feasibility and additive error
$\epsilon$.  Taking $0<\epsilon\le\min\{\varepsilon/8,1\}$ gives the required accuracy in line~\ref{line:upper-flow-packing} of Algorithm~\ref{alg:upper-flow}.

\subsection{Rounding and independent repetitions}

\begin{algorithm}[!htbp]
\caption{Rounding the fractional solution}
\label{alg:upper-completion}
\centering
\begin{minipage}{0.97\linewidth}
\textit{Input:} stored edges $S$, feasible pair $(x,y)$, and fixed accuracies.
\begin{pseudolines}
\codephase{First matching}
\codeline{0}{\textbf{if} $S=\varnothing$ \textbf{then return} $\varnothing$.}
\codeline{0}{For each vertex $v$, set $\ell_v\gets\sum_{e\ni v}y_e$. Treat missing edge coordinates as zero.}
\codeline{0}{\textbf{for each} $uv\in S$ \textbf{do}}
\codeline{1}{Accumulate $C_{uv}\gets\sum_{w\ne u,v}y_{uw}y_{vw}$ by adjacency scans without storing a triangle list.}
\codeline{1}{$q_{uv}\gets(1-\ell_u)(1-\ell_v)+y_{uv}(1-\ell_u-\ell_v+y_{uv})-C_{uv}$.}
\codeline{1}{Set $a_{uv}$ to $x_{uv}/q_{uv}$ rounded down to the common mesh $\varepsilon_0/(8|S|)$.}
\codeline{0}{Sample $T$ using the finite sampler for Shearer's distribution $\mu_y$ with edge marginals $y$, to total-variation error at most $\varepsilon_0/(4n)$.}
\codephase{Second matching}
\codeline{0}{For each vertex $v$, set $X_v\gets\mathbf1\{v\notin V(T)\}$.}
\codeline{0}{For each vertex $u$, set $Z_u\gets\sum_{uv\in S}a_{uv}X_v$.}
\codeline{0}{$\varepsilon_A\gets\varepsilon_R/4$.}
\codeline{0}{\textbf{for each} $uv\in S$ \textbf{do}}
\codeline{1}{$z_{uv}\gets\dfrac{a_{uv}}{1+\varepsilon_A}X_uX_v\mathbf1\{Z_u\le1+\varepsilon_A\}\mathbf1\{Z_v\le1+\varepsilon_A\}$.}
\codeline{0}{$\varepsilon_G\gets\varepsilon_R/2$, $\gamma\gets1-\varepsilon_G/4$.}
\codeline{0}{For each vertex $v$, set $b_v\gets1-\gamma z(\delta(v))$.}
\codeline{0}{For each $uv$ with $z_{uv}>0$, set $\lambda_{uv}\gets\gamma z_{uv}/(b_ub_v)$.}
\codeline{0}{\textbf{if} $z=0$ \textbf{then return} $T$.}
\codeline{0}{On the positive support of $z$, sample $N$ from the Gibbs distribution with activities $\lambda$ using the finite insert/delete/swap chain, to total-variation error at most $\varepsilon_0/(4n)$.}
\codeline{0}{\textbf{return} $T\cup N$.}
\end{pseudolines}
{\small The exact two-endpoint probability includes the triangle
correction.  Both endpoints of every edge available to $N$ are
unmatched by $T$, so the returned union is a matching.\par}
\end{minipage}
\end{algorithm}

The sampling algorithm for $\mu_y$ and the Gibbs chain are specified
in their respective proofs, including deterministic stopping rules
and finite random bits.  One trial per surviving instance suffices
for the expected guarantee.  For failure probability $\delta$, set
the total allowed probability of estimation failure to
$\omega=\min\{\varepsilon_0,\delta/2\}$ and use
$\lceil\pi_*^{-1}\log(2/\delta)\rceil$ independent trials, where
\[
 \pi_* =\frac{R_1-R_0}{R_0(R_1-1)}>0,
 \qquad \frac1{R_1}-\frac{\varepsilon_0}{2}>
             \frac1{R_{\rm target}}.
\]
Here $R_0$ is a valid expected factor conditional on the estimates
computed during the stream and the optimization, and $R_1>R_0$
allows for the final comparison error.  The appendix selecting the
constants fixes these numbers.  Every trial reuses the same stored
edges and fractional solutions.

\FloatBarrier
\section{Implementing the updates of vertex potentials}
\label{sec:joint-source-path}
An accepted edge raises both endpoint potentials over a prescribed amount of time. If the ratio
of the two potentials crosses an interval boundary, their rates change before the update ends.
The following condition ensures that boundaries are crossed in one direction. We can therefore
compute the entire update by visiting each crossed interval once and round the two increments
only at the end.
Fix rational rates $A_j\ge B_j>0$ and boundaries
$1/2=c_0<c_1<\cdots<c_k=1$, where $k$ is fixed.  Our rates satisfy
the following additional direction condition:
\begin{equation}
 A_1=B_1,\qquad
 A_j<c_{j-1}(A_j+B_j)\quad(2\le j\le k).
 \tag{SP0}\label{eq:source-drift}
\end{equation}
Appendix~\ref{sec:joint-exact-parameters} verifies this condition
exactly for the certified rates.  It is also preserved throughout
the approximation family, which keeps the rates fixed and reduces $p$.
Write $C_j=A_j+B_j$ and $C_{\max}=\max_j C_j$.
Condition (SP0) makes the share of each increment assigned to the
endpoint with larger potential smaller than its share of the current
potential sum, except when the two potentials are equal.  Thus both
potentials rise while the larger endpoint's fraction of their sum
moves toward $1/2$.

For an accepted edge, start from its actual potentials $P_H\ge P_L$
and use update length $p\widehat w_e$.  The labels $H,L$ remain fixed
during the update.  In interval $j$, after an update of length $s$,
the fraction of the potential sum at $H$ is
\[
 r(s)=\frac{P_H+A_js}{P_H+P_L+C_js}.
\]
If $P=P_H+P_L>0$ and $r_0=P_H/P$, then
\[
 r'(s)=\frac{P(A_j-r_0C_j)}{(P+C_js)^2}.
\]
For $j\ge2$ and $r_0\in(c_{j-1},c_j]$, condition (SP0) makes this
derivative negative.  The next boundary is $c_{j-1}$, reached after
an update of finite positive length
\begin{equation}
 s_{\rm hit}=
 \frac{P_H-c_{j-1}(P_H+P_L)}{c_{j-1}C_j-A_j}.
 \tag{SP1}
\end{equation}
The numerator is positive by the interval convention; the denominator
is positive by (SP0).  If the specified update ends earlier, the procedure stops within the
interval.  Otherwise it reaches the boundary exactly and, if part of
the update remains, continues in interval $j-1$.  At that boundary
the fraction is still nonincreasing, so no additional boundary rule
is needed.

In interval 1, the two rates are equal.  The potential difference therefore
stays fixed while the sum increases.  The ratio stays in
$[1/2,c_1]$ and moves toward $1/2$; when the starting potentials differ,
it does not reach $1/2$ during an update of finite length.  The
remaining update can consequently be performed in this interval.  If both starting potentials
are zero, the comparison $P_H\le c_j(P_H+P_L)$ selects interval 1,
and its equal increments keep the ratio $1/2$ thereafter.

This proves the correctness of $\textsc{UpdatePotentials}$ in
Appendix~\ref{sec:upper-algorithm-specification}.  It crosses at most
$k-1$ boundaries and uses $O(k)$ rational operations, including the
initial interval search.  Each derivative pair has $A_j\ge B_j>0$,
so the potential at $H$ remains at least the potential at $L$, and
both endpoint potentials increase
during every positive-length stage.

Finally, round each complete ideal endpoint increment upward to a
multiple of $\zeta>0$, adding less than $\zeta$ at either endpoint.
There is no rounding between stages.  The monotonicity just proved
concerns one ideal update; the final rounding can perturb its ratio.
Every later edge starts anew from its actual rounded endpoint potentials.

\section{Geometric guesses, estimation accuracy, and space}
\label{sec:joint-source-foundations}
The update rates satisfy \eqref{eq:source-drift} throughout this section.
Let $A_{\max}=\max_j A_j$, $A_{\min}=\min_j A_j$,
$B_{\min}=\min_j B_j$, and
$\kappa=\min\{A_{\min},2B_{\min}\}>0$, with $\kappa p\le1$.
Write $\kappa_{\rm src}=\kappa$.
\subsection{Geometric guesses and the stored edges}

If $n<2$ or all singleton values are zero, return the empty matching.
Monotone submodularity makes the latter case have objective
identically zero.  Otherwise retain a largest input singleton.
Let $V_i$ be the largest singleton value seen through arrival $i$,
and let $B>0$ be its final value.
Put $W=n/\varepsilon_0$.

Maintain geometric guesses with common ratio two, keeping exactly
those in $[V_i,WV_i]$.  The first positive singleton initializes the
reference value from which these guesses are generated.
If a new singleton maximum jumps from $B_{\rm old}$ to
$B_{\rm new}>WB_{\rm old}$, discard all old instances, reset the
reference value to $B_{\rm new}$, and create the eligible guesses.
Otherwise retain the reference value, discard guesses below $V_i$,
and create newly eligible guesses above the old interval.
Every new instance processes the edge that triggered its creation.
There are always at most
\[
 L=2+\lceil\log_2(n/\varepsilon_0)\rceil
\]
active guesses. Between resets of the reference value, each singleton-maximum increase
is at most a factor $W$, and there are at most $n^2$ arrivals.  Thus
the exponent of every guess relative to the reference value is
$O(n^2\log W)$ and occupies $O(\log n)$ bits for fixed accuracy.  Only $O(L)$ doublings and comparisons are needed
at an arrival, even after an arbitrarily large value jump.

For each active guess $\tau$, set
\[
 \theta_\tau=\varepsilon_0\tau/n,\qquad
 d_\tau=\varepsilon_0\tau/n^2.
\]
The instance starts with an empty set $S$ of stored edges and zero
vertex potentials.  At edge $e$, estimate
$w_e=\Ex{f(e\mid R_p(S_{<e}))}$ to accuracy $d_\tau$.
Subtract $d_\tau$, truncate at zero, and round down to the
$d_\tau$ grid, obtaining the weight estimate $\widehat w_e$.
On the event that all estimates made during the stream are accurate,
\begin{equation}
 0\le\widehat w_e\le w_e\le\widehat w_e+3d_\tau.
 \tag{SF1}
\end{equation}
With $P_e=\phi_u^-+\phi_v^-$, accept $e$ precisely when
\begin{equation}
 \widehat w_e\ge\theta_\tau,\qquad
 \widehat w_e\ge P_e+d_\tau.
 \tag{SF2}
\end{equation}
Set $r_e=p\widehat w_e$ and store the edge and its weight estimate.
Starting from the current potentials, apply the ideal update of
Appendix~\ref{sec:joint-source-path} over an interval of length $r_e$.
Ties between endpoint potentials are broken by vertex identifier.  That appendix
proves that the interval index only decreases during the ideal update,
including the initial update from zero potentials.
The update uses $O(k)$ rational operations and keeps the endpoint ratio
within $[c_{j-1},c_j]$ whenever interval $j$ is selected.

For this instance use the fixed rounding step
\begin{equation}
 \zeta_\tau=p\,d_\tau\varepsilon_0/n^2
                =p\varepsilon_0^2\tau/n^4.
 \tag{SF3}
\end{equation}
Round each complete ideal increment up to a multiple of
$\zeta_\tau$, and then add the rounded increments to the actual
potentials.  This is done only once per endpoint per edge, after all
changes of interval during that update.
Retain every accepted edge, in arrival order, until the instance is
discarded.  Every input edge appears at most once in $S$.

The operational inequality $r_e\ge pP_e$ holds on every
transcript, including inaccurate ones.
During each stage, the endpoint ratio belongs to $[c_{j-1},c_j]$,
where the pointwise inequality for the auxiliary potential functions holds.
Throughout the ideal update, the potential sum is at most
$(1+C_{\max}p)\widehat w_e$.
Rounding each increment upward cannot decrease the values of the
auxiliary potential functions;
no threshold condition on the updated vertex potentials is needed.
The proof of the bound on $W_D(O)$ therefore applies to these
rounded updates and to the estimates stored at arrival.

\subsection{Deterministic space and bounded numerical indices}
The space guarantee must hold even on a run in which some estimates are inaccurate. We first
bound how large and how small a positive potential can be at a fixed guess. Each accepted
incidence after the first multiplies the potential by a constant greater than one. Combining
these facts bounds the number of edges retained at a vertex and the number of bits needed for
its indices.
Every sample used to estimate a marginal value belongs to
$[0,f(\{e\})]\subseteq[0,\tau]$.
Therefore $0\le\widehat w_e\le\tau$ even on inaccurate
transcripts.
Before an accepted update each endpoint potential is at most $\tau$.
After the ideal update and rounding, it is at most
\[
 (1+A_{\max}p)\tau+\zeta_\tau
       \le[1+(A_{\max}+1)p]\tau.
\]
A vertex's first positive potential is at least
$B_{\min}p\theta_\tau$.
Throughout any ideal update, the rate at $H$ is at least
$A_{\min}$ and the rate at $L$ is at least $B_{\min}$.
Since $r_e\ge p(\phi_H^-+\phi_L^-)$, every later accepted
incidence multiplies a positive endpoint potential by at least
$1+\kappa p$.  Upward rounding only increases this factor.
Consequently each vertex has at most
\begin{equation}
 D_{\rm inc}
 =1+\frac2{\kappa p}
       \log\frac{[1+(A_{\max}+1)p]n}{B_{\min}p\varepsilon_0}
 \tag{SF4}
\end{equation}
accepted incidences in one instance.

For analysis only, take the endpoints of a maximum-cardinality
matching of the input.  They form a vertex cover of size at most
$2\nu$, so each set $S$ has at most $2\nu D_{\rm inc}$ edges.
The algorithm need not know or construct this cover.
Vertices with positive potential are kept in sparse dictionaries with
default value zero; each is incident to a stored edge.
All postprocessing arrays use consecutive identifiers for vertices incident to stored edges.
Across all guesses the word bound is
\begin{equation}
 O\left(
 \frac{\nu+1}{\kappa p}
       \log\frac n{B_{\min}p\varepsilon_0}
       \log\frac n{\varepsilon_0}\right)
       =O((\nu+1)\log^2n).
 \tag{SF5}
\end{equation}

Weight estimates are multiples of $d_\tau$, update lengths are
multiples of $pd_\tau$, and actual vertex potentials are multiples
of $\zeta_\tau$.
The respective indices are bounded by
$n^2/\varepsilon_0$, $n^2/\varepsilon_0$, and
$[1+(A_{\max}+1)p]n^4/(p\varepsilon_0^2)$.
All have $O(\log n)$ bits for the fixed parameters.
The update length measured in units of $\zeta_\tau$ has a fixed denominator
and magnitude at most $n^4/\varepsilon_0^2$.

Ideal boundary times and stage increments are rational expressions
with fixed rational
coefficients in the indices of the current potentials and the
remaining update length.
There are at most $O(k)$ stages per edge.
Since $k$ and all parameters are fixed, intermediate numerators
and denominators have $O(\log n)$ bits.
Compute each complete endpoint increment exactly, take its ceiling
in $\zeta_\tau$ units by integer division, and retain only that
integer.  Rounding is not performed between stages, and temporary
denominators do not accumulate across accepted edges.
This implements (SF3) exactly with bounded arithmetic.

When estimating weights, round each sample down to
$d_\tau/4$ and allow Hoeffding deviation $d_\tau/2$.
Draw each independent inclusion indicator with
\begin{equation}
 p_{\rm bit}=2^{-b_{\rm hist}}
                  \lfloor2^{b_{\rm hist}}p\rfloor,\qquad
 2^{b_{\rm hist}}\ge64n^4/\varepsilon_0.
 \tag{SF6}
\end{equation}
The resulting independent sample has total-variation error at most
$n^2 2^{-b_{\rm hist}}$.
Its contribution to the marginal mean error is at most
$d_\tau/64$.
Including sample rounding, the total error is at most
$(1/4+1/2+1/64)d_\tau<d_\tau$.
Each sample, subset, and accumulator is processed individually.
Rounding can be implemented by a bounded binary search among
integer grid indices.  Polynomial sample counts and bounded index
ranges give $O(\log n)$-bit accumulators relative to $\tau$.

\subsection{An upper bound on the optimum}
We now account for the finite-precision errors in Lemma~\ref{lem:main-charge}. There are three places
where the ideal charge can undercount the optimum: the guess may start late, its weight
estimates are conservative, and the rejection test and rounded potentials have slack. We bound
these costs at a single scale $\tau$. This keeps the total additive error proportional to the
best singleton even when the stream spans a large range of values.
For the final guess $B\le\tau<2B$ used in the analysis, every arrival
before its instance was created has singleton value below
$\theta_\tau=\varepsilon_0\tau/n$.  This follows from the previous
interval bound when a guess becomes newly eligible, and from the
rule for resetting the reference value after a large value jump.  Deleting the ignored
prefix from an optimum loses at most $\varepsilon_0\tau/2$.
On the event that all estimates made during the stream are accurate,
telescoping in the arrival order gives $0\le F(p\mathbf1_S)-G_S\le3\varepsilon_0\tau$.
For a rejected edge, the final value $g(\{e\})$ is at most
$P_e+\theta_\tau+4d_\tau$.
Let $O$ be the retained part of the remaining optimum.  The rejected
optimum edges have distinct endpoints outside $V(O)$, so their total
charge to vertex potentials at arrival is at most the sum of the final
vertex potentials outside $V(O)$.
The threshold and marginal allowances total at most
$(n/2)(\theta_\tau+4d_\tau)\le3\varepsilon_0\tau/2$.
Upward rounding adds at most $2n^2\zeta_\tau\le\varepsilon_0\tau/4$
to the sum of the vertex potentials.  With $\psi_v$ the sum of the
ideal increments at $v$, this yields
\[
 \operatorname{OPT}\le G_S+\sum_{v\notin V(O)}\psi_v+g(O)
                                    +6\varepsilon_0\tau.
\]
This bound is uniform over the accepted transcript and independent of
the auxiliary potential functions chosen for vertices in $V(O)$ and outside $V(O)$.
For an assigned failure probability $\delta_{\rm src}$ for one estimate, at most
$\lceil(2n^4/\varepsilon_0^2)\log(2/\delta_{\rm src})\rceil$
samples give the stated Hoeffding allowance.  There are at most
$n^2L$ estimates, including estimates made in discarded instances.  The overall
algorithm assigns a total failure probability to these estimates.

\section{\texorpdfstring{Local inequalities for the bound on $W_D(O)$}{Local inequalities for the linear objective}}
\label{sec:joint-inherited-potential}
For Lemma~\ref{lem:main-certificate}, the total charge to each vertex must fit its dual price,
and each accepted edge must pay for its increase in $X(O)$. The integral inequality below
enforces the first requirement over the whole history of a vertex. The local inequality
enforces the second at each point of one update. We derive the differential comparison from
these two inequalities here. Appendix~\ref{sec:joint-exact-parameters} checks them for the
stated rates and functions.

Fix rational update rates $A_j\ge B_j>0$ on ratio intervals
$[c_{j-1},c_j]$, with $c_0=1/2$ and $c_k=1$, satisfying
\eqref{eq:source-drift}.  Put
$C_{\max}=\max_j(A_j+B_j)$ and
$t_*=(1+pC_{\max})^{-1}$.  An accepted edge has weight estimate
$\widehat w_e\ge\phi_u^-+\phi_v^-$ and update length $r_e=p\widehat w_e$.  Its ideal
endpoint update rates are $A_j,B_j$ in interval $j$, and it follows the
finite boundary rule in Appendix~\ref{sec:joint-source-path}.  Each complete
endpoint increment is rounded upward only after the edge's ideal
path has finished.  Write $G_S=\sum_e r_e$, and let $\psi_v$ be the
sum of the ideal increments at $v$, excluding grid remainders.

Fix $D\in(0,1/4]$ and $\sigma>0$.  Suppose two nonnegative,
nondecreasing, bounded functions $H_0,H_1$, each zero near zero, obey
\begin{equation}
 \int_0^\infty H_i(a)a^{-2}\,da\le t_*\qquad(i=0,1).
 \tag{JI1}\label{eq:main-integral}
\end{equation}
These functions determine the auxiliary potential functions according to whether
a vertex belongs to $V(O)$; the algorithm never consults them.
The functions may differ because vertices in $V(O)$ and outside it
have different capacities and charges to vertex potentials.
For every interval, every endpoint ratio $c\in\{c_{j-1},c_j\}$,
every $a,d\ge0$, and every pair $(i_H,i_L)\in\{0,1\}^2$, require
\begin{equation}
 \begin{split}
 &(1-Di_H)A_j H_{i_H}(a)+(1-Di_L)B_j H_{i_L}(d)\\
 &\quad+[1-D(i_H+i_L)](1-ca-(1-c)d)_+\\
 &\hspace{3em}\ge
 \sigma[1+(1-i_H)A_j+(1-i_L)B_j].
 \end{split}
 \tag{JI2}\label{eq:main-local}
\end{equation}
The four cases are
\[
\begin{array}{c|c|c|c|c}
 (i_H,i_L)&H_{i_H}(a)\text{ coefficient}&H_{i_L}(d)\text{ coefficient}
       &\text{last coefficient}&\text{right-hand side}\\\hline
 (0,0)&A_j&B_j&1&\sigma(1+A_j+B_j)\\
 (1,0)&(1-D)A_j&B_j&1-D&\sigma(1+B_j)\\
 (0,1)&A_j&(1-D)B_j&1-D&\sigma(1+A_j)\\
 (1,1)&(1-D)A_j&(1-D)B_j&1-2D&\sigma.
\end{array}
\]
The local condition also implies the boundary inequality
\begin{equation}
 \min\{(1-Di_H)A_jH_{i_H}(\infty),
        (1-Di_L)B_jH_{i_L}(\infty)\}
 \ge\sigma[1+(1-i_H)A_j+(1-i_L)B_j].
 \tag{JI3}\label{eq:main-tail}
\end{equation}
Indeed, each interval has an endpoint ratio $c\in(0,1)$.  Substitute
$(a,d)=(1/c,0)$ and $(a,d)=(0,1/(1-c))$ into (JI2), using
$H_0(0)=H_1(0)=0$, and then use monotonicity.  Thus (JI3) follows
from the two stated conditions.  The exact checker also verifies
these additional inequalities.

\begin{theorem}[A lower bound on $W_D(O)$]
For every matching $O\subseteq S$, the linear program~\eqref{eq:main-comparator} on the
stored weight estimates satisfies
\begin{equation}
 W_D(O)\ge\sigma\left[G_S+\sum_{v\notin V(O)}\psi_v\right].
 \tag{JI4}
\end{equation}
Every accepted edge and every upward rounding increment is retained.
\end{theorem}
\begin{proof}
Fix $O$, write $I_v=\mathbf1_{v\in V(O)}$,
$s_v=1-DI_v$, and $k_{uv}=1-D(I_u+I_v)$.  The dual objective of
\eqref{eq:main-comparator}, after eliminating the dual variables for individual coordinate bounds, is
\[
 \sum_v s_v \lambda_v+p\sum_{uv\in S}k_{uv}(\widehat w_{uv}-\lambda_u-\lambda_v)_+,
 \qquad \lambda_v\ge0.
\]
It suffices to bound this from below by (JI4) for every $\lambda$.
Use the same auxiliary potential functions $\Phi_v$ as in the main proof:
\[
 \Phi_v(\phi)=s_v\int_0^\phi
 H_{I_v}\!\left(\frac{\lambda_v}{t_*u}\right)\,du
 \quad(\lambda_v>0),
 \qquad \Phi_v(\phi)=0\quad(\lambda_v=0).
\]
The integrand at $u=0$ is its limiting value.  These functions are
nondecreasing and initially zero.  Substitution in the integral and
(JI1) give the final bound
\[
 \Phi_v(\phi)
 \le\frac{s_v\lambda_v}{t_*}\int_0^\infty H_{I_v}(a)a^{-2}\,da
 \le s_v\lambda_v.
\]  During an ideal update with endpoint potentials
$\phi_H,\phi_L$, the derivative with respect to the update parameter is
\[
 s_H A_j H_{I_H}\!\left(\frac{\lambda_H}{t_*\phi_H}\right)
 +s_L B_j H_{I_L}\!\left(\frac{\lambda_L}{t_*\phi_L}\right).
\]
When both potentials are positive, set
$c=\phi_H/(\phi_H+\phi_L)$ and use the displayed two arguments
as $a,d$.  At every point of the edge path,
$t_*(\phi_H+\phi_L)\le \widehat w_e$.  Consequently
\[
 ca+(1-c)d=\frac{\lambda_H+\lambda_L}{t_*(\phi_H+\phi_L)}
              \ge\frac{\lambda_H+\lambda_L}{\widehat w_e}.
\]
Since the positive-part term decreases with its argument, (JI2) implies
\[
 \frac{d}{ds}(\Phi_H+\Phi_L)
 \ge\sigma[1+(1-I_H)A_j+(1-I_L)B_j]
           -k_e\frac{(\widehat w_e-\lambda_H-\lambda_L)_+}{\widehat w_e}.
\]
Checking only endpoint ratios is enough: the potential terms do
not depend on $c$, and the positive-part term is minimized when its
affine argument is largest, which occurs at an interval endpoint.
At an internal boundary, the next stage uses the interval on the left;
its endpoint inequality is among the conditions already checked.

Both ideal endpoint rates are strictly positive.  Hence both potentials
are positive away from the initial point of each edge update of positive
length.  The auxiliary potential functions $\Phi_v$ are Lipschitz because $H_0,H_1$ are bounded,
and the ideal paths are piecewise linear.  Their compositions are
therefore absolutely continuous.  The preceding derivative inequality
holds almost everywhere, which suffices for integration.  Zero dual
coordinates are already covered by $H_i(0)=0$; no separate case with zero endpoint potentials is needed.
The indicator $I_v$ is fixed throughout the sequence of stored edges.
Thus each vertex uses the same function $H_{I_v}$ even when it changes
from the endpoint with larger potential to the endpoint with smaller
potential, or conversely, between successive updates.

Integrate over the update interval of length $p \widehat w_e$.
Upward rounding of the vertex potentials cannot decrease the values of
the auxiliary potential functions.  Summing over all accepted edges and using the final
bounds on $\Phi_v$ gives
\[
 \sigma\left[G_S+\sum_{v\notin V(O)}\psi_v\right]
 \le\sum_v s_v\lambda_v+p\sum_e k_e(\widehat w_e-\lambda_u-\lambda_v)_+.
\]
Strong duality proves (JI4).
\end{proof}

\section{Solving linear packing problems in linear working space}
\label{sec:linear-space-positive-packing}
One step of continuous greedy maximizes a linear function over packing constraints. Storing an
explicit constraint matrix could already exceed our space bound. We therefore use a solver that
stores one price per constraint and an accumulated choice per variable. Each iteration needs
only products with the constraint matrix and its transpose. The next subsection shows how to
compute both products directly on the stored edges.

\begin{lemma}[Linear packing in linear working space]
\label{lem:linear-space-positive-packing}
Let $A\in\mathbb Q^{M\times N}$ have entries in $[0,1]$, let
$c\in\mathbb Q^N$ have entries in $[0,1]$, and suppose $M,N\ge1$.
For $0<\epsilon\le1$, consider
\[
 V=\max\{c\cdot t:At\le\mathbf1,\ 0\le t\le\mathbf1\}.
\tag{PK1}
\]
There is a deterministic algorithm returning a feasible $t^{\rm out}$
with $c\cdot t^{\rm out}\ge V-\epsilon$.
It stores $O(M+N)$ numbers, in addition to whatever representation is
used to multiply by $A$ and $A^\top$.
If all matrix entries lie on a common rational grid of denominator
$Q$, the working numbers have
$O(\log(MNQ/\epsilon))$ bits, up to the bit length of the supplied
objective coefficients.  The number of iterations is polynomial in
$M,N,\epsilon^{-1}$.
\end{lemma}

\paragraph{Algorithm.}
First replace the requested accuracy by a dyadic number in
$(\epsilon/2,\epsilon]$, and call that dyadic accuracy $\epsilon$
for the remainder of the algorithm and proof.  This can only strengthen
the requested guarantee, changes the stated polynomial bounds by a
constant factor in the accuracy parameter, and prevents an irrelevant
large denominator in a rational request from entering the update grid.
Set
\begin{equation}
 R=\left\lceil\frac{4N^2}{\epsilon}\right\rceil,
 \qquad \eta=\frac{\epsilon}{2MN^2},
 \qquad J=\left\lceil\frac{2R^2M}{\eta\epsilon}\right\rceil,
 \qquad \delta=\frac{\epsilon}{2N}.
\tag{PK2}
\end{equation}
Start with $u_1=0\in[0,R]^M$ and integer accumulators $b_i=0$.
For $j=1,\ldots,J$ do the following:
\begin{enumerate}
\item Compute $A^\top u_j$ and choose the following binary vector
\[
 t_{j,i}=\mathbf1\{c_i>(A^\top u_j)_i\}.
\]
Either tie choice is valid; the displayed rule chooses zero.
\item Add $t_{j,i}$ to $b_i$ and compute $At_j$.
\item For each row $h$, update
\[
 u_{j+1,h}=\min\{R,\max\{0,u_{j,h}
                         +\eta((At_j)_h-1)\}\}.
\]
\end{enumerate}
Return
\[
 t_i^{\rm out}=\frac{b_i}{J(1+\delta)}.
\tag{PK3}
\]
Every iteration and all rational comparisons are exact after the
objective coefficients have been supplied on their stated finite grid.

\begin{proof}
Put
$\mathcal L(t,u)=c\cdot t+u\cdot(\mathbf1-At)$.
The binary vector $t_j$ maximizes $\mathcal L(\cdot,u_j)$ over
$[0,1]^N$.  The dual update is gradient descent on the affine function
$u\mapsto\mathcal L(t_j,u)$ followed by projection onto $[0,R]^M$.
For every $u\in[0,R]^M$, nonexpansiveness of Euclidean projection
and a telescoping squared-distance calculation give
\begin{equation}
 \frac1J\sum_{j=1}^J
 [\mathcal L(t_j,u_j)-\mathcal L(t_j,u)]
 \le\frac{R^2M}{2\eta J}+\frac{\eta MN^2}{2}
 \le\frac\epsilon2.
\tag{PK4}
\end{equation}
Indeed, each coordinate of $\mathbf1-At_j$ has absolute value at
most $N$, and $\|u-u_1\|_2^2\le MR^2$.

If $t^*$ is a feasible optimum of (PK1), then for every $j$,
\[
 \mathcal L(t_j,u_j)\ge\mathcal L(t^*,u_j)
 \ge c\cdot t^*=V,
\]
because $u_j\ge0$ and $At^*\le\mathbf1$.
Let $\bar t=J^{-1}\sum_j t_j$.  It follows from (PK4) that,
simultaneously for every $u\in[0,R]^M$,
\begin{equation}
 V\le c\cdot\bar t+u\cdot(\mathbf1-A\bar t)
       +\epsilon/2.
\tag{PK5}
\end{equation}
Taking $u=0$ gives $c\cdot\bar t\ge V-\epsilon/2$.
Taking $u_h=R$ precisely on the violated rows and zero elsewhere gives
\[
 R\sum_h[(A\bar t)_h-1]_+
 \le c\cdot\bar t-V+\epsilon/2
 \le N+\epsilon/2.
\]
Since $N\ge1$ and $\epsilon\le1$, (PK2) implies
\begin{equation}
 \sum_h[(A\bar t)_h-1]_+\le\frac\epsilon{2N}=\delta.
\tag{PK6}
\end{equation}
Therefore $A\bar t\le(1+\delta)\mathbf1$, and (PK3) is feasible.
Finally $0\le V\le N$ gives
\[
 c\cdot t^{\rm out}
 \ge\frac{V-\epsilon/2}{1+\delta}
 \ge V-\epsilon/2-\delta N\ge V-\epsilon.
\]
The second inequality remains valid if $V-\epsilon/2<0$; alternatively
that case follows immediately from the nonnegative returned objective.

Only the current $M$ dual coordinates, the $N$ integer accumulators,
one binary vector, and the two matrix products are stored.
If $A$ has denominator $Q$ and the chosen dyadic accuracy is
$\epsilon=2^{-b}$, then each update increment has denominator
dividing $2^{b+1}MN^2Q$.
The clipping values $0,R$ are integers, so every dual iterate remains
on that common grid.  Its magnitude is at most $R$, and every
accumulator is at most $J$.  Formula (PK2) is polynomial in the
listed parameters, so these bounds prove the claimed bit lengths.
\end{proof}

\subsection{Matrix products for the direction linear program}
For a direction in our relaxation, a single edge constraint can involve all the edges incident
to either endpoint. Writing these coefficients out would repeat the same incident edges across
many rows. It suffices to compute totals at each vertex and read those totals when scanning the
edges. We give the formulas for both products required by the packing solver and bound the
sizes of their integer indices.
Let $S$ be the stored edges and $A_0$ the possible edges of the first
random matching $T$.  To
avoid confusing the edge set with the packing matrix, the latter is
denoted by $\mathcal A$ in this subsection.
The direction variables $(v_x,v_y)$ obey
\begin{equation}
 \begin{split}
 v_y(\delta(v))&\le D,\\
 v_x(\delta(v))+v_y(\delta(v))&\le1,\\
 (v_x)_{uv}+p\,[v_y(\delta(u))+v_y(\delta(v))]&\le p
       \qquad(uv\in S).
 \end{split}
\tag{PK7}
\end{equation}
All variables are nonnegative.  Scale
$(v_x)_e=p t_e^x$ and $(v_y)_f=D t_f^y$.
The upper bounds $t^x,t^y\le1$ are already implied by (PK7),
and may explicitly be included as coordinate constraints in (PK1).
There are $N=|S|+|A_0|$ variables and at most
$M=2|V(S\cup A_0)|+|S|$ rows.  The three row types become
\begin{equation}
 \sum_{f\in\delta_{A_0}(v)}t_f^y\le1,
 \quad p\sum_{e\in\delta_S(v)}t_e^x
       +D\sum_{f\in\delta_{A_0}(v)}t_f^y\le1,
 \quad t_{uv}^x+D\sum_{f\in\delta_{A_0}(u)}t_f^y
                  +D\sum_{f\in\delta_{A_0}(v)}t_f^y\le1.
\tag{PK8}
\end{equation}
An edge belonging to both sets contributes $2D$ to its own last row;
this is correct.  All coefficients lie in $[0,1]$ because $2D<1$,
and their common denominator depends only on the fixed rationals
$p,D$.

Write the dual arrays for the three row types as
$u_v^y,u_v^v,u_e^c$.  First compute
$q_v=\sum_{e\in\delta_S(v)}u_e^c$ by one scan of the stored edges.
The transpose products are then
\begin{equation}
 \begin{aligned}
 (\mathcal A^\top u)_e^x
   &=p(u_u^v+u_v^v)+u_e^c &&(e=uv\in S),\\
 (\mathcal A^\top u)_f^y
   &=u_u^y+u_v^y+D(u_u^v+u_v^v+q_u+q_v)
       &&(f=uv\in A_0).
 \end{aligned}
\tag{PK9}
\end{equation}
After choosing the binary vector, compute the incident counts
$P_v=\sum_{e\in\delta_S(v)}t_e^x$ and
$Q_v=\sum_{f\in\delta_{A_0}(v)}t_f^y$.
The three forward row products are
$Q_v$, $pP_v+DQ_v$, and $t_{uv}^x+D(Q_u+Q_v)$.
Thus every iteration takes $O(|S|+|A_0|+|V|)$ arithmetic operations
and the same linear amount of workspace; the potentially dense
matrix is never stored.

\paragraph{Objectives in continuous greedy.}
Suppose $\widehat w_e\le\tau$, the estimated multilinear derivatives
are clipped to $[0,\tau]$, and the coefficients $a,\kappa$ lie in
$[0,1]$.  The objective
\[
 \kappa\widehat w\cdot v_x+a\widehat\nabla F_g\cdot v_y
\]
has coefficients
$c_e^x=\kappa p\widehat w_e/\tau$ and
$c_f^y=aD\widehat\partial_fF_g/\tau$, all in $[0,1]$.
Applying the lemma with error $\epsilon$ therefore gives an exactly
feasible direction with additive objective error at most
$\epsilon\tau$.  It uses no value-oracle calls beyond the supplied
gradient estimates.  Coefficients need not be bounded away from zero,
and no inverse objective coefficient is used.

For a fixed set of stored edges and prescribed accuracy, $N,M,J,\delta$ do not
depend on the current gradient.  Every direction consequently lies
on the same rational grid.  Averaging a fixed polynomial number
of continuous-greedy Euler steps preserves a common grid and
all constraints in (PK7), which also supplies the finite
coordinates required when rounding $x$ and $y$.

\section{A finite algorithm for the multilinear relaxation}
\label{sec:joint-distorted-flow-oracle}
Lemma~\ref{lem:main-optimize} compares a continuous trajectory with every matching in the
stored graph. An implementation has only finitely many steps and can query $f$ on sets, while
the trajectory calls for derivatives of $F_g$. We discretize time, estimate each derivative by
differences of oracle answers, and account for the error of the packing solver in choosing a
direction. A common accuracy event then gives the comparison for every matching simultaneously.
Let $S$ be the edge set of a finite simple graph with $m$ edges.
Let $g:2^S\to\mathbb R_{\ge0}$ be monotone and submodular with
$g(\varnothing)=0$, let $F_g$ be its multilinear
extension~\cite{CCPV2011,CVZ}, and let $w_e\ge0$ be fixed weights
(the estimates $\widehat w$ in the streaming application).  Suppose
$w_e,g(\{e\})\le\tau$ for a known positive value $\tau$.  Fix rational $0<p\le1$ and $D\in(0,1/4)$, with
$D=r(1-r)$ for a rational $r>1/2$ when sampling from the distribution associated with Shearer's lemma.

For nonnegative vectors $x,y$ on $S$, write
$\ell_v(y)=\sum_{e\ni v}y_e$ for the fractional degree of $v$ in $y$.  Use the linear packing polytope
\begin{equation}
 \begin{split}
 \mathcal P_D=\{(x,y)\ge0:\quad
 &\ell_v(y)\le D,\\
 &x(\delta(v))+\ell_v(y)\le1\quad(v\in V),\\
 &x_{uv}+p[\ell_u(y)+\ell_v(y)]\le p\quad(uv\in S)\}.
 \end{split}
 \tag{JF1}
\end{equation}
It is bounded, convex, and down-monotone.  Every point satisfies
(GJ1), so Theorem~\ref{thm:general-joint-completion} applies.
The last inequality in (JF1) is part of the algorithm.  Replacing
its bound by $p$ times the product of the two remaining vertex
capacities is not justified on general graphs.

For a matching $O\subseteq S$, put
$I_v=\mathbf1_{v\in V(O)}$ and define
\begin{equation}
 W_D(O)=\max\left\{w\cdot z:
 z\ge0,\ z(\delta(v))\le1-DI_v,\
 z_{uv}\le p[1-D(I_u+I_v)]\right\}.
 \tag{JF2}
\end{equation}
These coordinate bounds are positive because $D<1/4$.  If $z^O$ attains this
maximum, then $(z^O,D\mathbf1_O)\in\mathcal P_D$.

\begin{theorem}[Optimizing the multilinear relaxation]
There is a finite randomized value-oracle procedure which, except
with prescribed probability $\delta$, returns a rational point
$(x,y)\in\mathcal P_D$ satisfying, simultaneously for every matching
$O\subseteq S$,
\begin{equation}
 w\cdot x+F_g(y)
 \ge W_D(O)+(1-e^{-D})g(O)-\varepsilon\tau/2.
 \tag{JF3}
\end{equation}
For fixed rational $D,p,\varepsilon>0$, it has polynomial worst-case
running time and query count and uses $O(m+|V(S)|)$ words, provided
the algorithm for the linear packing direction described below is
used.  Coordinates and auxiliary indices have polynomial bit length
and share a common rational grid.
\end{theorem}

\paragraph{Continuous greedy.}
We use continuous greedy~\cite{CCPV2011,CVZ} with the time-dependent
weighting of Feldman~\cite{FeldmanDistorted}.  Start from $(x(0),y(0))=0$ and, for $0\le t\le1$, choose a direction
$(v_x,v_y)\in\mathcal P_D$ maximizing
\[
 w\cdot v_x+a(t)\nabla F_g(y(t))\cdot v_y,
 \qquad a(t)=e^{D(t-1)}.
\]
The standard monotone-submodular gradient inequality is
$\nabla F_g(y)\cdot\mathbf1_O\ge g(O)-F_g(y)$.
For $\Phi(t)=w\cdot x(t)+a(t)F_g(y(t))$, compare the selected
direction to $(z^O,D\mathbf1_O)$.  Since $a'=Da$, this gives
\[
 \Phi'(t)\ge w\cdot z^O+Da(t)g(O).
\]
Integrating proves the ideal form of (JF3): the linear term has
coefficient one, and the submodular term has coefficient $1-e^{-D}$.
The argument does not require $F_g$ to be concave or require the
relaxation to be integral.

\paragraph{Finite Euler construction.}
For $m>0$, choose the following exact integer and rational values:
\begin{equation}
 K=\max\{1,\lceil2D^2m^2/\varepsilon\rceil\},\quad
 h=1/K,\quad
 \xi=\min\{1/4,\varepsilon/(16Dm)\}.
 \tag{JF4}
\end{equation}
Define analytical coefficients
$a_j=(1-Dh)^{K-j}$, so $a_K=1$ and
$a_{j+1}-a_j=Dh\,a_{j+1}$.  Starting from zero, at stage $j<K$
estimate the coefficients of the $y$ coordinates
$a_{j+1}\partial_eF_g(y_j)$ to coordinatewise accuracy $\xi\tau$.
Keep the coefficient $w_e$ of each $x$ coordinate exact.  Use the packing
solver with accuracy $\min\{\varepsilon/8,1\}$ to compute a feasible
direction in $\mathcal P_D$ whose objective for these approximate
coefficients is within $\varepsilon\tau/8$ of the maximum, and set
\[
 (x_{j+1},y_{j+1})=(x_j,y_j)+h(v_{x,j},v_{y,j}).
\]
Each partial point is feasible because it is a convex combination of
feasible directions and zero.  The comparison using the exact coefficients loses
at most $2D\xi m\tau+\varepsilon\tau/8$ per unit time: both a
direction and $(z^O,D\mathbf1_O)$ have $y$ coordinates with total sum at most
$Dm$.  Mixed second derivatives of $F_g$ lie in $[-\tau,0]$,
whereas unmixed second derivatives vanish.  Therefore
\[
 F_g(y_j+h v_{y,j})
 \ge F_g(y_j)+h\nabla F_g(y_j)\cdot v_{y,j}
              -\tfrac12h^2\tau D^2m^2.
\]
The potential $w\cdot x_j+a_jF_g(y_j)$ now gives the exact finite
cancellation
\begin{align*}
 w\cdot x_K+F_g(y_K)
 &\ge W_D(O)+\bigl[1-(1-D/K)^K\bigr]g(O)\\
 &\quad-2D\xi m\tau-\varepsilon\tau/8
                         -D^2m^2\tau/(2K).
\end{align*}
The three errors total at most $\varepsilon\tau/2$, and
$1-(1-D/K)^K\ge1-e^{-D}$.  This proves (JF3) on the event that all coefficient estimates have the stated accuracy.  For $m=0$, the zero point suffices.

\paragraph{Estimating derivatives.}
For the residual function $g$ in~\eqref{eq:main-residual}, we obtain one derivative sample from two value queries to $f$, taking the difference
\[
 f(R\cup Z\cup\{e\})-f(R\cup Z),
 \qquad R\sim R_p(S),\quad Z\sim R_{y_j}(S\setminus\{e\}),
\]
with independent samples $R$ and $Z$.  It lies in $[0,\tau]$ and has exactly
the required mean.  Quantize the sample downward on the
$\xi\tau/8$ grid, approximate each inclusion probability
downward to error at most $\xi/(16m)$, and reserve $\xi\tau/4$
for Hoeffding deviation.  These errors total at most $\xi\tau/2$.
An approximation of $a_{j+1}$ in $[0,1]$ to accuracy $\xi/2$
then supplies the required $\xi\tau$ coefficient accuracy.
To bound the approximation error explicitly, write
$\beta=1-D/K$, choose the least power of two $Q\ge2K/\xi$, and
start from $q_0=Q$.  Repeat $q_{t+1}=\lfloor\beta q_t\rfloor$
for $K-j-1$ steps and return $q_{K-j-1}/Q$.  Induction gives
\[
 0\le\beta^t-q_t/Q\le t/Q\le\xi/2.
\]
Every index is at most $Q$ and has $O(\log(K/\xi))$ bits.
Recomputing independently at each stage uses only constant additional
workspace; the exact analytical powers are never stored.

There are $Km$ derivative estimates.  Conditional Hoeffding bounds
with per-estimate failure $\delta/(Km)$, and a union bound, give
the common event with probability at least $1-\delta$.  For example
$\lceil8\xi^{-2}\log(2Km/\delta)\rceil$ independent samples per
estimate suffice for the stated Hoeffding allowance.  The adaptive
choice of earlier directions causes no change to this conditional
concentration argument.

\paragraph{Direction optimization and bounded representations.}
Normalize $x=p t_x$, $y=D t_y$.  The resulting direction problem has
$2m$ box variables $t\in[0,1]^{2m}$ and a nonnegative packing matrix
with $O(m+|V|)$ rows and coefficients at most one.  In the row bounding an individual coordinate, the coefficient of
the $y$ coordinate of the same edge is $2D$, which is still
at most one.  The objective coefficients, divided by $\tau$, lie
in $[0,1]$.  The lemma for linear packing gives the
required additive accuracy with linear workspace, deterministic
polynomial iteration count, and one common rational grid.

Both matrix products can be performed by adjacency scans: the
contributions of the dual variables for coordinate bounds to a
$y$ coordinate are the sums of the corresponding multipliers
incident to its endpoints.  No dense
constraint matrix is stored.  Fixing the packing iteration count and
step size throughout one instance gives a common denominator for all
directions, so the $K$ Euler averages do not accumulate unrelated
denominators.  All sampled subsets and accumulators are reused.

\paragraph{Rounding and probability amplification.}
Apply Theorem~\ref{thm:general-joint-completion} to $(x_K,y_K)$:
sample the first matching, perform the stated alteration, and sample
a Gibbs matching on the vertices left unmatched.  The expected value
of their union is at least
\[
 \vartheta_G[\kappa_A w\cdot x_K+F_g(y_K)]
 \ge\vartheta_G\kappa_A[W_D(O)+(1-e^{-D})g(O)
                                      -\varepsilon\tau/2],
\]
before the separate finite sampling and common-grid losses.
Both $w\cdot x_K$ and $F_g(y_K)$ are nonnegative.
Fix the estimates made during the stream and the optimization,
conditioned on their being accurate.  Independent repetitions of the
rounding algorithm give feasible matchings.  Comparing their values
under $f$, including the best singleton, gives the probability
amplification.  All estimation is finished before these random
matchings are sampled.

\section{A matching distribution with prescribed marginals}
\label{sec:referee-shearer-sampler}
The first rounding step requires a random matching with prescribed edge marginals. Both the expected submodular value and the second rounding step also depend on the conditional law of the remaining edges after one edge is selected. We construct a distribution in
which disjoint edges have product inclusion probabilities. A recursion on vertex deletions
gives a sampler without enumerating matchings.
Let $G=(V,E)$ be a finite simple graph, and let $x\in\mathbb Q_{\ge0}^E$.
Fix a rational $r\in(1/2,1)$ and put $D=r(1-r)$.
Assume that $\sum_{e\ni v}x_e\le D$ for every vertex $v$.
We use the distribution associated with Shearer's lemma on the line
graph of $G$; the inclusion--exclusion formula follows the notation of
Guo, Jerrum, and Liu~\cite{GJL2019}.  We give the distribution and its
sampling algorithm explicitly.  For a subgraph $H$ of $G$, define
\[
 Z_H=\sum_{T\text{ a matching of }H}(-1)^{|T|}\prod_{e\in T}x_e,
 \qquad
 R_H(v)=\frac{Z_H}{Z_{H-v}}.
\]
The polynomial $Z_H$ equals one when $H$ is empty.  All vertex deletions
below retain the original parameters $x_e$, which will be the marginal
probabilities of the distribution.

\paragraph{Positivity and ratios.}
Induction on $|V(H)|$ proves $Z_H>0$ and $R_H(v)\ge r$ for every $H,v$.
Indeed, the matching recurrence gives
\[
 R_H(v)=1-\sum_{u:uv\in E(H)}\frac{x_{uv}}{R_{H-v}(u)}
 \ge1-\frac{D}{r}=r.
\tag{RS1}
\]
The induction also shows $R_H(v)\le1$.

\paragraph{The target distribution.}
For every matching $M$ of $G$, set
\[
 \mu_x(M)=\Bigl(\prod_{e\in M}x_e\Bigr)Z_{G-V(M)}.
\tag{RS2}
\]
These numbers are nonnegative.  Expanding the polynomial $Z_{G-V(M)}$ and
collecting terms by the disjoint union of two compatible matchings gives
$\sum_M\mu_x(M)=1$: every nonempty matching $U$ contributes
$x^U\sum_{M\subseteq U}(-1)^{|U\setminus M|}=0$.
The same expansion gives the joint moments
\[
 \Pr_{\mu_x}(T\subseteq M)=
 \begin{cases}
 \prod_{e\in T}x_e,&T\text{ is a matching},\\
 0,&T\text{ is not a matching}.
 \end{cases}
\tag{RS3}
\]

\paragraph{Comparison with independent rounding.}
For an edge $e$ with $x_e>0$, condition (RS3) on $e\in M$.
The joint moments of all edges disjoint from $e$ remain unchanged,
while all edges meeting $e$ are absent.  Joint moments determine a
distribution on subsets of a finite set by inclusion--exclusion.  Hence the distribution
of $M\setminus\{e\}$ conditional on $e\in M$ is precisely the distribution
of $M\setminus N[e]$ without conditioning, where $N[e]$ contains $e$
and every edge meeting it.  In particular, the two remainders can be coupled so that the conditional
remainder is a subset of the unconditional remainder.

For completeness this coupling proves
\[
 \Ex[M\sim\mu_x]{f(M)}\ge F_f(x)
 \quad\text{for every submodular set function }f.
\tag{RS3a}
\]
Order the edges and replace independent Bernoulli coordinates by the
corresponding coordinates of $M$ one at a time.  At coordinate $e$,
let $S$ be the already replaced prefix and let $Q$ be the remaining
independent suffix.  The change in expected value is
\[
 x_e\left(\Ex{f(e\mid S\cup Q)\mid e\in M}
                   -\Ex{f(e\mid S\cup Q)}\right)\ge0.
\]
The inequality follows by coupling the conditional prefix as a subset
of the unconditional prefix and keeping $Q$ identical, then applying
diminishing returns.  Coordinates with $x_e=0$ give zero change.
Telescoping proves (RS3a).  This particular rounding inequality does
not require monotonicity or normalization.

\paragraph{Sampling a Bernoulli variable recursively.}
The following procedure $\operatorname{Coin}(v,H)$ returns one with
probability $t_H(v)=r/R_H(v)$.
The procedure draws each random choice when it is needed.
\begin{enumerate}
\item Independently stop and return one with probability $r$.
\item Otherwise choose a neighbor $u$ with probability $x_{uv}/D$ for
each $uv\in E(H)$, and choose a null symbol with the remaining
probability.
\item A null symbol returns zero.  For a nonnull $u$, recursively call
$\operatorname{Coin}(u,H-v)$ using fresh random bits.  Return zero if
that call returns zero, and otherwise return to step~1.
\end{enumerate}
Different calls use independent random bits, including calls with
identical arguments.  Depth is at most $|V|$, since each recursive call
deletes the parent vertex.  On an empty neighborhood the success
probability is $r$.
For the induction step, the probability of a successful continuation
in step~3 is
\[
 S=\sum_{u:uv\in E(H)}\frac{x_{uv}}D
       \frac r{R_{H-v}(u)}.
\]
Consequently the return-one probability is
\[
 \sum_{k\ge0}r\bigl((1-r)S\bigr)^k
 =\frac r{1-\sum_{u:uv\in E(H)}x_{uv}/R_{H-v}(u)}
 =\frac r{R_H(v)}.
\tag{RS4}
\]
The procedure requires no oracle for evaluating $Z_H$.

\paragraph{Sampling a matching.}
Process all edges in any fixed order.  Maintain a matching $A$ of
accepted edges and the graph $H$ of previously rejected edges whose
endpoints are both unmatched by $A$.  Initially both are empty.  Edges
incident to $A$ are skipped.  At a remaining edge $e=uv$, flip three
independent coins, with success probabilities
\[
 \frac{x_e}{r^2},\qquad
 \Pr(\operatorname{Coin}(u,H)=1),\qquad
 \Pr(\operatorname{Coin}(v,H-u)=1).
\tag{RS5}
\]
Accept $e$ if all three succeed; then insert it into $A$ and delete
$u,v$ from $H$.  Otherwise insert $e$ into $H$.
The first probability is valid because
$x_e/r^2\le D/r^2=(1-r)/r<1$.

To prove correctness, condition on the current accepted set $A$ and
all prior rejections.  From (RS3), conditioning on $A\subseteq M$
leaves the distribution $\mu_x$ on $G-V(A)$ with the same marginal
probabilities for the remaining edges.
Inclusion--exclusion on the prior rejected edges then yields
\[
 \Pr(e\in M\mid\text{current history})
 =x_e\frac{Z_{H-u-v}}{Z_H}
 =\frac{x_e}{r^2}\frac r{R_H(u)}\frac r{R_{H-u}(v)}.
\tag{RS6}
\]
This is exactly the probability in (RS5).  Histories have positive
probability because all the relevant $Z_H$ are positive.  Induction on
the processed edge order proves that the final $A$ has distribution $\mu_x$.

\paragraph{Expected running time and exponential truncation.}
The recursive computation is dominated by the family tree of a
Galton--Watson process whose offspring distribution is
\[
 \Pr(K=k)=r(1-r)^k,\qquad k=0,1,\ldots.
\]
Null choices and failed children can only reduce the explored part
of this tree.  Count requested children, including null
children, when bounding the number of recursive steps.  The expected full progeny is
$r/(2r-1)$, since $\Ex{K}=(1-r)/r<1$.
More explicitly, if $T$ is the full progeny, then the ordered-tree
count gives
\[
 \Pr(T=k)=C_{k-1}r^k(1-r)^{k-1}
 \le r(4D)^{k-1},
 \qquad
 \Pr(T\ge L)\le\frac{r(4D)^{L-1}}{1-4D},
\tag{RS7}
\]
where $C_j=\frac1{j+1}\binom{2j}{j}$.
For fixed $r>1/2$ this is an exponential tail.  For a prescribed
$0<\delta<1$, the following stopping rule gives total variation distance
at most $\delta$.
Write $m=|E|\ge1$, and choose the least integer $L\ge1$ satisfying
\[
 \frac{2mr(4D)^{L-1}}{1-4D}\le\frac\delta2.
\]
This integer is found by rational comparisons and is
$O_r(\log((m+1)/\delta))$.  Stop each of the at most $2m$ top-level
calls to $\operatorname{Coin}$ after $L$ requested children, counting null requests,
and return the empty matching if this stopping rule is invoked.  By (RS7) and a
union bound, the total variation distance between the resulting
distribution and $\mu_x$ is at most $\delta/2$.  Couple the procedure with this stopping rule and the original procedure
using the same random choices until the stopping rule is invoked.
This coupling remains valid when later inputs depend on earlier outcomes.

An adjacency representation of $G$, edge flags for $H$, matching flags,
and a depth-first recursion stack suffice.  Vertex exclusions are marked
on entry and unmarked on return.  One may sample neighbors from the
original graph with probabilities $x_{uv}/D$, treating an excluded
vertex or an edge not currently in $H$ as a null outcome.  Hence no
subgraph copies are needed.  Straight adjacency scans give polynomial
time, and the working space is $O(|V|+|E|)$ words.

To bound the number of random bits as well, observe that there are at most
$Q=10m(L+1)$ elementary rational choices in the procedure with this stopping rule,
including neighbor choices and final edge decisions.  Indeed, one
call to $\operatorname{Coin}$ uses at most $3L+1$ such choices, and there are at most
$2m$ calls and $m$ final edge decisions.  Implement each
Bernoulli or categorical choice by drawing a uniform integer from its
denominator: with denominator $q$, draw $\lceil\log_2q\rceil$ fair
bits and reject integers at least $q$.  Each attempt succeeds with
probability at least $1/2$.  Allow at most $B$ attempts per choice,
where $B$ is the least integer with $2^B\ge2Q/\delta$; on failure,
return the empty matching.  A union bound contributes at most
$Q2^{-B}\le\delta/2$ additional error.  The two stopping rules therefore
give total variation distance at most $\delta$ from $\mu_x$ and bounded
worst-case polynomial running time.  In the streaming application all edge probabilities
share a short rational grid, so the denominators and the neighbor
choice intervals are computed in linear workspace.

\section{Rounding small coordinates using the Gibbs distribution}
\label{sec:gibbs-small-coordinate}
After the first matching has used some vertices, the remaining vector can still have degree
close to one, but each individual edge receives little mass. We prove
Lemma~\ref{lem:main-diffuse} by comparing a random matching on these vertices to independent
edge choices. Deleting a vertex changes matched status along one alternating path. This
controls the effect of conditioning on one edge and lets us charge the comparison error to that
edge's marginal contribution. We then choose the edge activities so that the resulting
probabilities track the target vector. Throughout the graph is finite, loopless, and simple.
For positive edge activities $\lambda_e$, let
\[
 \mu_G(M)=Z_G^{-1}\prod_{e\in M}\lambda_e
\]
be the Gibbs distribution on all matchings, including the empty matching,
where $Z_G$ denotes the partition function of this distribution~\cite{JS1989,CLV2026}.  The parameters
$\lambda_e$ are edge activities, as in Jerrum, Sinclair, and
Vigoda~\cite{JSV2004}.  Edges of activity zero can be removed.  For a normalized monotone
submodular function $f$, write $F(y)=\Ex{f(R_y)}$, where
$R_y$ contains each edge independently with probability $y_e$.

Qiu and Singla~\cite{QS} give sufficient conditions for comparing the
expected value of a submodular function under dependent and independent
rounding.  Here we prove the comparison with an explicit error term for
the Gibbs distribution, together with the coupling and marginal
probability estimates needed in the proof.

\subsection{Vertex deletion and a variance bound}

Let $X_v$ be the indicator that $v$ is unmatched, and put
$a_v=\mathbb P(X_v=1)$.

\begin{lemma}[Coupling after deleting a vertex]\label{lem:one-monomer}
There is a coupling of $M\sim\mu_G$ and $N\sim\mu_{G-v}$ in which,
when the absent vertex $v$ is regarded as unmatched in $N$, the
two vectors of unmatched indicators agree outside $v$ except possibly at one other
vertex.  They agree everywhere whenever $v$ is unmatched in $M$.
Consequently,
\begin{equation}
 \sum_{u\ne v}|\operatorname{Cov}(X_u,X_v)|
 \le a_v(1-a_v).
 \tag{GM1}
\end{equation}
\end{lemma}
\begin{proof}
Induct on the number of vertices.  First sample the status of $v$
in $M$ according to $\mu_G$.  If $v$ is unmatched, sample
$M=N$ on $G-v$.  If $v$ is matched to $w$, put $vw$ in $M$ and
apply the inductive coupling at $w$ in the graph $G-v$.  This
couples a matching $N$ with distribution $\mu_{G-v}$ to a matching $P$
with distribution $\mu_{G-v-w}$; set $M=P+vw$.  In each branch the
comparison matching $N$ has distribution $\mu_{G-v}$, and the resulting
marginal distribution of $M$ is $\mu_G$.

In the matched branch, the symmetric difference is an alternating
simple path starting at $v$.  All its interior vertices are matched
in both matchings, so outside $v$ only its other endpoint can
change unmatched status.  In the unmatched branch the two matchings
coincide.

The distribution $\mu_{G-v}$ is the conditional distribution of $M$ on
$G-v$ given $X_v=1$.
For $u\ne v$,
\[
 \operatorname{Cov}(X_u,X_v)
 =a_v\bigl(\Ex{X_u\mid X_v=1}-\Ex{X_u}\bigr).
\]
The coupled vectors of unmatched indicators differ in at most one coordinate
outside $v$, and only on an event of probability $1-a_v$.
Summing absolute differences of their means and multiplying by
$a_v$ proves (GM1).
\end{proof}

\begin{lemma}[A variance bound for unmatched indicators]
\label{lem:monomer-weighted-variance}
For every real vector $c=(c_v)$,
\begin{equation}
 \operatorname{Var}\!\left(\sum_v c_vX_v\right)
 \le 2\sum_v c_v^2a_v(1-a_v).
 \tag{GM2}
\end{equation}
In particular, if $0\le c_v\le h$ and $L=\sum_vc_vX_v$, then
\begin{equation}
 \operatorname{Var}(L)\le2h\Ex{L},
 \qquad
 \frac{\operatorname{Var}(L)}{(1+\Ex{L})^2}
 \le\frac h2.
 \tag{GM3}
\end{equation}
\end{lemma}
\begin{proof}
Let $C$ be the covariance matrix of the unmatched indicators.
By (GM1), including the diagonal gives
$\sum_u|C_{vu}|\le2a_v(1-a_v)$.  Symmetry of $C$ and
$2|c_uc_v|\le c_u^2+c_v^2$ imply
\[
 \begin{aligned}
 c^TCc
 &\le\sum_{u,v}|c_uc_v|\,|C_{uv}|\\
 &\le\frac12\sum_{u,v}(c_u^2+c_v^2)|C_{uv}|\\
 &=\sum_vc_v^2\sum_u|C_{vu}|
 \le2\sum_vc_v^2a_v(1-a_v).
 \end{aligned}
\]
If the coefficients lie in $[0,h]$, the last expression is at
most $2h\sum_v c_va_v=2h\Ex{L}$.  Finally,
$t/(1+t)^2\le1/4$ for every $t\ge0$, proving the second bound
in (GM3).
\end{proof}

\subsection{Comparison with independent rounding}
To compare a Gibbs matching with independent edge choices, we replace one coordinate at a time.
For an edge $uv$, the comparison depends on the conditional distribution of the remaining edges given its inclusion. Conditioning on $uv$ deletes both its endpoints. We first bound
the effect of deleting one vertex and then use this estimate twice.

Suppose all edge activities are at most $h$. For $H\subseteq G$, write $\mu_G\downarrow H$ for
the law obtained by sampling a matching of $G$ and keeping its edges in $H$. The distribution
$\mu_H$ instead samples directly on $H$. The two can differ because edges outside $H$ may have
occupied vertices of $H$.

\begin{lemma}[Deleting a vertex]
\label{lem:gibbs-deletion-stability}
For every vertex $v$,
\[
 d_{\rm TV}(\mu_G\downarrow(G-v),\mu_{G-v})\le\sqrt{h/8}.
\]
\end{lemma}
\begin{proof}
Put $H=G-v$ and sample $N\sim\mu_H$.  Every extension of $N$ to
$G$ either leaves $v$ unmatched or matches it to one unmatched neighbor.
The total relative weight of these extensions is $1+L$, where
\[
 L=\sum_{a\in V(H)}\lambda_{va}X_a(N).
\]
Thus the probability of $N$ under $\mu_G\downarrow H$ is its
probability under $\mu_H$ multiplied by $(1+L(N))/(1+\Ex{L})$.  Simplicity ensures that every
coefficient in $L$ is at most $h$.  By (GM3),
\[
 \begin{aligned}
 d_{\rm TV}(\mu_G\downarrow H,\mu_H)
 &=\frac{\Ex{\lvert L-\Ex{L}\rvert}}{2(1+\Ex{L})}\\
 &\le\frac{\sqrt{\operatorname{Var}(L)}}{2(1+\Ex{L})}
 \le\sqrt{h/8}.
 \end{aligned}
\]
\end{proof}

Set
\begin{equation}
 D_h=\sqrt{h/2}.
 \tag{GM4}
\end{equation}
No bound on the sum of activities at a vertex is needed.

\begin{lemma}[Conditional expectations]
\label{lem:gibbs-approx-regression}
For every edge $e$ of positive activity and every nonnegative
decreasing function $g$ of the other edge indicators, with
$0\le g\le B$,
\begin{equation}
 \Ex{g\mid e\notin M}
 \le\Ex{g\mid e\in M}+D_hB.
 \tag{GM5}
\end{equation}
The same inequality applies to every projection retaining $e$,
with the same constant.
\end{lemma}
\begin{proof}
Write $e=uv$ and $H=G-u-v$.  Conditional on $e$ being absent,
the matching has distribution $\mu_{G-e}$.  Delete $u$, and then $v$.
Lemma~\ref{lem:gibbs-deletion-stability}, contraction of total
variation distance under restriction, and the triangle inequality give
\begin{equation}
 d_{\rm TV}(\mu_{G-e}\downarrow H,\mu_H)
 \le2\sqrt{h/8}=D_h.
 \tag{GM6}
\end{equation}
Indeed, the intermediate Gibbs distribution is $\mu_{G-u}$; deleting $u$
has already removed $e$.

Regard a matching on $H$ as a vector on all edges other than $e$,
with zeros on the omitted edges.  Restriction removes incident edges,
so it can only increase the decreasing function $g$.  Hence
\[
 \Ex{g\mid e\notin M}
 \le\Ex[\mu_{G-e}\downarrow H]{g}
 \le\Ex[\mu_H]{g}+D_hB.
\]
Conditional on $e$ being present, every other edge at $u$ or $v$
is absent and the remainder has distribution $\mu_H$.  Thus the last
expectation is $\Ex{g\mid e\in M}$, proving (GM5).
A decreasing function of a projection extends to the full vector
by ignoring omitted coordinates, which proves the final assertion.
\end{proof}

\begin{theorem}[Comparison with independent rounding]\label{thm:gibbs-submodular}
Let $m_e=\mathbb P(e\in M)$.  For every normalized monotone
submodular function $f$,
\begin{equation}
 \Ex{f(M)}\ge(1-D_h)_+F(m).
 \tag{GM7}
\end{equation}
\end{theorem}
\begin{proof}
Order the edges and independently sample Bernoulli variables
$R_e$ with means $m_e$, independently also of $M$.
Let $H_i=M_{\le i}\cup R_{>i}$.  Conditional on the independent
future $R_{>i}$, put
\[
 g_i=f(i\mid M_{<i}\cup R_{>i}),\qquad
 B_i=f(i\mid R_{>i}).
\]
As a function of the edge indicators other than $i$, $g_i$ is
nonnegative and decreasing, and submodularity gives $g_i\le B_i$.
The distribution of $M$ is unchanged when $R_{>i}$ is fixed, so (GM5)
applies with this fixed conditional bound.  Mixing its two
conditional expectations gives
\[
 \Ex{g_i}
 \le\Ex{g_i\mid i\in M}+D_h\Ex{B_i},
\]
after averaging over the independent future.  Indeed, before
averaging the error term can be bounded by
$(1-m_i)D_hB_i\le D_hB_i$.

The one-coordinate hybrid difference is exactly
\[
 \begin{aligned}
 \Ex{f(H_i)}-\Ex{f(H_{i-1})}
 &=m_i\Ex{g_i\mid i\in M}-m_i\Ex{g_i}\\
 &\ge-m_iD_h\Ex{B_i}.
 \end{aligned}
\]
The telescoping identity, with edges taken in reverse order, is
\[
 \sum_i m_i\Ex{B_i}
 =\sum_i m_i\Ex{f(i\mid R_{>i})}
 =F(m),
\]
using $f(\varnothing)=0$.  Summing all hybrid differences yields
$\Ex{f(M)}\ge(1-D_h)F(m)$.  Nonnegativity of $f$ supplies
the positive part when $D_h>1$.  In particular, the error is
multiplicative and contains no factor equal to the number of
edges.  The proof does not condition on an arbitrary assignment to the
preceding edge indicators.
\end{proof}

\subsection{Choosing the edge activities}
The previous comparison uses the actual Gibbs marginals. The algorithm is given a target vector
$x$, so it still needs a way to choose activities from $x$. Scale $x$ by $\gamma<1$ to leave
some free capacity at every vertex. An edge can be inserted only when both endpoints are free.
Dividing its target mass by the product of the two proposed free probabilities compensates for
this restriction. The estimates below bound the error in that product approximation and show
that the resulting marginals are close enough coordinate by coordinate.

Let $x\ge0$ satisfy $x(\delta(v))\le1$ and $x_e\le p$.
Fix $0<\gamma<1$, put $z=\gamma x$, and let
$q_v=1-z(\delta(v))\ge1-\gamma$.  Set
\begin{equation}
 \lambda_{uv}=\frac{z_{uv}}{q_uq_v},\qquad
 R=\frac{\gamma}{(1-\gamma)^2},\qquad h=Rp.
 \tag{GM8}
\end{equation}
Then $\lambda_e\le h$ and
$\sum_{e\ni v}\lambda_e\le R$ for every vertex.

\begin{lemma}[Edge marginals]\label{lem:gibbs-calibration}
With $K=5(1+R)h$, the edge marginal probabilities under $\mu_G$ satisfy
\begin{equation}
 e^{-K}z_e\le m_e\le e^Kz_e.
 \tag{GM9}
\end{equation}
\end{lemma}
\begin{proof}
For an induced subgraph $H$, write $a_v^H$ for the probability that $v$
is unmatched under $\mu_H$.  Lemma~\ref{lem:one-monomer} also gives
\[
 \sum_{u\in V(H)\setminus\{v\}}
 |a_u^{H-v}-a_u^H|\le1-a_v^H\le1.
\]
The exact partition-function recursion therefore yields
\[
 (a_v^H)^{-1}
 =1+\sum_{u\in N_H(v)}\lambda_{uv}a_u^{H-v},
 \qquad
 \left|(a_v^H)^{-1}
  -\left(1+\sum_{u\in N_H(v)}\lambda_{uv}a_u^H\right)\right|
 \le h.
\]
Both denominators are at least one.  For
$T_H(a)_v=(1+\sum_{u\in N_H(v)}\lambda_{uv}a_u)^{-1}$,
this implies the conservative bound
\[
 \|\log a^H-\log T_H(a^H)\|_\infty\le2h.
\]
In logarithmic coordinates $T_H$ contracts the maximum norm by
the factor $\rho=R/(1+R)$ on $(0,1]^{V(H)}$, because each row
of its logarithmic derivative has absolute sum
\[
 \frac{\sum_{u\in N_H(v)}\lambda_{uv}a_u}
 {1+\sum_{u\in N_H(v)}\lambda_{uv}a_u}\le\rho.
\]

On the full graph, $q$ is an exact fixed point:
\[
 1+\sum_u\lambda_{uv}q_u
 =1+\frac{z(\delta(v))}{q_v}=\frac1{q_v}.
\]
The triangle inequality and contraction thus give
\[
 \|\log a^G-\log q\|_\infty\le2(1+R)h.
\]
On $G-u$, the restricted vector $q|_{G-u}$ has logarithmic
fixed-point error at most $h$.  Indeed, the full denominator
$1/q_v$ loses at most the single term
$\lambda_{uv}q_u\le h$, while the remaining denominator is
still at least one.  Comparing logarithms therefore changes the
fixed-point equation by at most $h$.  Applying the same
contraction argument, now adding this error for $q$
to the error for the actual probabilities of being unmatched, gives
\[
 \|\log a^{G-u}-\log q|_{G-u}\|_\infty
 \le3(1+R)h.
\]
Finally, exact partition-function identities give
\[
 m_{uv}=\lambda_{uv}a_u^Ga_v^{G-u},
 \qquad z_{uv}=\lambda_{uv}q_uq_v.
\]
The sum of the two logarithmic errors is at most $K$, proving
(GM9) for edges of positive activity.  Edges with $z_e=0$ have
$m_e=0$, so the same inequalities hold for them.
\end{proof}

\begin{corollary}[Rounding small coordinates]
\label{cor:gibbs-small-coordinate}
For every $0<\gamma<1$, let $R,h$ and the activities be as in
(GM8).  Then every normalized monotone submodular $f$ satisfies
\begin{equation}
 \Ex{f(M)}
 \ge(1-D_h)_+\gamma e^{-5(1+R)h}F(x).
 \tag{GM10}
\end{equation}
In particular, for every fixed $\epsilon>0$ there is a constant
$c_\epsilon>0$, independent of the numbers of edges and vertices,
such that $\max_e x_e\le c_\epsilon$ guarantees that the Gibbs
distribution with activities (GM8) satisfies
$\Ex{f(M)}\ge(1-\epsilon)F(x)$ for every normalized monotone
submodular $f$.
\end{corollary}
\begin{proof}
The bound (GM9) gives $m\ge\gamma e^{-K}x$ coordinatewise.
By (GM7), monotonicity of $F$, and its concavity on nonnegative
rays from the origin,
\[
 \Ex{f(M)}
 \ge(1-D_h)_+F(m)
 \ge(1-D_h)_+F(\gamma e^{-K}x)
 \ge(1-D_h)_+\gamma e^{-K}F(x).
\]
The following explicit choice proves the final assertion for
$0<\epsilon<1$; for $\epsilon\ge1$ it is immediate from
nonnegativity.
\end{proof}

\paragraph{A sufficient coordinate bound.}
For $0<\epsilon<1$, take $\gamma=1-\epsilon/4$ and
\begin{equation}
 p\le c_\epsilon:=\epsilon^5/5120.
 \tag{GM11}
\end{equation}
Both $R$ and $1+R$ are at most $16/\epsilon^2$.  Therefore
\[
 h=Rp\le\frac{\epsilon^3}{320}\le\frac{\epsilon^2}{320},
 \qquad
 K=5(1+R)Rp\le\frac\epsilon4,
 \qquad
 D_h\le\sqrt h\le\frac{\epsilon}{\sqrt{320}}<\frac\epsilon4.
\]
Thus each of $\gamma$, $e^{-K}$, and $1-D_h$ is at least
$1-\epsilon/4$.  Their product in (GM10) is at least
$(1-\epsilon/4)^3\ge1-\epsilon$, proving the stated rounding
guarantee.  If the support is empty, both values are zero and the
claim is immediate.

\paragraph{Finite sampling.}
The theorem concerns the exact Gibbs distribution.  Sampling from a
distribution within total variation distance $\delta_{\rm samp}$ changes
the expected objective by at most
$\delta_{\rm samp}f(E)$.  Appendix~\ref{sec:gibbs-sampling-audit}
implements the sampler in polynomial time and linear workspace, and
bounds its objective loss using the range on feasible matchings.
The finite streaming algorithm uses the coordinate bound in (GM11).

\section{Polynomial-time sampling from the Gibbs distribution}
\label{sec:gibbs-sampling-audit}
The small-coordinate lemma specifies a probability distribution by assigning a weight to every
matching. Computing its normalizing constant by summing over matchings would take too long.
Instead, a Markov chain updates a single stored edge at a time. We bound the number of
transitions and random bits needed for a given sampling accuracy, then translate this accuracy
into a loss in the value of the returned matching.
\subsection{A sampler using only the stored edges}
Let the stored graph have $n$ vertices and $m$ edges of positive activity.
Vertices incident to no edge of positive activity can be omitted; thus
the vertex arrays in this section use compressed identifiers only for
vertices incident to the stored edges.  If $m=0$, return the empty matching.
The all-matchings Markov chain of Jerrum and Sinclair~\cite{JS1989},
with the analysis of Sinclair~\cite{sinclair1992},
uses the following local transitions.  Choose an edge $e$ uniformly.
Delete it, if present, with probability $1/(1+\lambda_e)$;
insert it, if both endpoints are unmatched, with probability
$\lambda_e/(1+\lambda_e)$; or, if exactly one endpoint is matched
by an edge $h$, swap $h$ for $e$ with probability
$\lambda_e/(\lambda_e+\lambda_h)$.  Otherwise stay put.
Add a holding probability $1/2$ to make the chain lazy.
Its stationary distribution is the desired Gibbs distribution.
The canonical-path bound in Section 3(i) of that paper gives a
spectral gap at least $1/(8nm c_{\max}^2)$ for the lazy version,
where $c_{\max}=\max(1,\max_e\lambda_e)$.

In the present application $c_{\max}=1$ once $p$ is sufficiently
small.  Start the chain from the empty matching.  Its stationary
probability satisfies
\[
 \log\mu(\varnothing)^{-1}=\log Z
 \le\sum_e\log(1+\lambda_e)
 \le\sum_e\lambda_e\le nR/2.
\]
For a prescribed total variation distance $0<\delta<1$, let $k$
be the least integer with $2^k\ge1/\delta$, and take
\begin{equation}
 T_{\rm mix}=\left\lceil
 8nm c_{\max}^2\left(\frac{nR}{4}+k\right)
 \right\rceil.
 \tag{GS1}
\end{equation}
The standard spectral-gap bound for a lazy reversible chain gives
\[
 d_{\rm TV}(\mathcal L(M_{T_{\rm mix}}),\mu)
 \le\tfrac12\exp\!\left(\frac{nR}{4}
             -\frac{T_{\rm mix}}{8nm c_{\max}^2}\right)
 \le\delta/2.
\]
Thus (GS1) specifies the number of transitions using integer and rational
operations, with half the error allowance reserved for approximating the
transition probabilities.
The resulting running-time bound is polynomial for fixed accuracy parameters.
There is no dependence on an inverse minimum positive activity.

Each transition uses an indexed stored edge and a mate array
indexed by vertices.  Insertion, deletion, and a one-edge swap
update only constantly many mate entries.  Thus the sampler uses
$O(m+n)$ stored words, including the stored edges and their activities.
It uses no additional pass over the input stream and no value
oracle calls during its Markov-chain transitions.  An oracle
query is needed only to compare the resulting matching against
other candidate solutions.

For a fractional matching on a common rational grid, write
$x_e=n_e/Q$ and $\gamma=a/b$ in lowest terms.  Put
$N_v=\sum_{e\ni v}n_e\le Q$.  The activities in (GM8) are
\[
 \lambda_{uv}
 =\frac{abQn_{uv}}{(bQ-aN_u)(bQ-aN_v)}.
\]
The denominator factors are positive.  The grid denominator and
indices have polynomial bit length; every transition uses only a
fixed number of such quantities.  

Finite random bits supply the other half of the same allowance
$\delta$.  For the $T_{\rm mix}$ transitions in (GS1), choose a binary
grid satisfying
$2^b\ge2T_{\rm mix}(m+1)/\delta$.
Map a $b$-bit uniform integer into $m$ nearly equal index intervals
for the stored edge, and round the rational acceptance probability
down to the $2^{-b}$ grid.  The total variation distance between the exact and approximate
transition distributions is at most $(m+1)2^{-b}$ at each step; use an exact fair bit for the holding probability.
Coupling successive transitions bounds the accumulated total variation
distance by $\delta/2$.
This gives bounded worst-case polynomial running time, with no
unbounded rejection loop, and linear workspace.

\subsection{Effect of approximate sampling on the objective}
An ideal and an approximate sample can be coupled so that their disagreement probability equals the total variation distance. Multiplying this probability by an upper bound on the value of a feasible matching bounds the loss in expected value. Let
$V=\max_{e\in E}f(\{e\})$, and retain a singleton attaining $V$
as an additional candidate solution.  Every matching $M$ obeys
\[
 0\le f(M)\le\sum_{e\in M}f(\{e\})\le(n/2)V.
\]
Hence total variation distance at most $\delta_{\rm samp}$ changes the
expected objective by at most
\begin{equation}
 \delta_{\rm samp}(n/2)V.
 \tag{GS2}
\end{equation}
This bound uses the range on matchings; it does not require
bounding the value of the full stored set, which need not be a matching.

\section{Rounding on arbitrary simple graphs}
\label{sec:general-joint-completion}
After sampling the first matching, we restrict the second fractional vector to the unmatched vertices. The argument below bounds the expected mass retained by each coordinate. Figure~\ref{fig:upper-triangle} illustrates why endpoint-free events can be
dependent. We compute their joint probability exactly, divide by that probability to restore
each edge's expected mass, and bound the loss at vertices where the rescaled mass is too large.
This gives the finite version of Lemma~\ref{lem:main-round}.
Let $G=(V,E)$ be a finite loopless simple graph.  Let $A\subseteq E$
be the set of edges used in the first rounding step, and let $S\subseteq E$
be the stored edges in arrival order.  These sets may overlap.
Let $f$ be normalized,
nonnegative, monotone, and submodular, and put
\[
 g(T)=\Ex{f(T\cup R_p(S))-f(R_p(S))}.
\]
For each stored edge, the expected marginal value at arrival is
$w_e:=\Ex{f(e\mid R_p(S_{<e}))}$.  Its stored estimate
$\widehat w_e$ satisfies $0\le\widehat w_e\le w_e$ and is not recomputed.

Fix $0<D\le1/4$ for which the distribution $\mu_y$ defined in
Appendix~\ref{sec:referee-shearer-sampler} is available.
For a nonnegative vector $y$ on $A$, write
$\ell_v=\sum_{e\in\delta_A(v)}y_e$ for its fractional degree at $v$,
and put $s_v=1-\ell_v$.
Assume $\ell_v\le D$.  Let $x\ge0$ on $S$ satisfy the linear constraints
\begin{equation}
 x(\delta_S(v))\le s_v,\qquad
 x_{uv}\le p(1-\ell_u-\ell_v).
 \tag{GJ1}
\end{equation}
For every absent edge, set its $y$ coordinate to zero.  Define
\[
 t_{uv}=\sum_{w\in V\setminus\{u,v\}}y_{uw}y_{vw},\qquad
 q_{uv}=s_us_v+y_{uv}(1-\ell_u-\ell_v+y_{uv})-t_{uv}.
 \tag{GJ2}
\]
The quantity $q_{uv}$ will be the exact probability that both endpoints
are unmatched by the first matching.  In particular it is at least
$1-\ell_u-\ell_v\ge1-2D$.

\begin{theorem}[Rounding in two stages]
\label{thm:general-joint-completion}
Set
\[
 \delta_\mu=\frac{pD^2}{1-D},\qquad
 V_*=\frac{pD(2-D)}{1-D}(1+\delta_\mu).
\]
For any $\varepsilon_A>\delta_\mu$, define
\[
 \kappa_A=\frac{1-2V_*/[(1-2D)(\varepsilon_A-\delta_\mu)^2]}
                   {1+\varepsilon_A}.
 \tag{GJ3}
\]
Assume $\kappa_A\ge0$.  Sample a matching $T$ from $\mu_y$.
There is an explicit alteration
$z(T)$ such that, on every outcome, it is supported on $S-V(T)$,
has fractional degree at most one at every vertex, and satisfies
$0\le z_e(T)\le p$.
Moreover,
\[
 \Ex{z_e(T)}\ge\kappa_A x_e\qquad(e\in S).
 \tag{GJ4}
\]
If rounding fractional matchings with coordinates at most $p$ has
coefficient $\vartheta_G$, as in Lemma~\ref{lem:main-diffuse},
then rounding $z(T)$ for the function $f_T$ and adjoining
$T$ gives a matching $M$ with
\[
 \Ex{f(M)}\ge
 \vartheta_G\bigl[F_g(y)+\kappa_A\widehat w\cdot x\bigr].
 \tag{GJ5}
\]
All calculations and the alteration use $O(|A|+|S|)$ workspace and
polynomial time, without storing a triangle list or a dense matrix.
\end{theorem}

\begin{proof}
Write $J_e=\mathbf1\{e\in T\}$, $I_v=\mathbf1\{v\in V(T)\}$,
and $X_v=1-I_v$.  The distribution $\mu_y$ has $\Ex{J_e}=y_e$,
$\Ex{J_eJ_{e'}}=y_ey_{e'}$ for disjoint distinct edges, and
$J_eJ_{e'}=0$ for conflicting edges.  Since $T$ is a matching,
$I_v=\sum_{e\in\delta(v)}J_e$.  For distinct vertices $u,v$, the
only conflicting pairs after removing the possible edge $uv$ share
a third vertex.  Consequently
\[
 \Ex{I_uI_v}
 =y_{uv}+(\ell_u-y_{uv})(\ell_v-y_{uv})-t_{uv},
\]
and therefore $\Ex{X_uX_v}=q_{uv}$ as defined in (GJ2).
Independently, the union bound gives
$q_{uv}\ge1-\ell_u-\ell_v$.  The same identity implies the useful
one-sided covariance bound
\begin{equation}
 \operatorname{Cov}(X_u,X_v)
 =y_{uv}(1-\ell_u-\ell_v+y_{uv})-t_{uv}\le y_{uv}.
 \tag{GJ6}
\end{equation}
Here $1-\ell_u-\ell_v+y_{uv}\le1$ and $t_{uv}\ge0$.

For each stored edge, divide its coordinate by the exact probability
that both endpoints are unmatched:
\[
 a_{uv}=x_{uv}/q_{uv}\le p.
\]
For each vertex $u$, sum over its neighbors that are unmatched by $T$,
whether or not $u$ itself is matched:
\[
 Z_u=\sum_{uv\in S}a_{uv}X_v,\qquad \mu_u=\Ex{Z_u}.
\]
The denominator $q_{uv}$ accounts for the term $t_{uv}$ in (GJ2).  The
effect on the mean of $Z_u$ is small because every $a_{uv}$ is at most $p$.
Indeed (GJ2), its nonnegative $y_{uv}$ term, and (GJ1) give
\begin{align*}
 s_u\mu_u
 &=\sum_{uv\in S}a_{uv}s_us_v\\
 &\le\sum_{uv\in S}a_{uv}(q_{uv}+t_{uv})
 \le x(\delta_S(u))+p\sum_{uv\in S}t_{uv}\\
 &\le s_u+p\sum_w y_{uw}\sum_{v\ne u,w}y_{vw}
 \le s_u+pD\ell_u.
\end{align*}
Thus $\mu_u\le1+pD\ell_u/s_u\le1+\delta_\mu$.

Simplicity ensures that $Z_u$ has at most one coefficient per
neighbor, and each coefficient is at most $p$.  Use (GJ6) to
bound its variance.  The diagonal terms satisfy
\[
 \sum_{uv\in S}a_{uv}^2\operatorname{Var}(X_v)
 \le p\sum_{uv\in S}a_{uv}s_v\ell_v\le pD\mu_u.
\]
For the off-diagonal terms, summing ordered pairs of distinct
neighbors of $u$ gives
\[
 \sum_{v\ne w}a_{uv}a_{uw}\operatorname{Cov}(X_v,X_w)
 \le\sum_{v\ne w}a_{uv}a_{uw}y_{vw}
 \le pD\sum_{uv\in S}a_{uv}
 \le\frac{pD}{1-D}\mu_u.
\]
Hence
\[
 \operatorname{Var}(Z_u)
 \le\frac{pD(2-D)}{1-D}\mu_u\le V_*.
 \tag{GJ7}
\]
No conditioning on $X_u$ has occurred in this calculation.

In the altered vector, set to zero all coordinates incident to vertices
with $Z_u>1+\varepsilon_A$.
Chebyshev's inequality gives
$\Pr(Z_u>1+\varepsilon_A)\le V_*/(\varepsilon_A-\delta_\mu)^2$.
Set
\[
 z_{uv}(T)=\frac{a_{uv}}{1+\varepsilon_A}X_uX_v
       \mathbf1\{Z_u\le1+\varepsilon_A\}\mathbf1\{Z_v\le1+\varepsilon_A\}.
 \tag{GJ8}
\]
A vertex with $Z_u>1+\varepsilon_A$, or one matched by $T$, has
fractional degree zero in $z(T)$.  Every other vertex has fractional
degree at most $Z_u/(1+\varepsilon_A)\le1$.
Also $z_e\le a_e\le p$.  For each stored edge,
\begin{align*}
 \Ex{z_{uv}}
 &\ge\frac{a_{uv}}{1+\varepsilon_A}
   \left(q_{uv}-\frac{2V_*}{(\varepsilon_A-\delta_\mu)^2}\right)\\
 &\ge\frac{x_{uv}}{1+\varepsilon_A}
   \left(1-\frac{2V_*}{(1-2D)(\varepsilon_A-\delta_\mu)^2}\right).
\end{align*}
This proves (GJ4).  The union bound involves only this edge's two
endpoints, so there is no factor depending on the graph size.

For each fixed $T$, the coordinate bound permits us to apply
Lemma~\ref{lem:main-history} to the stored edges in arrival order:
\[
 f(T)+F_{f_T}(z(T))\ge\widehat w\cdot z(T)+g(T).
\]
Conditional on $T$, rounding $z(T)$ using Lemma~\ref{lem:main-diffuse}
therefore gives
\[
 \Ex{f(T\cup N)\mid T}
 \ge\vartheta_G[\widehat w\cdot z(T)+g(T)]
             +(1-\vartheta_G)f(T).
\]
Average, use (GJ4), and apply the comparison with independent rounding
from (RS3a), $\Ex{g(T)}\ge F_g(y)$, to obtain (GJ5).
\end{proof}

\paragraph{Choosing the alteration accuracy.}
If $D\le1/4$, $p\le1$, and $p\le6\varepsilon_A$, then
$\delta_\mu\le p/12\le\varepsilon_A/2$ and
$V_*\le(7p/12)(1+p/12)<3p/4$.  Formula (GJ3) implies
\begin{equation}
 \kappa_A\ge1-\varepsilon_A-12p/\varepsilon_A^2.
 \tag{GJ9}
\end{equation}
Take $\varepsilon_A=\varepsilon_R/4$, $0<\varepsilon_R<1$, and
$p\le(\varepsilon_R/2)^5/5120=\varepsilon_R^5/163840$, exactly as
in (FJ1).  The side conditions above hold, and
\[
 12p/\varepsilon_A^2
 \le3\varepsilon_R^3/2560\le\varepsilon_R/4.
\]
Thus $\kappa_A\ge1-\varepsilon_R/2$.  Combined with
$\vartheta_G=1-\varepsilon_R/2$, this yields
$\vartheta_G\kappa_A\ge1-\varepsilon_R$, as required by the finite
streaming implementation.

\paragraph{Rational arithmetic and workspace.}
The vector $y$ is stored on one common rational grid.
Compute $\ell_v$ by adjacency scans.  For each stored edge $uv$,
compute $t_{uv}$ by a merge of the $y$ adjacency lists of its
endpoints, or by a nested scan using only a constant number of
indices.  All $y$ products have the same denominator, so the exact
sum and $q_{uv}$ have $O(\log n)$-bit integer indices relative to the
chosen scale, for fixed accuracy parameters.  Since $q_{uv}\ge1/2$, division is
well-conditioned, although no numerical approximation is necessary.
It is enough to retain one $a_e$ per stored edge; no list of
triangles is stored.  The total arithmetic cost is
polynomial even with the simplest nested scans.

Before sampling, round $a_e$ down to a common mesh $\zeta$ with
$|S|\zeta\le\varepsilon_f$, and put
$\widetilde x_e=q_e\zeta\lfloor a_e/\zeta\rfloor\le x_e$.
The vector $\widetilde x$ satisfies (GJ1), and, when
$\widehat w_e\le\tau$, its lost weighted value is at most
$|S|\zeta\tau\le\varepsilon_f\tau$, because $q_e\le1$.
Apply the proof to $\widetilde x$.  Row sums have the common mesh
$\zeta$, and $z(T)$ has mesh $\zeta/(1+\varepsilon_A)$.
The finite streaming implementation samples from the Gibbs distribution
and compares objective values using this vector on a common grid.


\section{Finite accuracy, running time, and amplification}
\label{sec:finite-joint-streaming-wrapper}
We now combine the finite routines into the algorithm of Theorem~\ref{thm:main-upper}. Errors
made while reading the stream or choosing a fractional point are shared by all later rounding
trials. We bound their joint failure probability first. Conditional on a successful transcript,
rounding can be repeated independently on the same stored graph, with a single best candidate
kept between trials. The best singleton absorbs the remaining additive error, so both the
expected and high-probability guarantees are multiplicative.

Fix rational interval boundaries and update rates satisfying
\eqref{eq:source-drift}, and
nonnegative, bounded, nondecreasing piecewise-affine functions
$H_0,H_1$ that vanish near zero and satisfy
(JI1) and (JI2), and hence (JI3), at a fractional degree bound
$D=r(1-r)<1/4$ and coefficient
$\sigma>0$.  The final coefficient $\eta$ is defined below.
Fix a rational $b>0$ with $b\le1-e^{-D}$.  Let
$0<\varepsilon_R<1$ and choose rational $p$ so that
\begin{equation}
 0<p\le\min\left\{1,\frac{(\varepsilon_R/2)^5}{5120}\right\},
 \qquad
 \int H_i(a)a^{-2}\,da\le(1+pC_{\max})^{-1}\quad(i=0,1).
 \tag{FJ1}
\end{equation}
Also impose the fixed growth condition for the potential updates
$\kappa_{\rm src}p\le1$.  Choose $\varepsilon_0\in(0,1/4)$ with
$2/\varepsilon_0$ an integer, and put $\Theta=1-\varepsilon_R$.
All these parameters are independent of $n$.

\begin{theorem}[Finite implementation]
Let the input be any finite simple graph whose distinct
edges arrive once in a fixed adversarial order.  Set
$\eta=\min\{\sigma,b\}$.  The algorithm below uses
$O((\nu+1)\log^2 n)$ words, polynomial computation and value queries,
and returns a matching with
\begin{equation}
 \Ex{f(M)}\ge
 \frac{(1-\varepsilon_0)\Theta\eta}{1+5\varepsilon_0}
             \operatorname{OPT}.
 \tag{FJ2}
\end{equation}
For any fixed confidence exponent, the same space bound, with
parameter-dependent constants, supports high-probability amplification
to any fixed approximation factor strictly larger than the factor given by
the conditional expectation bound, after allowing for the final
discretization of objective values.  Integer indices relative to the guessed
value use $O_c((\nu+1)\log^3 n)$ bits, under the convention for
representing oracle values stated below.
\end{theorem}

\subsection{Streaming estimates}

Normalize $f$ as in the main construction.  Maintain the best input
singleton $B$ and the dyadic guesses in
$[V_i,(n/\varepsilon_0)V_i]$.  There are at most
$L=2+\lceil\log_2(n/\varepsilon_0)\rceil$ active guesses.
The cases $n<2$ or $B=0$ are settled by the empty matching.
For each final guess, store the accepted edges in arrival order,
conservative estimates $\widehat w_e$, and vertex potentials.  

For guess $\tau$, use the same quantities
\[
 \theta_\tau=\varepsilon_0\tau/n,\quad
 d_\tau=\varepsilon_0\tau/n^2,\quad
 \zeta_\tau=p\varepsilon_0^2\tau/n^4.
\]
Estimate each expected marginal value $w_e$ at arrival to accuracy
$d_\tau$, subtract
$d_\tau$, truncate at zero, and round downward to that grid.
Accept precisely when
$\widehat w_e\ge\theta_\tau$ and
$\widehat w_e\ge\phi_u^-+\phi_v^-+d_\tau$.
Execute the same finite potential update with rational parameters and gain
$p\widehat w_e$, then round each complete endpoint increment upward
once to the $\zeta_\tau$ grid.

The analysis of the streaming selection rule, with its original sample
and probability grids, gives a simultaneous event on which
$0\le\widehat w_e\le w_e\le\widehat w_e+3d_\tau$.
For the final guess satisfying $B\le\tau<2B$, and the remaining part
$O$ of an optimum after deleting its ignored prefix, it gives
\begin{equation}
 \operatorname{OPT}\le X+g(O)+6\varepsilon_0\tau,
 \qquad
 X=G_S+\sum_{v\notin V(O)}\psi_v,
 \quad G_S=\sum_{e\in S}p\widehat w_e.
 \tag{FJ3}
\end{equation}
Here $\psi$ consists of ideal increments along the actual rounded
execution.  The comparison inequality gives
\begin{equation}
 W_D(O)\ge\sigma X.
 \tag{FJ4}
\end{equation}

The bounds on the number of stored edges, guesses, and integer
indices in Appendix~\ref{sec:joint-source-foundations} apply.  In
particular, the bound on the number of stored edges is deterministic
even when some estimates are inaccurate.  Vertices with positive potentials
use sparse dictionaries, and all final arrays use compressed identifiers
for the stored edges and their incident vertices.
Neither filtering nor postprocessing uses a bipartition.

\subsection{Discretized continuous greedy and sample bounds}

On each final stored set run (JF4) with $\varepsilon=\varepsilon_0$,
using the packing routine with projected dual updates at additive direction
accuracy $\varepsilon_0\tau/8$.  Gradients are estimated from
samples from two product distributions using the original value oracle,
as specified in Appendix~\ref{sec:joint-distorted-flow-oracle}.
For a prescribed failure probability $\omega$, allocate
$\omega/2$ to the estimates during the stream and $\omega/2$ to the
derivative estimates in continuous greedy.  There are at most $n^2L$
estimates during the stream,
including discarded guesses.

For an explicit deterministic bound on the number of iterations, let
$m_*=n^2$, let $K_*$
be (JF4)'s value at $m_*$, and put
$\xi_* =\min\{1/4,\varepsilon_0/(16Dm_*)\}$.
There are at most $LK_*m_*$ derivative estimates.  Assign failure
$\omega/(2LK_*m_*)$ to each and use
\begin{equation}
 N_{\rm grad}=
 \left\lceil8\xi_*^{-2}
              \log\frac{4LK_*m_*}{\omega}\right\rceil
 \tag{FJ5}
\end{equation}
samples, or a larger integer obtained by repeated-doubling logarithm
bounds.  The sample bound for the streaming selection rule applies with
failure probability $\omega/(2n^2L)$ per estimate.  Conditional Hoeffding
bounds and a union bound show that all estimates are accurate with
probability at least $1-\omega$.

The packing routine is deterministic and always returns a feasible
direction; it introduces no additional probability of failure.  Its bound on the number of iterations,
step size, and output denominator are fixed for each guess.  Together
with the Euler average, this supplies a common polynomial-size grid
for the final $x,y$.  All coefficient approximations and sample sums
have $O_c(\log n)$-bit indices for fixed parameters and
$\omega=n^{-c}$ or fixed $\omega$.  There are at most
$2LK_*m_*N_{\rm grad}$ value-oracle queries for the gradients,
in addition to the polynomial number of queries during the stream.
When all estimates are accurate, the output point satisfies
\begin{equation}
 \widehat w\cdot x+F_g(y)
 \ge\sigma X+b g(O)-\varepsilon_0\tau/2.
 \tag{FJ6}
\end{equation}

\subsection{The rounding guarantee}

Set $\varepsilon_A=\varepsilon_R/4$ for alteration and
$\varepsilon_G=\varepsilon_R/2$ for rounding using the Gibbs distribution.
The coefficient in the second step is $\vartheta_G=1-\varepsilon_G$ under (FJ1).
For all $D\le1/4$, the general-graph estimate (GJ9) gives
\[
 \kappa_A\ge
 1-\varepsilon_A-\frac{12p}{\varepsilon_A^2}
 \ge1-\varepsilon_R/2.
\]
Indeed (FJ1) implies
$12p/\varepsilon_A^2\le3\varepsilon_R^3/2560
\le\varepsilon_R/4$ and
$pD^2/(1-D)\le\varepsilon_A/2$.
Thus the error from the terms $t_{uv}$ is covered by the stated rounding
allowance.
Consequently
\begin{equation}
 \vartheta_G\kappa_A\ge(1-\varepsilon_R/2)^2
                              \ge\Theta.
 \tag{FJ7}
\end{equation}

Before sampling, compute the fractional degrees and exact probabilities
$q_{uv}$ in (GJ2), and floor $a_{uv}=x_{uv}/q_{uv}$
to a common grid of mesh
$\varepsilon_0/(8m)$, with the case $S=\varnothing$ treated separately.
Writing $\bar a_e$ for the rounded value and
$\bar x_e=q_e\bar a_e$, we have $\bar x\le x$; the pair
$(\bar x,y)$ remains feasible.  The normalization identity becomes
$\Ex{X_uZ_u}=\bar x(\delta(u))\le s_u$, and the lost weighted
value is at most $\varepsilon_0\tau/8$.
The row sums and altered vector are now on one common rational grid.
Sample $T$ using the sampler for $\mu_y$ with the stopping rules above.
Construct $z(T)$ as in (GJ8), setting coordinates to zero at vertices
with $Z_u>1+\varepsilon_A$.
Then round this feasible vector on $S-V(T)$ using the Gibbs distribution
with activities (GM8) for $f_T$, and adjoin $T$.

Require total variation distance at most $\varepsilon_0/(4n)$ from
$\mu_y$, including errors from stopping the recursive procedure and from
stopping rejection sampling for rational probabilities.  Require total variation distance at most
$\varepsilon_0/(4n)$ from the Gibbs distribution, including the mixing
error and the error from approximating transition probabilities.
Every resulting matching has value at most $(n/2)B$, so these
two errors together cost at most $\varepsilon_0B/4$ in expected
original objective.  These bounds apply in expectation; no event
requiring simultaneous concentration of all $Z_u$ is imposed.

The edge-marginal estimate (GM9) applies to the vector $z$ on the common
grid because it is nonnegative, has fractional degree at most one at
every vertex, and has coordinate bound $p$.  Row sums on the common grid
have short denominators; each
activity and each local transition uses only a fixed number of
these rational quantities.  The activity bound, spectral gap, mixing estimate, and approximation
of the transition probabilities therefore give deterministic polynomial
time and linear workspace.

\subsection{Expectation and amplification}

Finish the streaming computations and continuous greedy before drawing
the random choices used for rounding.  Conditional on any execution in
which all estimates are accurate, apply (GJ5), (FJ6), and
(FJ7).  Since $X,g(O)\ge0$ and $\eta=\min\{\sigma,b\}$, a
candidate for the guess satisfying $B\le\tau<2B$ has expectation at least
\[
 \Theta\eta\operatorname{OPT}
   -(6\eta+5/8)\varepsilon_0\tau-\varepsilon_0B/4.
\]
Compare all candidate matchings and the best singleton by flooring
their $f$-values on the $\varepsilon_0B/2$ grid.  This costs
at most $\varepsilon_0B/2$ against the best true candidate.  The
best singleton is itself a grid point, so the returned value is
always at least $B$.

Let $Y$ be its conditional expectation over the random choices used for
rounding.
Because $\eta\le b<1/4$ and $\tau<2B$, the preceding allowances
give
\begin{equation}
 Y\ge B,\qquad Y+5\varepsilon_0B
                          \ge\Theta\eta\operatorname{OPT}.
 \tag{FJ8}
\end{equation}
The first inequality absorbs the additive term in the second:
\[
 (1+5\varepsilon_0)Y\ge Y+5\varepsilon_0B
                       \ge\Theta\eta\operatorname{OPT}.
\]
Thus keeping one singleton converts the entire additive error budget
into a multiplicative loss.  The event that all estimates are accurate
precedes the random choices used for rounding; its complement still gives a nonnegative feasible output.
Taking $\omega=\varepsilon_0$ and averaging proves (FJ2).

For high probability choose the failure probability for the estimates $\omega=\min\{\varepsilon_0,\delta/2\}$.
Conditional on each execution with accurate estimates, repeat the two
rounding steps and the intervening alteration independently, and include
the best singleton.
The fixed sampler errors have already been paid in each trial's
expectation; they need not be divided by the number of trials.
Every trial lies in $[0,\operatorname{OPT}]$.  If a trial expectation
is at least $\operatorname{OPT}/R_0$, then for any fixed $R_1>R_0$
its probability of reaching $\operatorname{OPT}/R_1$ is at least
$(R_1-R_0)/(R_0(R_1-1))$.  Choose $R_1$ with a positive margin for
the final comparison grid and repeat a fixed constant times
$\log(2/\delta)$ trials.  Total failure is at most $\delta$.
All trials reuse the same stored sets and fractional vectors and are processed
sequentially, so no extra stream pass or asymptotic space is required.

\subsection{An explicit polynomial running-time bound}
\label{sec:explicit-upper-runtime}

All potential-update rates, the number of intervals for $c$, and accuracy parameters are fixed
constants.  We count value-oracle calls and arithmetic operations
separately.  The oracle model permits arithmetic and comparisons with
the returned values; the implementation stores a bounded number of
original scale values and short integer indices relative to those scales.
This convention does not assert that an arbitrary real oracle value has
a finite bit representation.  If the cost of representing the original
oracle values is charged, it is added to the bounds below.

Write $s$ for the number of edges in one stored set.  The packing
problem has $N=2s$ variables and $M=O(s)$ rows.  For fixed direction
accuracy, its parameters obey
\[
 R=O(s^2),\qquad \eta^{-1}=O(s^3),\qquad J=O(s^8).
\]
An iteration requires $O(s)$ arithmetic operations through the implicit
matrix products.  Discretized continuous greedy has $K=O(s^2)$ steps, so its direction
optimization requires $O(s^{11})$ arithmetic operations.  Computing
all scalar coefficients by the recurrence on a fixed grid takes only
$O(K^2)=O(s^4)$ additional operations.  We use this elementary solver because its working space is linear.

For prescribed failure $\delta=n^{-\lceil c\rceil}$ and fixed $c>0$, the global
bound on the number of gradient samples is $O_c(n^4\log n)$ samples per derivative.
There are $O(s^3)$ derivatives per stored set and $O(\log n)$ final
stored sets.  Since $s\le n^2$, all gradient estimates use
$O_c(n^{10}\log^2 n)$ value-oracle calls.  There are at
most $n^2L$ estimates during the stream, each with $O_c(n^4\log n)$ samples, so their
query count is smaller.  Each sample set is assembled and discarded in
$O(s)$ operations; these costs are below the direction-optimization
bound.  The two matching samplers, scans for computing $t_{uv}$, and independent
rounding trials also have smaller polynomial costs.

Thus a conservative bound for the stated implementation is
\[
 O_c(n^{22}\log n)\ \text{arithmetic operations}
 \quad\text{and}\quad
 O_c(n^{10}\log^2 n)\ \text{value-oracle calls}.
\]
The sharper bound on the number of stored edges,
$s=O((\nu+1)\log n)$ for one stored set, can be substituted in the
intermediate expressions.  No near-linear processing-time claim is
needed for the approximation theorem.  The space bound remains
$O_c((\nu+1)\log^2 n)$ words throughout, because directions, sample
sets, and rounding trials are processed sequentially.

\section{Choice of parameters and exact verification}
\label{sec:joint-exact-parameters}
To obtain the factor $4.524656$, the same choice of parameters must make the charging,
optimization, and rounding inequalities hold. In particular, increasing the degree bound $D$
improves the coefficient $1-e^{-D}$ but leaves less vertex capacity for the linear term. We
give one rational choice, check the four endpoint cases of the charging inequality and the two
integral bounds, and then show that the finite errors still leave enough slack for
amplification. Only feasibility of these inequalities is needed.

\subsection{The parameters}

\begin{center}
\begin{tabular}{@{}ll@{}}
\toprule
Parameter & Exact value\\
\midrule
Coefficient $\sigma$ of $X(O)$ & $10000000/45246551$\\
Sampler parameter $r$ & $128882864927/250000000000$\\
Coefficient $b$ of $g(O)$ & $110505663541/500000000000$\\
Coordinate bound $p$ & $10^{-48}$\\
Rounding allowance $\varepsilon_R$ & $10^{-8}$\\
Finite allowance $\varepsilon_0$ & $10^{-10}$\\
Number of intervals for $c$ & $27648$\\
Common knots of $H_0,H_1$ & $1136$\\
\bottomrule
\end{tabular}
\end{center}
The algorithm and the verification use the same degree bound:
\[
 D_0=D=r(1-r)
 =\frac{15609923359958673284671}{62500000000000000000000}
 <\frac14.
\]
Both functions $H_0,H_1$ are piecewise linear on the common rational knot
sequence, are zero near zero, and are constant after the final knot.
The intervals for $c$ have rational boundaries from $1/2$ to $1$ and
rational rates $A_j\ge B_j>0$ satisfying \eqref{eq:source-drift}.

\begin{figure}[!htbp]
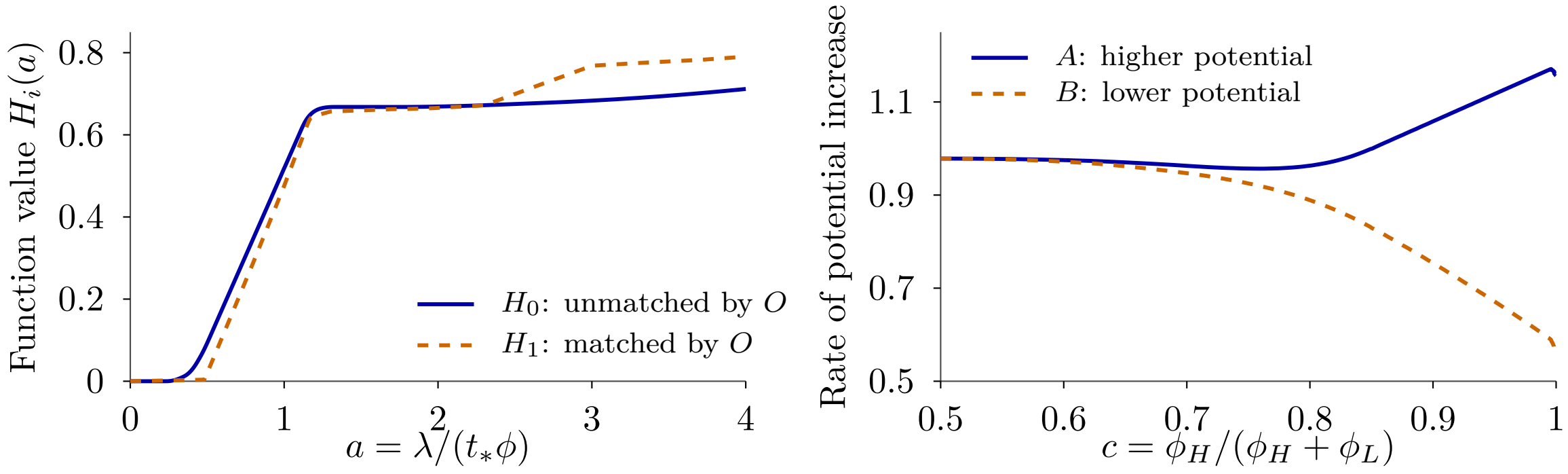

\centering
%
\caption{The functions $H_0,H_1$ and the update rates. Left: the functions for vertices unmatched and matched by $O$, shown for $0\le a\le4$, where
$a=\lambda/(t_*\phi)$.  Right: the rates at the higher and lower
endpoint vertex potentials as functions of $c=\phi_H/(\phi_H+\phi_L)$.  These curves are
illustrative; the proof uses the complete rational arrays and their
exact verification.}
\label{fig:source-profiles}
\end{figure}

Figure~\ref{fig:source-profiles} illustrates the two functions and the update rates.

\FloatBarrier

\subsection{Complete exact data and interpretation}
The complete parameter file \nolinkurl{source_parameters.json} contains
$\sigma,D_0,p$, the 27648 pairs $(A_j,B_j)$, the 27648+1
boundaries of the intervals for $c$, and the 1136 common knots and values of $H_0,H_1$.
Every number is a string representing an exact integer or a rational
fraction.  The functions are affine between successive knots and
constant after the final knot.  They vanish on a neighborhood of zero.
The local inequalities are checked on $[c_{j-1},c_j]$, with
$c_0=1/2$ and $c_k=1$.  Operationally, an internal boundary belongs
to the interval on its left, and the first interval also contains $1/2$.

The ancillary files include the complete parameter table, the verification
programs, and the reference certificates. The driver \texttt{verify.py}
checks the SHA-256 digest of the parameter table before using it.
The parameter-file hash is
\begin{center}\footnotesize\ttfamily
338f62f4ba1c0fcb0edb8841e56d2e2c\\
b3e0da4e137c8e8ed70ad787551c7970
\end{center}

The validation checks strictly positive rates, $A_j\ge B_j$,
$A_1=B_1$, ordered boundaries and knots, nonnegative nondecreasing functions, $\sigma(1+A_j+B_j)<1$, and the required range on which $H_0,H_1$ are constant.  The driver \texttt{verify.py} additionally checks (SP0) by exact
rational comparisons on all 27648 intervals for $c$.  Thus the direction of each
ideal update to vertex potentials is certified as well.

\subsection{Verification over the entire continuous domain}
For every interval $[c_{j-1},c_j]$ and both of its endpoints $c$, and every
$(i,j)\in\{0,1\}^2$, the checker proves
\begin{align*}
 &(1-D_0i)A H_i(a)+(1-D_0j)B H_j(d)\\
 &\quad+[1-D_0(i+j)](1-ca-(1-c)d)_+\\
 &\hspace{2em}\ge\sigma[1+(1-i)A+(1-j)B]
 \qquad(a,d\ge0).
\end{align*}
For fixed $(a,d)$ the only dependence on $c$ is through the monotone function $(1-ca-(1-c)d)_+$,
so the two interval endpoints suffice.  For fixed $c<1$, monotonicity of $H_i,H_j$ reduces the half-plane above $ca+(1-c)d=1$
to its boundary.  Below that line the expression is affine on each
rectangle between successive knots.  It is therefore enough to check grid vertices
and all intersections of grid lines with the line $ca+(1-c)d=1$.

The implementation evaluates all boundary intersections explicitly.
It checks the pairs of knots by exact prefix minima.
Writing $U(a)=(1-D_0i)A H_i(a)$,
$V(d)=(1-D_0j)B H_j(d)$ and $q=1-D_0(i+j)$, the expression where $ca+(1-c)d\le1$ is
\[
 [U(a)-qca]+[V(d)-q(1-c)d]+q
       -\sigma[1+(1-i)A+(1-j)B].
\]
For every $a$-knot, the eligible $d$-knots form a prefix, and its
exact minimum certifies every pair in that prefix.  The case $c=1$
explicitly includes $(a,d)=(1,0)$ even if $1$ is not a knot.
The derived inequalities for zero endpoint potentials are also checked.

The complete run covers 46\,483\,643\,836 grid pairs and evaluates
368\,163\,240 separation points, with 110\,592 checks with zero endpoint potentials.
Every exact slack is positive.  In particular the minimum is at least
$\frac{6187}{500000000000}$; its full rational value and witness occur in
\nolinkurl{source_certificate.json}.  These counts concern universal
scalar inequalities, not a sample of finite graph instances.

All operations in the checker for the local inequalities use arbitrary-precision GMP
rationals.  The runner compiles the included C++ source, records the
source and binary hashes and the compiler version, and partitions the index range
into disjoint intervals whose union is the entire index set.

\subsection{Checking the integral inequalities}
The local inequalities certify individual updates. We must still show that the total charge at
a vertex fits its dual budget, which is the integral condition~\eqref{eq:main-integral}. The
functions are piecewise linear, so the integral reduces to a finite sum of slopes times
logarithms. Its upper bound must be rigorous even when the numerical margin is small. We
therefore bound each logarithm from above using rational arithmetic.
Let $a_j$ be the common knots and let
$s_{i,j}=[H_i(a_{j+1})-H_i(a_j)]/(a_{j+1}-a_j)$ be the slope
on $[a_j,a_{j+1}]$.  Integration by parts gives
\[
 \int_0^\infty H_i(a)a^{-2}\,da
 =\sum_j s_{i,j}\log(a_{j+1}/a_j),
\]
where only positive-slope intervals contribute.  Range reduction writes
each logarithm using $z\in[1,2]$.  With $u=(z-1)/(z+1)$, the
checker uses the rational upper bound
\[
 \log z\le2\sum_{j=0}^{35}\frac{u^{2j+1}}{2j+1}
                   +\frac{2u^{73}}{73(1-u^2)}.
\]
Each nonnegative summand is rounded upward to the $10^{-18}$ grid.
The resulting exact bounds are
\[
 I_0\le \frac{999999950956391491}{1000000000000000000},\qquad I_1\le \frac{990582544907118101}{1000000000000000000},
\]
and both are strictly below $(1+pC_{\max})^{-1}$,
where $C_{\max}=\max_j(A_j+B_j)$.

\subsection{Finite factor and probability margin}
The separate finite-constant checker proves
\[
 b<\sum_{j=1}^{24}(-1)^{j+1}\frac{D^j}{j!}\le1-e^{-D},
 \qquad b>\sigma,
\]
as well as the coordinate bound (FJ1), the growth condition for the vertex potentials, and the general-graph mean, variance and alteration
inequalities used by the finite implementation.  Its conditional coefficient
is
\[
 \Theta\min\{\sigma,b\}=\frac{99999999}{452465510}.
\]
Consequently (FJ2) gives the finite expected approximation factor
\[
 R_{\rm fin}
 =\frac{1+5\varepsilon_0}{(1-\varepsilon_0)\Theta\sigma}
 =\frac{1508218367420775850}{333333329966666667}.
\]
The finite-constant checker certifies the slightly larger bound
\[
 \overline\rho
 =\frac{1/[(1-\varepsilon_R)\sigma]+100\varepsilon_0}
        {1-\varepsilon_0}
 =\frac{4524655109999999900}{999999989900000001}
 <4.524656.
\]
Indeed, $1/(\Theta\sigma)<10$ for these parameters, so
\[
 \overline\rho-R_{\rm fin}
 =\frac{\varepsilon_0}{1-\varepsilon_0}
       \left(100-\frac5{\Theta\sigma}\right)>0.
\]
This conservative bound also bounds the conditional rounding-trial
factor, and we use its certified probability margin below.
Choose $\rho_{\rm hp}=4.5246552$.
The rounding-trial success bound in the finite proof satisfies
\[
 \frac{\rho_{\rm hp}-\overline\rho}
      {\overline\rho(\rho_{\rm hp}-1)}
 \ge\frac1{359988611}>0,
\]
and the final comparison-grid margin is
\[
 \frac1{\rho_{\rm hp}}-\frac{\varepsilon_0}2
            -\frac1{4.524656}=\frac{1248400585289171}{31988294216580000000000}>0.
\]
Use transcript failure budget $\omega=\min\{\varepsilon_0,\delta/2\}$.
Run $\lceil 359988611\log(2/\delta)\rceil$ independent
sequential rounding trials on the stored edge sets.  This gives
factor $4.524656$ with failure at most $\delta$.  Taking
$\delta=n^{-\lceil c\rceil}$ proves the high-probability theorem.  The trials
are postprocessing and do not revisit the input stream.

\subsection{An approximation family with fixed update rates}

The explicit theorem uses fixed small error allowances. The same update rates and functions $H_0,H_1$ also give a family of guarantees.
\begin{corollary}\label{cor:fixed-source-family}
For every fixed $\epsilon>0$, there is a one-pass
$(4.5246551+\epsilon)$-approximation on general simple graphs,
both in expectation and with high probability. Its space is
$O_{\epsilon,c}((\nu+1)\log^2 n)$ words, and its running time and
number of oracle calls are polynomial for fixed $\epsilon,c$.
\end{corollary}
\begin{proof}
Write $p_0$ for the displayed coordinate bound. Decreasing $p\le p_0$ can only
relax the integral bound $\int H_i(a)a^{-2}\,da\le(1+pC_{\max})^{-1}$;
the local inequalities do not otherwise depend on $p$. Keep $D$ and
$b>\sigma$ fixed. Choose positive rational
$\varepsilon_R,\varepsilon_0$ tending to zero, with
$2/\varepsilon_0$ integral, and then
\[
 0<p\le\min\{p_0,(\varepsilon_R/2)^5/5120\}.
\]
The factor in~\eqref{eq:main-finite-factor} tends to
$1/\sigma=4.5246551$. Choose the allowances to leave a strict
margin below $4.5246551+\epsilon$. Reducing $\varepsilon_0$ further
leaves room for an intermediate amplification threshold and the final
comparison grid. The finite amplification argument applies without
another pass. All choices are constants for fixed accuracy.
\end{proof}

\subsection{Reproducing the numerical certificates}
\label{sec:reproduce-certificates}
The ancillary files contain the rational inputs and verification programs
for the monotone and nonmonotone guarantees. In a writable directory
containing those files, run the following commands with Python~3, a
C++17 compiler (\texttt{g++}), and the GMP development libraries installed.
\begin{verbatim}
python3 verify.py
python3 check_nonmonotone.py
\end{verbatim}
The first command checks all $27{,}648$ parameter intervals, both integral
bounds, and the finite-factor and amplification arithmetic in this appendix.
It records the minimum local slack and a witness attaining it. To check
only the parameter digest, the update direction, and the finite constants,
use \texttt{python3 verify.py --finite-only}. The second command verifies
the nonmonotone integral bounds, the four endpoint cases, and the factor
in Appendix~\ref{sec:nm-asymmetric-certificate}. Both programs read the
rational parameter tables directly from the accompanying JSON files.

\section{Proofs of the extensions}
\label{sec:extension-proofs}

\subsection{Exact reduction for monotone submodular \texorpdfstring{$b$}{b}-matching}
\label{sec:exact-monotone-reduction-proof}

\begin{proof}[Proof of Proposition~\ref{prop:exact-monotone-capacity}]
Cap each $b_v$ at $n-1$ and discard edges with a zero-capacity endpoint.
Represent vertex $v$ by slots $(v,1),\dots,(v,b_v)$. Maintain a greedy
$b$-matching $B$, initially empty. When $e=uv$ arrives, if both
endpoints have room in $B$, add $e$ to $B$ and emit one edge joining
the next unused slot of $u$ to the next unused slot of $v$. These slot
assignments are permanent. Otherwise emit every pair consisting of a
slot of $u$ and a slot of $v$. Each emitted edge carries its source
label $e$, which its endpoints identify because the source graph is simple.

For any set $A$ of emitted edges, let $\pi(A)$ be its set of distinct
source labels and give $A$ value $g(A)=f(\pi(A))$. This objective is
monotone and submodular. Indeed, the marginal of a first copy is the
original marginal and a repeated copy has marginal zero. For nested
sets, the first-copy marginals decrease by submodularity of $f$,
and the transition from a first to a repeated copy respects diminishing
returns because $f$ is monotone.

Every ordinary matching projects to a feasible $b$-matching since each
slot is used at most once. Conversely, given a source $b$-matching $S$,
seat the edges of $S\cap B$ in their permanent slots. At each vertex
assign distinct unused slots to its remaining incidences in $S$.
Every edge of $S\setminus B$ has all slot pairs available, so these
assignments lift $S$ to an ordinary matching of value $f(S)$.
Thus the optima agree and projection preserves the value returned by
any ordinary solver.

To bound space, let $C$ be the vertices saturated by the final $B$.
Every rejected edge touches $C$, as $B$ only grows, and
$\sum_{v\in C} b_v\le 2|B|$. In an ordinary matching each selected copy
of a rejected edge consumes a distinct slot at a vertex of $C$.
Thus at most $2|B|$ such copies and at most $|B|$ single copies of
accepted edges can occur. The target matching number is at most
$3|B|\le 3\nu_b$. The reduction stores endpoint counts only for vertices
incident to $B$, using $O(\nu_b+1)$ words. It enumerates the slot pairs
one at a time, emitting at most $n^2$ edges per arrival on at most $n^2$
virtual vertices. Each target oracle query needs one original query
on its distinct source labels, using space linear in the query size.
These bounds give the claimed resources. Applying
Theorem~\ref{thm:main-upper} on the virtual graph proves the monotone
part of Theorem~\ref{thm:extensions}, with its expectation and
high-probability guarantees.
\end{proof}

\subsection{Tools for nonmonotone objectives}
\label{sec:ext-nonmonotone}

We first record how random augmentation changes a nonnegative objective. If $h$ is
nonnegative and submodular and $R$ is a random set with coordinate
marginals $r_e$, then
\begin{equation}\label{eq:ext-lovasz}
 \Ex{h(R)}\ge \int_0^1 h(\{e:r_e\ge t\})\,dt.
\end{equation}
To see this, order the coordinates by decreasing $r_e$ and let $P_i$
be the first $i$ elements in this order. Submodularity bounds $h(A)$
below by $h(\varnothing)+\sum_{e_i\in A}[h(P_i)-h(P_{i-1})]$.
Expectations give the integral. In particular, adding a random set
with maximum inclusion probability $p$ to a fixed set preserves at
least a $1-p$ fraction of its value in expectation. This follows by
applying~\eqref{eq:ext-lovasz} to the function with the fixed set
already included.

\paragraph{Charging the optimum.}
Use the same streaming rule and potential updates, with $f$ unshifted.
Let $f_0=f(\varnothing)$ and set
\[
 h(T)=\Ex{f(T\cup R_p(S))},\qquad
 h_0=h(\varnothing),\qquad
 g(T)=h(T)-h_0,\qquad
 G=p\sum_{e\in S}\widehat w_e .
\]
Thus $h$ is nonnegative and submodular, while $g$ may be signed.
For the ideal analysis put $P=\sum_{v\notin V(O)}\psi_v$ and
$X=G+P$, with $O=M^*\cap S$ for an optimal matching $M^*$.
By~\eqref{eq:ext-lovasz}, $h(M^*)\ge(1-p)\OPT$. Submodularity
bounds $h(M^*)\le h(O)+\sum_{e\in M^*\setminus S}g(\{e\})$.
The charge to rejected edges is the same as in
Lemma~\ref{lem:main-charge}: each final singleton gain of a rejected
edge is at most its arrival marginal, and distinct rejected optimum
edges have disjoint endpoints outside $V(O)$. Since $h_0=f_0+G$ in
the ideal analysis, we obtain
\begin{equation}\label{eq:ext-charge}
 (1-p)\OPT\le f_0+X+g(O).
\end{equation}

\paragraph{Gibbs rounding for a nonnegative submodular objective.}
The following lemma allows negative marginals in the second
rounding step.

\begin{lemma}\label{lem:ext-sparse}
For every $\xi>0$, there is a constant $p_\xi>0$ such that the Gibbs
construction of Lemma~\ref{lem:main-diffuse}, with suitable $\gamma$
and $p\le p_\xi$, satisfies
$\Ex{k(N)}\ge (1-\xi)F_k(z)$ for every nonnegative submodular function
$k$, including unnormalized functions.
\end{lemma}
\begin{proof}
Omit coordinates of zero activity. Write $m_i=\Pr(i\in N)$,
$m_*=\max_i m_i<1$, $\lambda_*=\max_i\lambda_i$, and
$D_h=\sqrt{\lambda_*/2}$. If there are no positive activities,
the claim holds with $N=\varnothing$.
For any set $A$, nonnegativity and submodularity imply
\begin{equation}\label{eq:ext-negative}
 \sum_{i\notin A}[-k(i\mid A)]_+\le k(A),
\end{equation}
by adjoining simultaneously all elements with negative marginal.
Order the edges and independently draw a product set $R$ with
probabilities $m$. In the coordinate hybrid from $R$ to $N$, put
$L_i=k(i\mid N_{<i}\cup R_{>i})$ and $B_i=k(i\mid R_{>i})$.
The hybrid change is $m_i(\Ex{L_i\mid i\in N}-\Ex{L_i})$.
Apply (GM5) to $[L_i]_+$ with upper bound $[B_i]_+$ for fixed
$R_{>i}$. For the negative part keep only the possible loss upon
inclusion. Thus the total loss is at most
\[
 D_h\sum_i m_i\Ex{[B_i]_+}
   +\sum_i m_i\Ex{[-L_i]_+\mid i\in N}.
\]
The signed sum of the $m_i\Ex{B_i}$ telescopes to
$F_k(m)-k(\varnothing)$. On $i\notin R$,
$[-B_i]_+\le[-k(i\mid R)]_+$, so~\eqref{eq:ext-negative}
bounds the negative part of this sum by
$m_* F_k(m)/(1-m_*)$. Consequently the first displayed loss
is at most $D_h F_k(m)/(1-m_*)$.

For the second loss, the injection $N\mapsto N\setminus\{i\}$
divides a Gibbs weight by $\lambda_i$ and leaves $L_i$ unchanged.
Hence $m_i\Ex{[-L_i]_+\mid i\in N}$ is at most
$\lambda_i\Ex{\mathbf1_{i\notin N}[-L_i]_+}$.
We may further restrict to $i\notin R$ at a cost of $1/(1-m_i)$,
since $R_i$ is independent of the integrand. On the resulting event,
$[-L_i]_+\le[-k(i\mid N\cup R)]_+$.
Summing and using~\eqref{eq:ext-negative} gives the bound
$\lambda_*\Ex{k(N\cup R)}/(1-m_*)$, which is at most
$\lambda_*(\Ex{k(N)}+F_k(m))/(1-m_*)$ by nonnegative
submodularity. Rearranging, we obtain
\[
 \Ex{k(N)}\ge
 \frac{1-(D_h+\lambda_*)/(1-m_*)}
      {1+\lambda_* /(1-m_*)}F_k(m).
\]
For a fixed $\gamma<1$,
the activities in (GM8) and calibration (GM9) have
$\lambda_*=O_\gamma(p)$ and $m_*=O_\gamma(p)$.
Writing $\alpha=\gamma e^{-K}$ with $K$ from (GM9), we have
$m\ge\alpha z$. Couple $R_m$ to a superset of $R_{\alpha z}$
by adding independent coordinates whose conditional inclusion
probabilities are at most $m_*$. By~\eqref{eq:ext-lovasz} and concavity
along the ray from zero,
$F_k(m)\ge(1-m_*)F_k(\alpha z)\ge(1-m_*)\alpha F_k(z)$.
Take $\gamma$ sufficiently close to one and then $p$ sufficiently
small. The sampling algorithm and its accuracy are those in
Appendix~\ref{sec:gibbs-sampling-audit}.
\end{proof}

\paragraph{Recovering the recorded value.}
Let $T$ be an initial matching and let $z(T)$ be a diffuse
fractional matching on the remaining vertices.
Set $k(A)=f(T\cup A)$.
The history calculation in Lemma~\ref{lem:main-history} gives
\[
 F_k(z(T))\ge f_0+g(T)+\widehat w\cdot z(T).
\]
Indeed, the telescoping identity is unchanged and
$w_e-w_e^T\ge0$ for every stored edge. When $e\in T$ this uses
$w_e^T=0$ and the positive acceptance criterion on an accurate
transcript. When $e\notin T$ it follows from submodularity.

\paragraph{Finite streaming implementation.}
Use the guessing and comparison grids of
Appendix~\ref{sec:finite-joint-streaming-wrapper} with scale
$B=\max\{f_0,\max_e f(\{e\})\}$, and retain both the empty set
and the best singleton. Initialize the running scale at $f_0$ and
open the corresponding guesses before the first arrival if $f_0>0$.
If $B=0$, nonnegativity and submodularity imply the objective is zero. For a guess $\tau$, every sample
marginal is in $[-(n^2+1)\tau,\tau]$, since on the queried ground
set $f(A)\le f_0+|A|\tau$. The estimator uses the same lower
rounding and clipping at zero, with the sample count and product
probability precision increased polynomially for this range.
On the simultaneous accuracy event it gives
$\widehat w_e\le w_e$ on accepted edges, and
$w_e\le\widehat w_e+3d_\tau$ on all edges, including those with
negative true marginal. The deterministic upper bound
$\widehat w_e\le\tau$ still holds. Thus the deterministic space
analysis in Appendix~\ref{sec:joint-source-foundations} applies.
Ignoring the prefix loses at most the sum of its optimum singleton
values. The charge, upper rounding, and streaming sampling errors
in that appendix add at most $6\varepsilon_0\tau$ to the right
side of~\eqref{eq:ext-charge}; also
$0\le h_0-f_0-G\le3\varepsilon_0\tau$.

\subsection{Nonmonotone matching}
\label{sec:nm-two-layers}

We prove the result first for ordinary matching. The capacitated
construction below uses the same optimization and rounding
parameters. Keep the signed history, charging, and diffuse rounding
from Appendix~\ref{sec:ext-nonmonotone}. Here we use two
independent rounding layers and a fractional optimizer based on
a linear comparison with the retained optimum. The streaming
certificate uses separate coefficients for $G$ and $P$.

After the pass, we optimize over the retained edges and use the resulting
fractional solution to sample two Shearer matchings independently. We keep
a sampled copy if it precedes every conflicting copy in the priority order
and its estimated marginal gain against the earlier disjoint copies is
positive. We reserve every vertex touched by either sample, including the
endpoints of rejected copies. On the free vertices, we rescale and alter
the remaining fractional solution and use Gibbs rounding to complete the
matching. We repeat the rounding independently and include the empty set
and the best singleton in the final comparison.

\paragraph{The joint constraints and the linear certificate.}
Write $D_0$ for the certified degree from
Appendix~\ref{sec:joint-exact-parameters}. Fix a degree $d>D_0$
as in Appendix~\ref{sec:nm-asymmetric-certificate} and set
\[
 \theta_1=D_0/d,\qquad \theta_2=1-\theta_1,\qquad
 s=d-D_0.
\]
We use the rational lower bound $\gamma$ on priority-rounding
retention specified in Appendix~\ref{sec:nm-asymmetric-certificate}.
The coefficients needed from the linear certificate are
\[
 \beta=\frac{\gamma d}{(1+d/2)^2},\qquad
 \alpha=\beta(1+d/4).
\]
For a vector $Y$ on the retained edges let $\ell_v=Y(\delta(v))$.
Use one diffuse vector $x$ and the nonnegative packing region
\begin{equation}\label{eq:nm-two-packing}
 \ell_v\le d,\quad Y_e\le d,\qquad
 x(\delta(v))+B_v\ell_v\le A_v,\qquad
 \frac{x_{uv}}p+B_e(\ell_u+\ell_v)\le A_e .
\end{equation}
Here the constants $A_v,B_v,A_e,B_e$ are the same at all vertices
and edges, respectively. Put $a=\theta_1\theta_2$.
They are specified by two tangents to $Q(L)=1-L+aL^2$:
\begin{equation}\label{eq:nm-cap-tangents}
 A_v=1-a t_v^2,\quad B_v=1-2a t_v,\qquad
 A_e=1-a t_e^2,\quad B_e=1-2a t_e .
\end{equation}
We choose rational tangent points in
Appendix~\ref{sec:nm-asymmetric-certificate}.
In particular, the upper bounds on $x$ in~\eqref{eq:nm-two-packing}
are no greater than the corresponding values of $Q$.
Define $\mathcal L(x)=\widehat w\cdot x$. For binary flags
$i_v=\mathbf1[v\in V(O)]$, let $W(O)$ maximize $\mathcal L$
when $\ell_v=d i_v$. Write $r_i=A_v-B_vdi$ and
$k_{il}=A_e-B_e d(i+l)$. The same dual argument as in
Lemma~\ref{lem:main-certificate} gives
\begin{equation}\label{eq:nm-asymmetric-source}
 W(O)\ge\alpha G+\beta P
\end{equation}
from the local inequality
\begin{equation}\label{eq:nm-asymmetric-local}
 r_i A_j' H_i^\star(u)+r_l B_j' H_l^\star(v)
 +k_{il}(1-cu-(1-c)v)_+
 \ge \alpha+\beta[(1-i)A_j'+(1-l)B_j'].
\end{equation}
Here $c$ is the high endpoint's fraction of the potential sum,
and $A_j',B_j'$ are the update rates in its interval.
Multiply the integral vertex potentials by $r_i$ and the edge
slack by $pk_{il}$ in the dual objective of the linear program.
Integrating the constant $\alpha$ over each update pays
$\alpha G$. Integrating the terms weighted by $\beta$ pays
$\beta P$ at unflagged endpoints.
Appendix~\ref{sec:nm-asymmetric-certificate} gives the functions
$H_i^\star$, checks the integral budget, and verifies
\eqref{eq:nm-asymmetric-local} with a strict margin.
It also specifies the positive multipliers $\lambda_j$ of the
original rates, so $A_j'=\lambda_j A_j$ and $B_j'=\lambda_j B_j$.

\paragraph{Fractional optimization by a linear comparison.}
The union of the two independent nonlinear layers, before removing
conflicts, has marginal
$m_e(Y_e)=1-(1-\theta_1Y_e)(1-\theta_2Y_e)$ on an ordinary edge.
This map satisfies $m_e(0)=0$ and $m'_e\ge1-m_e$ on $[0,d]$.
The capacitated analogue below satisfies the same inequalities.

Use a feasible state $(x,du)$ in
\eqref{eq:nm-two-packing}, so $u\in[0,1]^S$, and put $c=1/d$.
The action uses nonlinear marginals
\[
 a_e=\frac{u_e}{c+u_e}.
\]
Indeed $a_e\le1-e^{-du_e}\le m_e(du_e)$, so we may round at the
unique $Y_e^*\in[0,du_e]$ with $m_e(Y_e^*)=a_e$. Decreasing the
nonlinear coordinates preserves the linear constraints.

Let $H=F_h$ be the multilinear extension of the nonnegative
submodular function $h$. We use the two coordinate chains of
Aggarwal and Lu~\cite{AggarwalLu26}. Order the edges by increasing
$\mu_e=u_e(1-u_e)/c$. Start a lower chain at $\mathbf0$ and an
upper chain at $\mathbf1$. Process coordinates in that order,
moving each to $a_e$ in both chains. Write $A_e$ and $C_e$ for
the corresponding partial derivatives of $H$ on the lower and
upper chains before moving coordinate $e$. A partial derivative
of a multilinear extension does not depend on its own coordinate.
Let $q_e=(1-a_e)A_e+a_e C_e$. Then for every $o\in[0,1]^S$,
\begin{equation}\label{eq:nm-linear-comparison}
 H(o)+cH(\mathbf0)
 \le \left(1+c+\frac1{4c}\right)H(a)
            +q\cdot(o-u).
\end{equation}
For completeness, moving $o$ to $a$ in the same coordinate order
gives a value decrease at coordinate $e$ of at most
$(1-a_e)o_e A_e-a_e(1-o_e)C_e$. This follows from the lower and
upper bounds on the intermediate derivative and the identity
$o_e-a_e=(1-a_e)o_e-a_e(1-o_e)$. Summing and adding and subtracting
$q\cdot u$, the remaining terms are
$u_e(1-a_e)A_e-a_e(1-u_e)C_e$. The lower-chain terms sum to
$c[H(a)-H(\mathbf0)]$ because $u_e(1-a_e)=c a_e$.
The upper-chain terms are the changes in $H$ along that chain
weighted by $\mu_e$, since $a_e(1-u_e)=\mu_e(1-a_e)$.
These nondecreasing weights and $H\ge0$ give an upper bound of
$\max_e\mu_e\,H(a)\le H(a)/(4c)$ by summation by parts.
Adding the terms proves~\eqref{eq:nm-linear-comparison}.

Put $\beta=\gamma/(1+c+1/(4c))$ and
$\alpha=\gamma-\beta c=\beta(1+1/(4c))$.
Consider successive feasible states of projected gradient ascent,
using the vector with linear function
\[
 (v_x,v_u)
     \longmapsto
 \mathcal L(v_x)+\beta q\cdot v_u
\]
at each state. The feasible region for $(x,u)$
is \eqref{eq:nm-two-packing} with $Y=du$.
These are linear payoffs for the ascent iterations. The usual
squared-distance projection inequality bounds the average
advantage of any one fixed comparator by an arbitrarily small
additive tolerance. Compare to an optimizer of $W(O)$ together
with $o=\mathbf1_O$. Multiply~\eqref{eq:nm-linear-comparison} by
$\beta$ and average over the states. Some state's action has
\begin{equation}\label{eq:nm-two-flow}
 \mathcal L(x)+\gamma F_g(a)
 \ge W(O)+\beta g(O)-(\alpha-\beta)h_0
\end{equation}
up to the additive tolerance. The algorithm keeps the state with
the largest estimated left side and decodes $Y^*$ for that state.

\paragraph{Rounding in priority order.}
Independently draw Shearer matchings $T_1,T_2$ at the two layer
vectors. The expected $h$-value of the union is at least $F_h(a)$,
by the Shearer comparison (RS3a) for ordinary matching and
Lemma~\ref{lem:nm-projected-shearer} for $b$-matching.
Give copies in the first layer independent uniform times in
$[0,D_0]$ and those in the second independent uniform times in
$[0,s]$. An edge may survive only if its time is earlier than
the time of every other sampled copy incident to either endpoint.
Equalities have probability zero. In the capacitated case the
endpoints here are slots.

For a sampled copy $e$ at time $t$, let $w_e(t)$ be the positive
part of the $h$-marginal of its label against the labels of all
earlier sampled copies disjoint from its endpoints. Retain a
surviving copy if this score is positive. Write $T$ for the
projected output. The sum of all scores is at least
$h(\pi(T_1\cup T_2))-h_0$ by telescoping in time order and
diminishing returns. The sum of retained scores is at most $g(T)$,
since earlier retained copies are disjoint and therefore their labels
belong to the context used for each retained score. If a label has
several disjoint copies, a later copy has zero score once the label
is in its context. Here $\pi$ is the identity on original labels
for ordinary matching.

Fix a copy and its time $t$ conditional on being drawn in one
layer. Its conditional score has the same expected value in
either layer. Indeed, conditioning on a copy being drawn leaves
the Shearer law outside its closed neighborhood unchanged by
(RS3), and the other layer is independent. The expected score
is nonincreasing in $t$: in a coupling with the same sampled
copies and times, the context outside that neighborhood increases.

Given the exact set of earlier copies outside the neighborhood
in the opposite layer, the chance of having no earlier conflict
in that layer is at least $r(z)^2$, where
$r(z)=(1+\sqrt{1-4z})/2$ and $z=\min(t,s)$ for the time schedule
above. To see this, the earlier copies in the opposite layer have
the Shearer law with their layer probabilities multiplied by the
probability that their time precedes $t$. Condition on the
observed outside copies being present, which deletes their
endpoints, then on all other outside copies being absent. The
conditional probability that the remaining neighbors of $e$ are
also absent is $Z_H/Z_{H-\{u,v\}}$ for the resulting graph $H$
and endpoints $u,v$ of $e$. This ratio is at least $r(z)^2$
by applying the deletion recurrence (RS1) twice. During
$[0,s]$ the two earlier degree caps are both $t$; after $s$
only first-layer times occur and the opposite cap is $s$.
The conditioning makes the score independent of the remaining
test against opposite-layer conflicts in the sense that the
score is determined by the two outside contexts.

For every copy, the total time density from both layers, including
the probability of drawing the copy, is proportional to two on
$[0,s]$ and to one on $(s,D_0]$. The conditional mean score and
the lower bound on survival probability are both nonincreasing
in time. Their covariance under this time measure is
nonnegative, by averaging the product of differences over two
independent times. Thus the expected surviving score is at least
the expected score times
\begin{align*}
 \gamma_{\rm pr}
 &= 1-\frac{2J(s)+(D_0-s)k(s)}{d},
   \qquad k(s)=1-r(s)^2, \\
 J(s)&=\int_0^s k(z)\,dz
     =\frac{s+s^2}{2}-\frac{1-(1-4s)^{3/2}}{12}.
\end{align*}
Summing over copies and using the two telescoping bounds gives
$\Ex{g(T)}\ge\gamma(F_h(a)-h_0)=\gamma F_g(a)$ for the rational
lower bound $\gamma\le\gamma_{\rm pr}$ in the appendix.
For this last inequality when $F_g(a)<0$, use $g(T)\ge0$, which
follows from retaining only positive scores.

\paragraph{Adding the diffuse matching.}
Reserve all vertices used by either layer,
even if some of its edges are subsequently dropped. Let $X_v$
indicate a free vertex
and put $p_v=\Ex{X_v}$, $q_{uv}=\Ex{X_uX_v}$.
Independence between layers gives $p_v=Q(\ell_v)$ and
$q_{uv}\ge(1-\theta_1 L)(1-\theta_2 L)=Q(L)$ for
$L=\ell_u+\ell_v$, by the union bound within each layer.
It follows from~\eqref{eq:nm-two-packing} that
$x(\delta(v))\le p_v$ and $x_{uv}/q_{uv}\le p$.
For the degree alteration use $Z_u=\sum_{uv} (x_{uv}/q_{uv})X_v$
with mean $\mu_u$. If $X_v^j$ is its corresponding free
indicator in layer $j$, write $p_v^j=\Ex{X_v^j}$, so $X_v=X_v^1 X_v^2$.
As in (GJ2),
$\operatorname{Cov}(X_v^j,X_w^j)\le y^j_{vw}p_v^j$ for
distinct vertices. Expanding the difference of the products
of the two joint moments gives
$\operatorname{Cov}(X_v,X_w)\le (y^1_{vw}+y^2_{vw})p_v$.
Also $\operatorname{Var}(X_v)\le d p_v$.
These bounds imply
$\operatorname{Var}(Z_u)\le 2pd\mu_u$ since the graph is simple
and the summed marginal degrees of the two layers are at most $d$.
Finally $\Ex{X_u Z_u}=x(\delta(u))\le p_u$ implies
\[
 p_u(\mu_u-1)\le
 \sqrt{p_u(1-p_u)\operatorname{Var}(Z_u)} .
\]
Since $p_u\ge1-d$, this gives $\mu_u\le1+O(\sqrt{p})$.
Apply the alteration of Lemma~\ref{lem:main-round}, with the
two-layer free indicators, to obtain a diffuse vector $z$ on
free vertices with $\Ex{z_e}\ge(1-\xi)x_e$ for any prescribed
$\xi>0$ by taking $p$ sufficiently small. In particular, the
overload probability at each endpoint is $O(p/\xi^2)$ and
$q_{uv}\ge1-2d$ bounds the cost of the alteration per coordinate.

Round the altered diffuse vector by Lemma~\ref{lem:ext-sparse}
and add the resulting matching to $T$. The signed history
inequality from Appendix~\ref{sec:ext-nonmonotone}, the priority
estimate, and $\mathcal L(x)\le G$ give
\begin{equation}\label{eq:nm-two-combine}
 \Ex{f(M)}\ge(1-\xi)
 \left[f_0+\mathcal L(x)+\gamma F_g(a)-\xi G\right].
\end{equation}

\paragraph{Finishing the approximation and the finite algorithm.}
Combine~\eqref{eq:nm-two-flow} and~\eqref{eq:nm-asymmetric-source}.
In the ideal analysis, before rounding losses the bound inside
\eqref{eq:nm-two-combine} is at least
\[
 (1-\alpha+\beta) f_0+\beta[G+P+g(O)]
 \ge \beta[f_0+G+P+g(O)].
\]
The coefficient of $g(O)$ is the same throughout, even for
negative $g(O)$. By~\eqref{eq:ext-charge}, the last quantity is
at least $\beta(1-p)\OPT$. The strict slack in the checked
local inequalities allows $\xi G$ to be absorbed by choosing
a sufficiently small constant $\xi$. This follows by retaining,
for example, half the minimum checked slack as an extra
coefficient of $G$ in~\eqref{eq:nm-asymmetric-source}.

The finite streaming implementation in
Appendix~\ref{sec:ext-nonmonotone} applies with the rescaled rates.
For the postprocessing, scale the $x$ coordinates by $pA_e$
so the feasible region lies in a unit cube and the normalized
packing rows are in $[0,1]$. The $u$ coordinates already lie in $[0,1]$.
Dividing the linear payoff by the singleton scale bounds its
coordinates by a polynomial in $n$. Estimate each $A_e$ or $C_e$ by product-sample
marginal queries at its corresponding chain prefix. The action's value is estimated by product sampling
from $a$. All quantities are bounded by a polynomial times the
singleton scale, so polynomially many queries and a union bound
give inverse-polynomial additive errors simultaneously, including
when the queries depend on prior estimates.

We detail the projection implementation to account for space.
Let $K$ be the normalized packing region and $z_t$ the current
feasible state. With step size $\eta$, project
$z_t+\eta\widehat g_t$ to $K$, where $\widehat g_t$ is the vector of
normalized estimated payoff coefficients. Approximate this projection
by minimizing the squared distance on $K$ with conditional gradient
steps. Each step uses the linear packing oracle of
Appendix~\ref{sec:linear-space-positive-packing}. A direction
can set every coordinate with a negative objective coefficient
to zero because $K$ is downward closed.
Round each new state downward on a fixed dyadic grid to keep it
feasible. If the fixed step is $\lambda$, the projection objective
gap after one step is at most $(1-\lambda)$ times the old gap
plus $O(\lambda^2\operatorname{diam}(K)^2)$, the linear oracle
error times $\lambda$, and the grid error times a polynomial.
The inequality follows from the quadratic expansion of the distance
function. Taking $\lambda$ and all errors inverse-polynomial and
iterating polynomially many times gives any prescribed
inverse-polynomial objective error. Strong convexity then gives
inverse-polynomial Euclidean error in the projection.
For the outer iterations the squared-distance projection
inequality gives average regret at most
\[
 \frac{\operatorname{diam}(K)^2}{2\eta T}
 +\frac{\eta L^2}{2}
 +O\left(\frac{\operatorname{diam}(K)\epsilon_{\rm proj}}{\eta}
          +\operatorname{diam}(K)\epsilon_{\rm grad}\right)
\]
after $T$ steps, where $L$ bounds the norm of the estimated vector
and $\epsilon_{\rm proj},\epsilon_{\rm grad}$ bound the respective
Euclidean errors. Choose $T$ polynomially large, $\eta$ proportional
to $T^{-1/2}/L$, and the errors polynomially small. State vectors,
gradients, and projection iterates use linear working space and
grid indices of logarithmic bit length. If packing coefficients
are rounded upward because of capacities, the comparator may be
scaled down as in the capacitated proof, losing only an
inverse-polynomial additive amount.

The action probabilities can be decoded using binary search on
a common grid because $m_e'$ is bounded below by $1-d$ and above
by one. In the capacitated case use lower and upper evaluations
of the probability polynomial. Round the decoded value downward,
so it remains at most $du_e$; the error in $H$ is controlled by
the same derivative bound. Keeping the best estimated action
introduces only another additive estimation loss.

For priority rounding, sample times on a common dyadic grid.
Coupling to continuous times changes the ordering only if two
times share a grid cell. There are polynomially many potential
copies and the time densities are constant, so inverse-polynomial
precision makes this error arbitrarily small. Estimate the score
marginals to inverse-polynomial additive accuracy relative to the
guess. Keeping surviving copies with positive estimated score
can lose at most the marginal error per copy, by the same
diminishing returns comparison. A separate inverse-polynomial
failure allowance covers inaccurate estimates within a trial.
Feasible values are polynomially bounded relative to $B$, so
all these errors can be made arbitrarily small relative to $B$.
The strict numerical margin in
Appendix~\ref{sec:nm-asymmetric-certificate} allows the
guarantee in expectation and the same threshold with high
probability by independent postprocessing trials.

\subsection{Nonmonotone \texorpdfstring{$b$}{b}-matching}
\label{sec:nm-direct-b}

We first establish that projecting a Shearer matching to edge labels
preserves its submodular comparison at the \emph{projected}
marginals. This will let us use several copies of a stored edge
while evaluating the objective only on original edges.

\begin{lemma}[Shearer comparison after projection]\label{lem:nm-projected-shearer}
Let $M$ have a Shearer distribution in the interior of its region of
positivity, and partition its coordinates into fibers with distinct
labels. Let $\pi(M)$ be the set of labels with some coordinate in $M$,
and put $m_e=\Prb(e\in\pi(M))$. Then every submodular function $h$ on
the labels satisfies
\[
 \Ex{h(\pi(M))}\ge F_h(m).
\]
The same conclusion holds for the union of the label projections of
independent Shearer matchings.
\end{lemma}
\begin{proof}
Use $\preceq$ for stochastic domination with respect to set
inclusion. For any set of forbidden coordinates $B$ and a coordinate
$i\notin B$, we claim that, on the other coordinates outside $B$,
\[
 \mathcal L(M\mid B\cap M=\varnothing,\ i\in M)
 \preceq \mathcal L(M\mid B\cap M=\varnothing)
 \preceq \mathcal L(M\mid B\cap M=\varnothing,\ i\notin M).
\]
The restriction to those other coordinates is implicit. We omit
zero-marginal coordinates, and all absence events used here have
positive probability.

Induct on $|B|$. By the Shearer inclusion moments (RS3),
conditioning on $i\in M$ leaves the unconditioned law outside
$N[i]$ and removes all neighbors $N(i)$ of $i$ in the conflict
graph. After imposing $B\cap M=\varnothing$, the retained part
outside $N[i]$ therefore has the law of the unconditioned sample
restricted outside $N[i]$, conditioned only on the absence of
$B'=B\setminus N(i)$. If $B'=B$, this is obtained by deletion
from the law conditioned on the absence of $B$. If $B'\ne B$,
forbid the coordinates of $B\setminus B'$ one at a time. At each
step the conditioning set has size smaller than $|B|$, so the
inductive claim shows that this can only increase the law on
coordinates outside $B\cup N[i]$. After reinserting the
coordinates in $N(i)\setminus B$, which are zero in the law with
$i$ included, this proves the first comparison in either case.
The second comparison then follows because the middle law is
a mixture of the laws with $i$ included and excluded. This
also covers $B=\varnothing$.

Fix a label $e$ and order its fiber. Conditional on $e\in\pi(M)$,
disintegrate by the first selected coordinate $i$ in that order.
This specifies that the earlier coordinates of the fiber are absent
and $i$ is present. The claim shows that, on the other fibers, this
law is dominated by conditioning only on the earlier absences.
Forbidding the rest of the fiber one coordinate at a time can only
increase the law on those other fibers. Hence the law conditioned
on $e\in\pi(M)$ is dominated on other labels by the law conditioned
on $e\notin\pi(M)$. A coordinate hybrid from the independent
Bernoulli sample at marginals $m$ to $\pi(M)$ now proves the
inequality. In each hybrid step, conditioning the projected
prefix on inclusion of the current label decreases the prefix
and thereby increases the current marginal of $h$ by submodularity.
The independent suffix is the same in both terms. For the last
assertion, the independent Shearer matchings together are the
Shearer distribution on the disjoint union of their conflict
graphs. Group the copies of each label across all layers.
\end{proof}

\paragraph{Streaming on slots.}
Discard edges with a zero-capacity endpoint and cap each other
capacity at $n-1$. Give vertex $v$ slots $1,\ldots,b_v$, initially
with zero potentials. On arrival of $e=uv$, find the slot of least
potential at each endpoint with ties broken by index. Run the same
streaming test and update on these two slots. Upon acceptance store
one copy, called the assigned copy of $e$, with its original label
and weight. The history sample includes each retained original
edge independently with probability $p$. In particular, its
ground set is the set $S$ of original labels, with each label
appearing once. Slots without positive potentials are implicit and
the smallest unused index can be found by a scan.

For a $b$-matching of original edges, let $O$ consist of its
retained labels and put $I_v=|O\cap\delta(v)|/b_v$. Let
$\psi_{v,a}$ be the cumulative ideal increment to slot $(v,a)$.
In the comparison with a fixed optimal $b$-matching, each
unretained optimum edge at $v$ is charged at most the final
minimum slot potential there, apart from the streaming
allowances. There are at most $b_v-|O\cap\delta(v)|$ such
incidences. The minimum is at most the average of the slot
potentials, so the charge to ideal increments is at most
\[
 P_b(O)=\sum_v(1-I_v)\sum_{a=1}^{b_v}\psi_{v,a}.
\]
The augmentation argument from
Appendix~\ref{sec:ext-nonmonotone} thus gives
$(1-p)\OPT\le f_0+G+P_b(O)+g(O)$, up to its finite additive
allowance.

Define $W^b(O)$ as the optimum of $\mathcal L$ on assigned
copies when the load at each slot of $v$ is $dI_v$ in
\eqref{eq:nm-two-packing}.
At each vertex independently inject the incidences of $O$ into
its slots uniformly. For every resulting injection, the source
certificate applies with the binary flags given by the occupied
slots. Its local inequality~\eqref{eq:nm-asymmetric-local}
depends only on these flags and on the updates of the assigned
copies, so the injection need not agree with the assigned
copies of $O$. Averaging the comparator vectors over the injections gives
the affine bounds defining $W^b(O)$.
The average of the unoccupied-slot potential charge is $P_b(O)$.
Consequently $W^b(O)\ge\alpha G+\beta P_b(O)$, with the same
slack in the $G$ coefficient as in the ordinary comparison.

\paragraph{The fractional solution and its projected probabilities.}
Use the diffuse vector on assigned copies and $Y$ on the
original labels, with total degree bound $d$ for the two layers.
The load of each slot of $v$ is
$\ell_v(Y)=\sum_{e\ni v}Y_e/b_v$. In the joint packing constraints,
use these loads in place of the ordinary vertex loads. Also impose
$Y_e\le d$. The comparator uses $d\mathbf1_O$ on the labels and
the linear comparison vector on the assigned copies. The load of
$d\mathbf1_O$ at each slot of $v$ is $dI_v$. Thus the same
comparison of degree and coordinate bounds as in ordinary matching
applies with the averaged flags $I_v$.

The two independent Shearer layers use proportions $\theta_j$ of
$Y$, with $\theta_1+\theta_2=1$. For $e=uv$ put $a=b_u$, $b=b_v$.
In layer $j$, every pair of slots over $e$ has marginal
$\theta_jY_e/(ab)$. Thus the slot-degree bound of the layer is
at most $\theta_j d$. Write
\[
 Z_{a,b}(s)=\sum_{k=0}^{\min(a,b)}
       (-1)^k\binom ak\binom bk k!s^k .
\]
The probability of no $e$-copy in layer $j$ is
$Z_{a,b}(\theta_jY_e/(ab))$ by (RS3) and inclusion--exclusion.
For the projection of the union before removing conflicts, the
marginal of $e$ is consequently
\[
 m_e(Y_e)=1-\prod_{j=1}^2 Z_{a,b}(\theta_jY_e/(ab)).
\]
The derivative of $Z_{a,b}(Y/(ab))$ with respect to $Y$ is
$-Z_{a-1,b-1}(Y/(ab))$, by the sum formula or by deleting the
endpoints of a copy. Under the Shearer law, the probability of
avoiding the full fiber is at most the probability of avoiding
only the copies disjoint from this pair of endpoints. The product
rule thus gives $m_e'(Y_e)\ge 1-m_e(Y_e)\ge1-d$.
It also gives $|m'_e|,|m''_e|\le1$, as the corresponding
avoidance probabilities lie in $[0,1]$ and
$\sum_j\theta_j=1$.

Consequently the same decoding step finds the layer mass
$Y^*_e\le du_e$ for each label whose union marginal is
$a_e=u_e/(1/d+u_e)$. By Lemma~\ref{lem:nm-projected-shearer},
the unaltered union of the label projections has the required
comparison to its independent multilinear value.

\paragraph{Reserving slots for the linear term.}
Reserve all slots occupied in either layer. The resulting
slot matching projects to a $b$-matching. A label
can occur in both layers or in several disjoint copies in a
layer, and projection counts it only once. In the diffuse step
there is exactly one assigned copy per original label. Hence
for each fixed projected first-stage set $T$, the objective on
the diffuse support is $U\mapsto f(T\cup\pi(U))$, a nonnegative
submodular function. The history inequality applies to the
same original labels even if $T$ intersects this support.

Each layer's matching is on a simple graph of slots with the
same degree bound as in ordinary matching. Thus the conflict
bound, the product occupancy bounds, and the alteration apply
with the slots as vertices. The signed Gibbs step uses only
the assigned copies and the projected objective above.

\paragraph{Finite implementation.}
The complete slot graph is implicit throughout. Its edges can
be enumerated in original-edge order and then slot-pair order.
In the Shearer sampler the current rejection graph consists
of the edges earlier in this order whose endpoints are unused
by the current partial matching. A prefix index, the list of
matched slots, and the recursion's exclusion stack therefore
replace a materialized rejection graph. At most $n^2|S|$
implicit edges need scanning in any one layer. If the partial
matching reaches a cap $K$, abort the sample. The inclusion
moments give
\[
 \Prb(|T^j|\ge K)\le
 \frac{(\sum_e\theta_jY_e)^K}{K!}
 \le \left(\frac{\mathrm e\,d|S|}{K}\right)^K .
\]
A cap $K=O(|S|+\log(1/\delta))$ bounds this event by any
prescribed $\delta>0$. Independently truncate every
$\operatorname{Coin}$ call as in (RS7) and use a union bound
over the polynomial number of implicit edges. This maintains
a worst-case recursion stack of $O(\log((n+1)/\delta))$ words.
Neighbor choices use dyadic probabilities obtained by scanning
the incident waiting copies, rounding each probability
downward, and reserving the residual probability for a null
choice. With inverse-polynomial grid width, the sum of
per-choice errors over all implicit scans and truncated calls
is at most $\delta$ after adjusting the polynomial precision.
Consequently each approximate sample has arbitrarily small
inverse-polynomial total variation error and uses
$O(|S|+\log((n+1)/\delta))$ words on every transcript.

The other postprocessing steps also require no dense slot
arrays. Store $Y$ by label and $x$ by assigned copy. For example,
when $i$ is a slot of $u$ and $k$ is a slot of $v\ne u$,
the common-neighbor term for layer $j$ in (GJ2) equals
\[
 t^j_{ik}=\frac{\theta_j^2}{b_ub_v}
        \sum_{w\ne u,v}\frac{Y_{uw}Y_{vw}}{b_w},
\]
with zero coordinates for absent original edges. This and the
individual slot loads can be computed by scans of the retained
labels. Compute upper approximations of the pair-availability
denominators $q_{ik}$ on an inverse-polynomial grid and round
the corresponding diffuse coordinate divided by $q_{ik}$
conservatively downward. Since the exact denominators are at least
$1-2d$ and the approximations can be made arbitrarily close, this preserves the cap $p$ and
changes the weighted alteration bound by arbitrarily small
additive error relative to the active guess. Only slots
incident to assigned copies need the alteration load array.

The packing matrix has $O(|S|+1)$ rows and columns after
discarding rows for unassigned slots that repeat identical
original-vertex degree bounds. Its coefficients involving
$b_v^{-1}$ need not share an exact polynomial-size denominator.
Round positive coefficients upwards on a common
inverse-polynomial dyadic grid. With $N$ columns and entry
error $\epsilon_{\rm row}$, dividing any feasible vector by
$1+N\epsilon_{\rm row}$ makes it feasible for the rounded matrix
after normalization of the rows. This changes its linear payoff
by at most an inverse-polynomial additive amount when
$\epsilon_{\rm row}$ is sufficiently small. The avoidance
polynomials and their derivatives can be estimated on the
same type of grid. Successive unsigned terms in $Z_{a,b}(s)$
satisfy the recurrence
$T_{k+1}=T_k(a-k)(b-k)s/(k+1)$, starting from $T_0=1$, and
are bounded by $d^k/k!$ for all needed parameters.
The derivative lower bound $m'_e\ge1-d$ ensures that fixed-grid
evaluations give the desired accuracy in the decoding step.
Projection and all oracle estimates operate on the original
labels and assigned copies.

Finally, each slot stores $O(\log n)$ assigned incidences per
guess by the potential-growth bound. If $K^*$ is a
maximum-cardinality original $b$-matching, every edge outside
$K^*$ meets a vertex saturated by $K^*$, and those vertices
have at most $2\nu_b$ slots in total. The stored labels outside
$K^*$ charge to these slots, while at most $\nu_b$ stored
labels lie in $K^*$. This gives $O((\nu_b+1)\log n)$ words
per guess, deterministically. Using $n^2$ as an upper bound
on the virtual vertex count in the guessing and precision
schedules leaves $O((\nu_b+1)\log^2 n)$ total space.
Every remaining singleton and the empty set are feasible,
and all feasible values are at most $(n^2+1)B$, so the
finite errors are absorbed by the same singleton scale.
There is no new randomness in streaming beyond the estimates.
The independent final rounding trials and probability
amplification therefore use the same space as in ordinary
matching.

\subsection{Exact verification of the nonmonotone factor}
\label{sec:nm-asymmetric-certificate}

In Appendix~\ref{sec:nm-two-layers} use $d=\NMDegree$,
$t_v=\NMVertexTangent\,d$ and $t_e=\NMEdgeTangent\,d$.
Both layer degrees $D_0$ and $d-D_0$ are strictly below $1/4$.
To certify a lower bound on $\gamma_{\rm pr}$, set $\Delta=1-4s$
and enclose $\sqrt{\Delta}$ between adjacent multiples $r_-,r_+$
of $10^{-12}$ using integer square root. In the priority-rounding loss
$k(s)=\tfrac12+s-\sqrt{\Delta}/2$, use $r_-$; in its integral
$J(s)=(s+s^2)/2-(1-\Delta\sqrt{\Delta})/12$, use $r_+$.
This gives the rational upper bounds
\[
 k_+=\tfrac12+s-r_-/2,\qquad
 J_+=(s+s^2)/2-(1-\Delta r_+)/12.
\]
Since $0<s\le D_0$, the retention coefficient
$\gamma=1-[2J_++(D_0-s)k_+]/d$ satisfies
$\gamma\le\gamma_{\rm pr}$.

The ancillary file \texttt{nonmonotone\_profiles.json} gives the
rational knot positions and heights of $H_0^\star,H_1^\star$
and a table of 64 positive rational rate multipliers.
Number the $J$ original rate intervals from $0$ to $J-1$.
On interval $j$, multiply both rates by the table entry
of index $\lfloor 64j/J\rfloor$. Multiplying both by a
positive number preserves~\eqref{eq:source-drift}.

The ancillary driver \texttt{check\_nonmonotone.py} and its C++ checker
verify the local inequalities, the integrals, and the ratio. Run the second
command in Appendix~\ref{sec:reproduce-certificates}.

The driver verifies that both integrals
$\int_0^\infty H_i^\star(u)u^{-2}\,du$ are at most $0.999990001$.
On each affine segment the integral equals a rational contribution
and a nonnegative slope times a logarithm of a ratio of knots.
The driver bounds this logarithm above using the series for
$2\operatorname{arctanh}((z-1)/(z+1))$ with the geometric
tail bound, after reducing the ratio to $[1,2]$.
Choosing $p$ sufficiently small therefore gives
\eqref{eq:main-integral} under the new rates.

The check of the local inequalities uses a grid of denominator
$Q=10^8$.
\begin{enumerate}
\item For the local check, drop knots beyond $1000$. Round remaining
knot positions up and heights down, interpolate linearly, and extend
the last retained height constantly. These modified
profiles are pointwise lower bounds on the true profiles, as
the corresponding point in the original affine segment has the
same interpolation fraction and lies weakly to the left.
The driver checks that the rounded knots are strictly increasing.
\item For each rate interval separately, round all left-side
coefficients down and the rates on the right up. Round the
enclosing endpoints of the ratio interval outwards. For fixed
$u,v$, the positive part is monotone as a function of the endpoint
ratio, so the minimum over the enlarged interval occurs at an
endpoint. The right-side constants $\alpha,\beta$ are rational
and are rounded up on the same grid.
\item In the retained rectangle, check the horizontal and vertical
grid vertices at or below the line $cu+(1-c)v=1$ and the
intersections of the line with the grid. The expression is affine
on each such cell, and above the line it is coordinatewise
nondecreasing. The checker uses prefix minima for the grid
vertices and rounds the interpolated heights at line intersections
down. Outside the rectangle the single last height of either
profile suffices by itself, as checked separately.
\end{enumerate}
The checker subtracts an upper bound on the right side
of~\eqref{eq:nm-asymmetric-local} from each local lower bound,
using integer units of $Q^{-3}$. In order
$(0,0),(0,1),(1,0),(1,1)$, the lower slack bounds are at least
\[
 10^{-6}\NMSlacks.
\]
With the two-sided square-root enclosure above, the driver verifies
$1/\beta<4.726494<\NMFactor$, leaving a strict margin for the finite losses.


\begin{thebibliography}{FNFSZ23}

\bibitem[AASV21]{AASV}
Yeganeh Alimohammadi, Nima Anari, Kirankumar Shiragur, and Thuy-Duong Vuong.
\newblock Fractionally log-concave and sector-stable polynomials: Counting
  planar matchings and more.
\newblock In {\em Proceedings of the 53rd Annual ACM SIGACT Symposium on Theory
  of Computing (STOC 2021)}, pages 433--446. ACM, 2021.

\bibitem[ABK{\etalchar{+}}26]{ABKNS2026}
Sepehr Assadi, Soheil Behnezhad, Christian Konrad, Kheeran Naidu, and Janani
  Sundaresan.
\newblock Settling the pass complexity of approximate matchings in dynamic
  graph streams.
\newblock {\em Journal of the ACM}, 73(4), 2026.

\bibitem[AL26]{AggarwalLu26}
Vaneet Aggarwal and Yiyang Lu.
\newblock Geometry-dependent online and offline approximation for
  non-monotone {DR}-submodular maximization.
\newblock arXiv:2610.00545v3, 2026.

\bibitem[AJX26a]{ajx1}
Sepehr Assadi, Max Jiang, and Mars Xiang.
\newblock Semi-streaming matching in a single pass {I}: A new framework for
  lower bounds via blueprints.
\newblock In {\em Proceedings of the 58th Annual ACM Symposium on Theory of
  Computing (STOC 2026)}. ACM, 2026.
\newblock Full version: arXiv:2607.14644v1.

\bibitem[AJX26b]{ajx2}
Sepehr Assadi, Max Jiang, and Mars Xiang.
\newblock Semi-streaming matching in a single pass {II}: Greedy is optimal,
  2026.
\newblock arXiv:2607.14656v2.

\bibitem[BBG{\etalchar{+}}25]{DynamicMSM}
Kiarash Banihashem, Leyla Biabani, Samira Goudarzi, MohammadTaghi Hajiaghayi,
  Peyman Jabbarzade, and Morteza Monemizadeh.
\newblock Dynamic algorithms for submodular matching.
\newblock In {\em 52nd International Colloquium on Automata, Languages, and
  Programming (ICALP 2025)}, volume 334 of {\em Leibniz International
  Proceedings in Informatics (LIPIcs)}, pages 19:1--19:21. Schloss Dagstuhl --
  Leibniz-Zentrum f{\"u}r Informatik, 2025.

\bibitem[BFLS26]{RandomOrder2026}
Niv Buchbinder, Moran Feldman, Siyue Liu, and Sherry Sarkar.
\newblock Semi-streaming algorithms for submodular maximization under random
  arrival order, 2026.
\newblock arXiv:2605.14296.

\bibitem[BZ22]{bz}
Simon Bruggmann and Rico Zenklusen.
\newblock An optimal monotone contention resolution scheme for bipartite
  matchings via a polyhedral viewpoint.
\newblock {\em Mathematical Programming}, 191:795--845, 2022.

\bibitem[CCPV11]{CCPV2011}
Gruia Calinescu, Chandra Chekuri, Martin P{\'a}l, and Jan Vondr{\'a}k.
\newblock Maximizing a monotone submodular function subject to a matroid
  constraint.
\newblock {\em SIAM Journal on Computing}, 40(6):1740--1766, 2011.

\bibitem[CGQ15]{CGQ}
Chandra Chekuri, Shalmoli Gupta, and Kent Quanrud.
\newblock Streaming algorithms for submodular function maximization.
\newblock In {\em 42nd International Colloquium on Automata, Languages, and
  Programming (ICALP)}, volume 9134 of {\em Lecture Notes in Computer Science},
  pages 318--330. Springer, 2015.

\bibitem[CK15]{CK}
Amit Chakrabarti and Sagar Kale.
\newblock Submodular maximization meets streaming: Matchings, matroids, and
  more.
\newblock {\em Mathematical Programming}, 154(1--2):225--247, 2015.
\newblock Preliminary version in IPCO 2014.

\bibitem[CLV26]{CLV2026}
Zongchen Chen, Kuikui Liu, and Eric Vigoda.
\newblock Optimal mixing of {Glauber} dynamics: Entropy factorization via
  high-dimensional expansion.
\newblock {\em SIAM Journal on Computing}, 55(4):STOC21--104--STOC21--153,
  2026.
\newblock First published electronically in 2023.

\bibitem[CVZ14]{CVZ}
Chandra Chekuri, Jan Vondr{\'a}k, and Rico Zenklusen.
\newblock Submodular function maximization via the multilinear relaxation and
  contention resolution schemes.
\newblock {\em SIAM Journal on Computing}, 43(6):1831--1879, 2014.

\bibitem[Fel21]{FeldmanDistorted}
Moran Feldman.
\newblock Guess free maximization of submodular and linear sums.
\newblock {\em Algorithmica}, 83(3):853--878, 2021.
\newblock Preliminary version in WADS 2019; correction in Algorithmica 84,
  3101--3102 (2022).

\bibitem[FKK18]{FKK}
Moran Feldman, Amin Karbasi, and Ehsan Kazemi.
\newblock Do less, get more: Streaming submodular maximization with
  subsampling.
\newblock In {\em Advances in Neural Information Processing Systems},
  volume~31, 2018.

\bibitem[FKM{\etalchar{+}}05]{FKMSZ}
Joan Feigenbaum, Sampath Kannan, Andrew McGregor, Siddharth Suri, and Jian
  Zhang.
\newblock On graph problems in a semi-streaming model.
\newblock {\em Theoretical Computer Science}, 348(2--3):207--216, 2005.

\bibitem[FNFSZ22]{fnsz}
Moran Feldman, Ashkan Norouzi-Fard, Ola Svensson, and Rico Zenklusen.
\newblock Submodular maximization subject to matroid intersection on the fly.
\newblock In {\em 30th Annual European Symposium on Algorithms (ESA 2022)},
  volume 244 of {\em Leibniz International Proceedings in Informatics
  (LIPIcs)}, pages 52:1--52:14. Schloss Dagstuhl -- Leibniz-Zentrum f{\"u}r
  Informatik, 2022.
\newblock Full version: arXiv:2204.05154.

\bibitem[FNFSZ23]{FNSZ2023}
Moran Feldman, Ashkan Norouzi-Fard, Ola Svensson, and Rico Zenklusen.
\newblock The one-way communication complexity of submodular maximization with
  applications to streaming and robustness.
\newblock {\em Journal of the ACM}, 70(4):24:1--24:52, 2023.

\bibitem[FNSW11]{FNSW}
Moran Feldman, Joseph~(Seffi) Naor, Roy Schwartz, and Justin Ward.
\newblock Improved approximations for {$k$}-exchange systems.
\newblock In {\em 19th Annual European Symposium on Algorithms (ESA 2011)},
  volume 6942 of {\em Lecture Notes in Computer Science}, pages 784--798.
  Springer, 2011.

\bibitem[GJL19]{GJL2019}
Heng Guo, Mark Jerrum, and Jingcheng Liu.
\newblock Uniform sampling through the {Lov{\'a}sz} local lemma.
\newblock {\em Journal of the ACM}, 66(3):18:1--18:31, 2019.

\bibitem[GJS23]{GJS}
Paritosh Garg, Linus Jordan, and Ola Svensson.
\newblock Semi-streaming algorithms for submodular matroid intersection.
\newblock {\em Mathematical Programming}, 197:967--990, 2023.
\newblock Preliminary version in IPCO 2021.

\bibitem[GKPS06]{GKPS2006}
Rajiv Gandhi, Samir Khuller, Srinivasan Parthasarathy, and Aravind Srinivasan.
\newblock Dependent rounding and its applications to approximation algorithms.
\newblock {\em Journal of the ACM}, 53(3):324--360, 2006.

\bibitem[GW19]{GW}
Mohsen Ghaffari and David Wajc.
\newblock Simplified and space-optimal semi-streaming
  {$(2+\varepsilon)$}-approximate matching.
\newblock In {\em 2nd Symposium on Simplicity in Algorithms (SOSA 2019)},
  volume~69 of {\em Open Access Series in Informatics (OASIcs)}, pages
  13:1--13:8. Schloss Dagstuhl -- Leibniz-Zentrum f{\"u}r Informatik, 2019.

\bibitem[HS21]{HS}
Chien-Chung Huang and Fran{\c{c}}ois Sellier.
\newblock Semi-streaming algorithms for submodular function maximization under
  {$b$}-matching constraint.
\newblock In {\em Approximation, Randomization, and Combinatorial Optimization.
  Algorithms and Techniques (APPROX/RANDOM 2021)}, volume 207 of {\em Leibniz
  International Proceedings in Informatics (LIPIcs)}, pages 14:1--14:18.
  Schloss Dagstuhl -- Leibniz-Zentrum f{\"u}r Informatik, 2021.
\newblock The expanded arXiv version also treats matroid and matchoid
  constraints.

\bibitem[HTW20]{HTW}
Chien-Chung Huang, Theophile Thiery, and Justin Ward.
\newblock Improved multi-pass streaming algorithms for submodular maximization
  with matroid constraints.
\newblock In {\em Approximation, Randomization, and Combinatorial Optimization.
  Algorithms and Techniques (APPROX/RANDOM 2020)}, volume 176 of {\em Leibniz
  International Proceedings in Informatics (LIPIcs)}, pages 62:1--62:19.
  Schloss Dagstuhl -- Leibniz-Zentrum f{\"u}r Informatik, 2020.

\bibitem[JS89]{JS1989}
Mark Jerrum and Alistair Sinclair.
\newblock Approximating the permanent.
\newblock {\em SIAM Journal on Computing}, 18(6):1149--1178, 1989.

\bibitem[JSV04]{JSV2004}
Mark Jerrum, Alistair Sinclair, and Eric Vigoda.
\newblock A polynomial-time approximation algorithm for the permanent of a
  matrix with nonnegative entries.
\newblock {\em Journal of the ACM}, 51(4):671--697, 2004.

\bibitem[LW21]{LW}
Roie Levin and David Wajc.
\newblock Streaming submodular matching meets the primal-dual method.
\newblock In {\em Proceedings of the 2021 ACM-SIAM Symposium on Discrete
  Algorithms (SODA)}, pages 1914--1933. SIAM, 2021.

\bibitem[NV25]{NV}
Pranav Nuti and Jan Vondr{\'a}k.
\newblock Towards an optimal contention resolution scheme for matchings.
\newblock {\em Mathematical Programming}, 210:761--792, 2025.
\newblock Preliminary version in IPCO 2023.

\bibitem[PS19]{pazschwartzman}
Ami Paz and Gregory Schwartzman.
\newblock A {$(2+\epsilon)$}-approximation for maximum weight matching in the
  semi-streaming model.
\newblock {\em ACM Transactions on Algorithms}, 15(2):18:1--18:15, 2019.
\newblock First published electronically in December 2018; preliminary version
  in SODA 2017.

\bibitem[QS22]{QS}
Frederick Qiu and Sahil Singla.
\newblock Submodular dominance and applications.
\newblock In {\em Approximation, Randomization, and Combinatorial Optimization.
  Algorithms and Techniques (APPROX/RANDOM 2022)}, volume 245 of {\em Leibniz
  International Proceedings in Informatics (LIPIcs)}, pages 44:1--44:21.
  Schloss Dagstuhl -- Leibniz-Zentrum f{\"u}r Informatik, 2022.

\bibitem[She85]{Shearer}
James~B. Shearer.
\newblock On a problem of {Spencer}.
\newblock {\em Combinatorica}, 5(3):241--245, 1985.

\bibitem[Sin92]{sinclair1992}
Alistair Sinclair.
\newblock Improved bounds for mixing rates of {Markov} chains and
  multicommodity flow.
\newblock {\em Combinatorics, Probability and Computing}, 1(4):351--370, 1992.

\end{thebibliography}
\end{document}